\documentclass[preprint,11pt,nofootinbib]{revtex4-1}

	\usepackage{amssymb}
	\usepackage{amsmath}
	\usepackage{hyperref}
	\usepackage{graphicx}
	\usepackage{color}
	\usepackage{comment}
	\newcommand{\nn}{\nonumber}
	\newcommand{\lam}{\lambda}
	\newcommand{\rbra}[1]{\left(#1\right)}

	\newcommand{\re}{\textrm{Re}}
	\newcommand{\im}{\textrm{Im}}
	\newcommand{\eq}[1]{\begin{align}#1\end{align}}

\allowdisplaybreaks[4]

\begin{document}

\title{Probing double-aligned two Higgs doublet models at LHC}

\author{Shinya Kanemura}
\email{kanemu@het.phys.sci.osaka-u.ac.jp}
\author{Michihisa Takeuchi}
\email{m.takuechi@het.phys.sci.osaka-u.ac.jp}
\author{Kei Yagyu}
\email{yagyu@het.phys.sci.osaka-u.ac.jp}
\affiliation{Department of Physics, Osaka University, Toyonaka, Osaka 560-0043, Japan}


\begin{abstract}
We consider two Higgs doublet models (THDMs) with both the Higgs potential and Yukawa interactions being aligned, which we call ``double-aligned THDMs".
In this scenario, coupling constants of the discovered Higgs boson to the Standard Model (SM) particles are 
identical to those of the SM Higgs boson, and flavor changing neutral currents via neutral Higgs bosons do not appear at tree level.
We investigate current constraints and future prospects of the model by using measurements from flavor experiments and data of multi-lepton final states at LHC.
Especially, we focus on the electroweak pair production of the additional Higgs bosons with their masses below $2m_t$.
We find that the most of the parameter space are already excluded by the current LHC data 
when the leptonic decays of the additional Higgs bosons are dominant, which can be interpreted to the scenario in the Type-X THDM as a special case. 
We also clarify the parameter region where the high-luminosity LHC can explore, and demonstrate the reconstruction of the masses of additional Higgs bosons from the $b\bar{b}\tau^+\tau^-$ final states
in a few benchmark points.

\end{abstract}

\preprint{OU-HET-1116}

\maketitle


\section{Introduction}

The nature of the electroweak (EW) symmetry breaking is one of the biggest mysteries in the elementary particle physics even after the discovery of the Higgs boson at LHC. 
Although the Higgs mechanism in the standard model (SM) can parameterize the EW symmetry breaking and can explain currently measured properties of the Higgs boson, 
the true structure of the Higgs sector is still unknown as it is not determined by a fundamental principle. 
In fact, various possibilities of non-minimal structures of the Higgs sector can be considered without any contradiction with the current experimental data. 
Therefore, comprehensive studies of extended Higgs sectors are important to understand the nature of EW symmetry breaking as a bottom-up approach. 

Two Higgs doublet models (THDMs) are one of the simplest examples of the extended Higgs sector, which naturally satisfy the electroweak rho parameter to be unity at tree level. 
In addition, THDMs are well motivated, because they appear in various new physics scenarios such as supersymmetry~\cite{Haber:1984rc}, 
a neutrino-philic model~\cite{Ma:2000cc}, 
models with radiative generations of neutrino masses~\cite{Zee:1980ai,Ma:2006km,Aoki:2008av}, those with dark matter~\cite{Barbieri:2006dq}, those with new sources of CP violation (CPV) and those with 
strongly first order EW phase transitions~\cite{Kanemura:2004ch,Fromme:2006cm,Funakubo:1993jg}. 
In particular, such CPV and first order EW phase transitions can be used in the successful scenarios of the electroweak baryogenesis~\cite{Turok:1990zg,Funakubo:1993jg,Cline:1995dg,Fromme:2006cm,Modak:2018csw,Basler:2021kgq}. 
Thus, among the many possibilities of non-minimal Higgs sectors, THDMs should phenomenologically be examined with a special attention. 

It has been well known that THDMs generally give rise to flavor changing neutral currents (FCNCs) via neutral Higgs boson exchanges at tree level. 
In order to avoid such FCNCs, a natural flavor conservation scenario~\cite{Glashow:1976nt} has often been considered, where 
only one of the Higgs doublets couples to each type of charged fermions, i.e., up-type quarks, down-type quarks and charged leptons. 
This scenario can naturally be realized by introducing a discrete $Z_2$ symmetry to the Higgs sector, in which four independent type of Yukawa interactions appear~\cite{Grossman:1994jb,Aoki:2009ha,Barger:1989fj}, 
the so-called Type-I, Type-II, Type-X and Type-Y.  
As an alternative possibility to avoid tree level FCNCs, 
the Yukawa alignment scenario has also been discussed~\cite{Pich:2009sp}, in which two Yukawa matrices for each type of charged fermions are assumed to be proportional with each other. 
In this case, Yukawa couplings can generally be complex, so that we can consider two independent CPV sources in the Yukawa interactions and the Higgs potential. 

Apart from the Yukawa alignment, current LHC data show that measured properties of the discovered Higgs boson agree with those of the SM Higgs boson within the uncertainty. 
This suggests that one of the neutral Higgs bosons in the THDM, denoting $h$,  must have SM-Higgs like properties. 
There are two ways to realize such a situation, i.e., taking decoupling or alignment regions of the parameter space. 
The former corresponds to the case with all the masses of additional Higgs bosons being much higher than the EW scale, in which 
properties of $h$ approach to those of the SM Higgs boson according to the decoupling theorem~\cite{Appelquist:1974tg,Gunion:2002zf}. 
On the other hand, the latter can be realized by taking the $h$ state being aligned to be one of the states in the Higgs doublet which possesses the vacuum expectation value (VEV) same as in the SM \footnote{Such a Higgs doublet can be defined by taking the so-called Higgs basis~\cite{Davidson:2005cw} without loss of generality. }. 
Unlike the decoupling case, the scenario with the ``Higgs alignment'' provides phenomenologically interesting consequences, because the additional Higgs bosons can be at the EW scale, which is often referred as 
``alignment without decoupling''~\cite{Aiko:2020ksl}. \footnote{In this case, non-decoupling effects of additional Higgs bosons can be significant, 
which provide a sizable deviation in the triple Higgs boson coupling, see e.g., Refs.~\cite{Kanemura:2004mg,Kanemura:2015mxa,Kanemura:2021fvp,Braathen:2019pxr,Braathen:2019zoh}. 
Such a large deviation can be probed in the double Higgs boson production at future colliders~\cite{Plehn:1996wb, Baur:2002rb, Baur:2002qd, Asakawa:2010xj,Barr:2014sga, Papaefstathiou:2015iba, Cao:2016zob,Goncalves:2018qas,Biekotter:2018jzu,FCC:2018byv,FCC:2018bvk,ILC:2019gyn, Klamka:2021cjt, Han:2020pif,Chiesa:2020awd}. }
From the above mentioned reasons, THDMs with the Yukawa and Higgs alignment provide a phenomenologically interesting scenario.
We call this type of scenario as ``double-alignment''. 
Recently in Ref.~\cite{Kanemura:2020ibp}, such a double-aligned scenario has been studied, and found that sizable CPV phases can be taken without contradiction to bounds 
from current data of electric dipole moments (EDMs)~\cite{ACME:2018yjb,nEDM:2020crw} as well as flavor experiments and LHC experiments. 
See also Refs.~\cite{Cheung:2020ugr,Abe:2013qla,Jung:2013hka,Altmannshofer:2020shb,Low:2020iua} for discussions of EDMs in THDMs. 
In addition, it has been shown in Ref.~\cite{Enomoto:2021dkl} that these CPV phases can be used to realize a successful scenario of the EW baryogenesis. 
Furthermore, effects of CPV can directly be detected by measuring azimuthal angle distributions in the decay product of additional Higgs bosons~\cite{Kanemura:2021atq} at 
future lepton colliders such as the International Linear Collider (ILC)~\cite{Baer:2013cma,Asai:2017pwp,Fujii:2017vwa}, the Circular Electron Positron Collider (CEPC)~\cite{CEPC-SPPCStudyGroup:2015csa} and the Future Circular Collider (FCC-ee)~\cite{Gomez-Ceballos:2013zzn}.

In this paper, we study the possibility of direct searches for additional Higgs bosons at LHC in the THDM with the double-alignment. 
In particular, we focus on the EW pair production of the additional Higgs bosons whose cross sections are simply determined by the masses of the additional Higgs bosons~\cite{Kanemura:2001hz,Cao:2003tr,Belyaev:2006rf}. 
\footnote{In Ref.~\cite{Eberhardt:2020dat}, 
global fits in the Yukawa aligned THDM have been performed, where various flavor data, signal strengths of the discovered Higgs boson and single productions of the additional Higgs bosons have been combined. }
We apply the data for multi-lepton final states at the LHC Run-II experiment to constrain the parameter space of the THDM, and combine 
constraints from flavor experiments such as $B \to X_s \gamma$, $B_s \to \mu\mu$ and leptonic tau decays. 
Furthermore, we extrapolate the Run-II data to obtain the parameter space expected to be explored at the High-Luminosity LHC (HL-LHC). 
We find that a large portion of the parameter space has already been excluded by the current LHC data, especially the region being able to be regarded as the Type-X (or lepton specific) THDM. 

\newpage

This paper is organized as follows.
In Sec.~\ref{sc:model}, we give the Lagrangian of our THDM, and define the double-alignment, i.e., the Yukawa alignment and the Higgs alignment. 
In Sec.~\ref{sc:LHC}, we discuss the decays of the additional Higgs bosons. 
Sec.~\ref{sc:flavor} summarizes various constraints from flavor experiments. 
Sec.~\ref{sc:LHC} is devoted to show our main results, i.e., parameter regions excluded by the current LHC data and expected to be explored at the HL-LHC. 
We also demonstrate the reconstruction of masses of additional Higgs bosons from the $b\bar{b}\tau^+\tau^-$ state. 
Conclusions and discussions are given in Sec.~\ref{sc:summary}. 
In Appendix, we show the constraint on the parameter space for the special case with $\zeta_u = 0$. 

\section{Model}\label{sc:model}

We consider a model with two isospin Higgs doublets without introducing
any symmetry other than the SM gauge symmetry.
In this case, the Lagrangian is invariant under the global $U(2)$ transformation
between the two doublets by using an appropriate redefinition of parameters in the potential and Yukawa interactions.
Using this invariance, we can write down the Lagrangian in terms of the Higgs basis~\cite{Davidson:2005cw} $\Phi$ and $\Phi'$ without loss of generality, which is  defined as 
\begin{align}
  \Phi =
 \begin{pmatrix}
  G^+\\\frac{1}{\sqrt{2}}(v+h^0_1+iG^0) \end{pmatrix}
  ,\quad
 \Phi' =  \begin{pmatrix}
   H^+\\\frac{1}{\sqrt{2}}(h^0_2+ih^0_3) \end{pmatrix}
  \label{eq:higgsbasis}
  ,\end{align}
where $v$ is the VEV related to the Fermi constant $G_F$ through $v = (\sqrt{2}G_F)^{-1/2}$.
In Eq.~(\ref{eq:higgsbasis}), $G^\pm$ and $G^0$ are the Nambu-Goldstone bosons which are absorbed into the longitudinal component of
the $W^\pm$ and $Z$ boson, respectively, while
$H^\pm$ and $h_i^0~(i = 1,2,3)$ are the physical charged and neutral Higgs bosons, respectively.

The potential is written in the Higgs basis as
\begin{align}
 V=  &	m^2 |\Phi|^2  + M^2 |\Phi'|^2 - \left(\mu^2 \Phi^\dagger\Phi'+ \text{h.c.}\right)
 +\frac{\lam_1 }{2} |\Phi|^4 + \frac{\lam_2}{2} |\Phi'|^4
  +\lam_3 |\Phi|^2|\Phi'|^2 + \lam_4 |\Phi^\dagger\Phi'|^2
  \notag\\
  & + \left[\frac{\lam_5}{2}(\Phi^\dagger\Phi') +\lam_6 |\Phi|^2 +\lam_7 |\Phi'|^2\right] (\Phi^\dagger\Phi')+\text{h.c.}
  \label{eq:potential2},
\end{align}
where $\mu^2$ and $\lambda_{5,6,7}$ are complex in general. 
The tadpole conditions, vanishment of the linear term of $h_i^0$, provide
\begin{align}
  m^2 = \frac{1}{2}\lam_1v^2,\quad \mu^2 = \frac{1}{2}\lam_6v^2. 
\end{align}
We note that the second equation is given by the conditions with respect to $h_2^0$ or $h_3^0$. 
By imposing the tadpole condition, the squared mass of $H^\pm$ is given by 
\begin{align}
  m_{H^\pm}^2 = M^2 + \frac{1}{2}\lam_3v^2.
\end{align}
The squared-mass matrix for the neutral Higgs bosons in the basis of $(h^0_1,h^0_2,h^0_3)$ is given by
\begin{align}
  \mathcal{M}^2
  =	v^2
    \begin{pmatrix}
      \lam_1	&\re[\lam_6]	&-\im[\lam_6]	\\
      \re[\lam_6]	&\frac{M^2}{v^2} + \frac{1}{2}(\lam_3+\lam_4+\re[\lam_5])	&-\frac{1}{2}\im[\lam_5] \\
      -\im[\lam_6]	&-\frac{1}{2}\im[\lam_5]	&\frac{M^2}{v^2}+\frac{1}{2}(\lam_3+\lam_4-\re[\lam_5])
    \end{pmatrix}. 
    \label{eq:massmatrix}
    \end{align}
The mass eigenstates $H_i^0~(i=1,2,3)$
can be defined by introducing the orthogonal matrix $R$ as $h_i^0 = O_{ij}H_j^0$
with the eigenvalues $\text{diag}(m_{H_1^0}^2,m_{H_2^0}^2,m_{H_3^0}^2) \equiv O^T{\cal M}^2 O$.
We define $m_{H_1^0}\leq m_{H_2^0} \leq m_{H_3^0}^2$, 
and identify the $H_1^0$ state with the discovered Higgs boson with the mass of 125 GeV. 

In this paper, we impose the alignment condition:
\begin{align}
  \lambda_6 = 0,     \label{eq:alignment}
\end{align}
by which the $h_1^0$ state coincides with the mass eigenstate $H_1^0$, and as a result 
couplings of $H_1^0$ to the gauge bosons and fermions agree with those of the SM Higgs boson at tree level.
We refer this condition as ``Higgs alignment''. 
By rephasing $\Phi'$,
the complex phase of $\lambda_5$ can be removed without loss of generality,
so that the mass matrix becomes diagonal form.
We note that at this stage we cannot identify the additional Higgs boson $H_2^0$ ($H_3^0$) with
the CP-even (CP-odd) state, because its CP property depends on the structure of the Yukawa interaction.

In this scenario, there are 7 free parameters\footnote{The number of parameters can consistently be counted as follows:
14 (initial number of real parameters in the potential) $-$ 3 (tadpole conditions) $-$ 2 (Higgs alignment) $-$ 1 (rephasing) = 7 [Eq.~(\ref{eq:parameters})] + 1 ($m_{H_1^0} = 125$ GeV). } which can be chosen as follows
		\eq{
		  M^2,~m_{H^0_2},~m_{H^0_3},~m_{H^\pm},~\lam_2,~|\lam_7| ~\textrm{and}~ \theta_7,
                  \label{eq:parameters}
		}
where $\theta_7\equiv\arg[\lam_7]\in(-\pi,\pi]$. 

The most general Yukawa interactions are given in the mass eigenstates of fermions as 
\begin{align}
  \mathcal{L}_{\rm Y}=
  &	-\bar{Q}_L^u \rbra{
    \sqrt{2}\frac{M_u}{v}\tilde{\Phi} +\rho_u\tilde{\Phi}'
  } u_R
  -\bar{Q}_L^d \rbra{
    \sqrt{2}\frac{M_d}{v}\Phi +\rho_d\Phi'
  }d_R
  \nn\\
  &	-\bar{L}_L \rbra{
    \sqrt{2}\frac{M_e}{v}\Phi +\rho_e\Phi'
  } e_R
  + \text{h.c.},
\end{align}
where $Q_L^u=(u_L,V_{\rm CKM} d_L)^T$,
$Q_L^d = (V_{\rm CKM}^\dagger u_L,d_L)^T$
and $L_L = (\nu_L, e_L)$ are left-handed doublet fermions
with $V_{\rm CKM}$ being the Cabibbo-Kobayashi-Maskawa (CKM) matrix, while
$u_R$, $d_R$ and $e_R$ are right-handed up-type quarks,
down-type quarks and charged leptons, respectively.
The charge conjugation of the Higgs doublets is denoted as $\tilde{\Phi}^{(\prime)} = i\tau_2\Phi^{(\prime) *}$.
In the first term of each parentheses, $M_f$ ($f=u,d,e$) denote diagonalized mass matrices, while
in the second term $\rho_f$ represent arbitrary $3\times 3$ complex matrices. 
In this expression, we do not explicitly show flavor indices.

It is clear that the $\rho_f$ terms give rise to tree-level FCNCs mediated by neutral
Higgs bosons, particularly via $H_2^0$ and/or $H_3^0$
in the Higgs alignment limit defined in Eq.~(\ref{eq:alignment}).
In order to avoid such FCNCs, we impose so-called the Yukawa alignment~\cite{Pich:2009sp}, i.e.,
\begin{align}
  \rho_f = \sqrt{2}\zeta_f \frac{M_f}{v}, 
\end{align}
where $\zeta_f$ are arbitrary complex parameters.
The Yukawa alignment can also be described in the general basis of the two doublets as
the assumption that two Yukawa matrices for a fermion type $f$ are proportional to each other.
We note that the $\zeta_f$ parameters are flavor universal, e.g., $\zeta_\mu = \zeta_\tau = \zeta_e$, due to the above assumption. 
Because the $\zeta_f$ parameters are complex, new sources of the CPV appear in addition to the potential parameter $\lambda_7$.
In Ref.~\cite{Kanemura:2020ibp}, it has been shown that 
we can take sizable CPV phases, while severe constraints from
experiments of the electron EDM~\cite{ACME:2018yjb} and the neutron EDM~\cite{nEDM:2020crw} can be avoided by using cancellation among Barr-Zee diagrams~\cite{Barr:1990vd} 
with fermion and scalar boson loops due to the independent phases from Yukawa interactions
and the Higgs potential. Such a scenario is compatible with constraints from the other flavor experiments as well as the current LHC data. 
We note that in THDMs with a softly-broken $Z_2$ symmetry~\cite{Glashow:1976nt}
these $\zeta_f$ parameters are determined by one parameter $\tan\beta$ (the ratio of the two Higgs VEVs)
depending on the type of Yukawa interactions~\cite{Barger:1989fj,Aoki:2009ha} by 
\begin{align}
\begin{split}
  (\zeta_u,\zeta_d,\zeta_e) &= (\cot\beta, \cot\beta,\cot\beta) \quad\quad\quad \text{in Type-I}, \\
  (\zeta_u,\zeta_d,\zeta_e) &= (\cot\beta, -\tan\beta,-\tan\beta) \quad \text{in Type-II}, \\
  (\zeta_u,\zeta_d,\zeta_e) &= (\cot\beta, \cot\beta,-\tan\beta) \quad\quad \text{in Type-X}, \\
  (\zeta_u,\zeta_d,\zeta_e) &= (\cot\beta, -\tan\beta,\cot\beta) \quad\quad \text{in Type-Y}.  
\end{split}\label{eq:types}
\end{align}

In the Higgs and Yukawa alignment, the double-alignment, the Higgs boson couplings are expressed as follows
\begin{align}
  {\cal L}_{\rm int} & =
  -\sum_{f = u,d,e}\bar{f}\frac{M_f}{v}\left[H_1^0+ |\zeta_f|(\cos\theta_f + i\sin\theta_f\gamma_5)H_2^0 + 2I_f|\zeta_f|(\sin\theta_f -i \cos\theta_f\gamma_5)H_3^0\right]f \notag\\
  & - \frac{\sqrt{2}}{v}\bar{u}\left(P_R \zeta_d V_{\rm CKM}M_d - P_L \zeta_u^* M_u^\dagger V_{\rm CKM}^\dagger\right)  d H^+
   - \frac{\sqrt{2}}{v}\bar{\nu}\left(P_R \zeta_eM_e \right)  e H^+ + \text{h.c.}, 
\end{align}
where $\theta_f\equiv\arg[\zeta_f]\in(-\pi,\pi]$, $I_u=1/2$, $I_d=I_e=-1/2$ and $P_L$ ($P_R$) is the projection operator for left- (right) handed fermions. 
  It is clear that non-zero phases $\theta_f$ result in the CP mixing in the Yukawa sector.
  We note that the $H_1^0VV$ ($V = W,Z$) couplings coincide with the SM values at tree level due to
  the Higgs alignment while  $H_{2,3}^0VV$ couplings vanish.
  On the other hand, there are Higgs-Higgs-Gauge type interactions for the additional Higgs bosons such as $H_2^0H_3^0Z$ and $H_{2,3}^0 H^\pm W^\mp$,
  which is phenomenologically important for the EW pair productions of the additional Higgs bosons and their decays as discussed in the following sections.

\section{Decays of the additional Higgs bosons}\label{sc:decay}

We discuss decays of additional Higgs bosons in the double-alignment limit, 
where they can mainly decay into a fermion pair or a lighter additional Higgs boson with a gauge boson as long as 
these are kinematically allowed. 
In the following discussion, we focus on the case where
the charged Higgs boson mass is degenerate with one of the additional neutral Higgs bosons in order to avoid the $T$ parameter constraint~\cite{Peskin:1990zt,Bertolini:1985ia,Peskin:2001rw,Grimus:2008nb,Kanemura:2011sj}. 
In this case, there are two possible scenarios denoted as 
the light $H^\pm$ scenario: $m_{H^\pm} = m_{H_2^0}$ and the heavy $H^\pm$ scenario: $m_{H^\pm} = m_{H_3^0}$, where we define $m_{H_3^0} \geq m_{H_2^0} \geq m_{H_1}$ as mentioned in Sec.~\ref{sc:model}. 

The decay rates of the Higgs bosons into a fermion pair are given by 
\begin{align}
\Gamma(H_{2,3}^0 \to f\bar{f})&=
\frac{\sqrt{2}G_Fm_{H_{2,3}}^3}{8\pi} N_f^c x_f |\zeta_f|^2[1 - 2x_f(1 \pm \cos2\theta_f) ]\lambda^{1/2}\left(x_f,x_f\right), \\
\Gamma(H^\pm\to ff') & =\frac{\sqrt{2}G_Fm_{H^\pm}^3}{8\pi}|V_{\rm CKM}^{ff'}|^2N_f^c\lambda^{1/2}(x_f,x_{f'}) \notag\\
& \times \left[(x_f |\zeta_f|^2+x_{f'}|\zeta_{f'}|^2)(1-x_f-x_{f'})+4x_fx_{f'} \text{Re}(\zeta_f^* \zeta_{f'})\right], 
\end{align}
where $\lambda(x,y) = (1-x-y)^2 - 4xy$, $x_i = m_i^2/m_{\cal H}^2$ with $m_{\cal H}$ being the mass of the decaying Higgs boson and $N_f^c = 3 (1)$ for $f$ being quarks (leptons). 
As we mentioned in Sec.~\ref{sc:model}, $\zeta_f$ factors are assumed to be flavor universal, e.g., $\zeta_\mu = \zeta_\tau = \zeta_e$. 
For the case with non-zero mass differences among the additional Higgs bosons, 
the following decay rates have to be added to their total widths: 
\begin{align}  
\Gamma(H^0_3 \to \phi V^{(\ast)})
=
\begin{cases}
	\frac{ \sqrt{2} G_Fm_{H_3^0}^3}{16\pi}
\,\lambda^{3/2}(x_\phi,x_V)\quad (m_{H_3^0} - m_\phi \ge  m_V) \\
\frac{9 m_{H_3^0}}{16\pi^3} G_F^2m_V^4
\delta_V\,G(x_\phi,x_V)\quad (m_{H_3^0} - m_\phi <  m_V)
\end{cases}, \label{eq:HWA}\\
\Gamma(H^\pm \to H_2^0 W^{\pm(\ast)})
=
\begin{cases}
	\frac{ \sqrt{2} G_Fm_{H^\pm}^3}{16\pi}
\,\lambda^{3/2}(x_{H^\pm},x_W)\quad (m_{H^\pm} - m_{H_2^0} \ge  m_W) \\
\frac{9 m_{H^\pm}}{16\pi^3} G_F^2 m_W^4G(x_{H_2},x_W)\quad (m_{H^\pm} - m_{H_2^0} <  m_W)
\end{cases}, 
\end{align}
where \footnote{In Eq.~(\ref{eq:HWA}) with $\phi = H^\pm$ and $V = W$, this expression does not take the sum over two possible states, i.e., $H^+W^{-(*)}$ and $H^-W^{+(*)}$. }
$\phi=H^0_{2}\,(H^\pm)$ for $V^{(*)}=Z^{(*)}\,(W^{\mp (*)})$, 
$\delta_W=1$ and $\delta_Z=\frac{7}{6}-\frac{20}{9}\sin^2\theta_W+\frac{80}{27}\sin^4\theta_W$. 
The function $G$ is the phase space function for three body decays, see e.g., Ref.~\cite{Aoki:2011pz} for the explicit form of this function: 
\begin{align}
G(x,y)&=\frac{1}{12y}\Bigg\{2\left(x - 1\right)^3-9\left(x^2 - 1\right)y+6\left(x - 1\right)y^2-3\left[1+\left(x-y\right)^2-2y\right]y\log x\notag\\
&+6\left(1+x-y\right)y\sqrt{-\lambda(x,y)}
\left[\arctan\left(\frac{-1+x-y}{\sqrt{-\lambda(x,y)}}\right)+\arctan\left(\frac{-1+x+y}{\sqrt{-\lambda(x,y)}}\right)\right]\Bigg\}. 
\end{align}
We note that loop induced decays of the additional neutral Higgs bosons, i.e., $H_{2,3}^0 \to gg/\gamma\gamma/Z\gamma$ are also taken into account in numerical evaluations, while 
those of the charged Higgs boson, i.e., $H^\pm \to W^\pm \gamma$~\cite{CapdequiPeyranere:1990qk} and $H^\pm \to W^\pm Z$~\cite{Kanemura:1997ej} are neglected because of their tiny partial widths. 

As we mentioned in the above, the decay modes of $H_3^0$ can be classified into two categories, i.e., 
fermionic modes ($H_3^0 \to f\bar{f}$) and bosonic modes ($H_3^0 \to \phi V^{(*)}$). 
The relative size of the branching ratios for the fermionic and bosonic modes can be 
expressed by introducing the following ratio $R$:
\begin{align}
R     &\equiv \frac{\sum_f\Gamma(H_3^0 \to f\bar{f})}{\sum_f\Gamma(H_3^0 \to f\bar{f}) + \sum_V\Gamma(H_3^0 \to \phi V^{(*)})} \simeq \sum_f {\cal B}(H_3^0 \to f\bar{f}). \label{eq:R}
\end{align}
In addition, we introduce 
\begin{align}
R_\tau &\equiv \frac{\Gamma(H_3^0 \to \tau^+\tau^-)}{\sum_f \Gamma(H_3^0 \to f\bar{f})},  \label{eq:Rtau}
\end{align}
by which we can parameterize the relative magnitude of the branching ratio of $H_3^0 \to \tau^+\tau^-$ among the fermionic modes. 
These $R$ parameters can simply be rewritten as 
\begin{align}
R =  \frac{1}{1 + r/\zeta^2}, \quad 
R_\tau  = \frac{|\zeta_e|^2}{\zeta^2}, \label{eq:R2}
\end{align}
where 
\begin{align}
\zeta^2 &= \frac{\sum_f\Gamma(H_3^0 \to f\bar{f})}{\Gamma_0}, \quad \Gamma_0 =\frac{\sqrt{2}G_F}{8\pi}m_{H_3^0}m_\tau^2, \\
r &= 
\begin{cases}
\frac{m_{H_3^0}^2}{2m_\tau^2} \sum_{V,\phi}^{} \lambda^{3/2}\left(\frac{m_\phi^2}{m_{H_3^0}^2}, \frac{m_V^2}{m_{H_3^0}^2}\right) \quad (m_{H_3^0} - m_\phi \ge  m_V)\\
\frac{9}{2\sqrt{2}\pi^2}\frac{G_F}{m_\tau^2} \sum_{V,\phi}^{} m_V^4\delta_V G\left(\frac{m_\phi^2}{m_{H_3^0}^2}, \frac{m_V^2}{m_{H_3^0}^2}\right) \quad (m_{H_3^0} - m_\phi <  m_V), 
\end{cases}
\end{align}
with the summation $\sum_{V,\phi}^{}$ being taken to be $(V,\phi) = (Z,H_3^0)$ and $(W,H^\pm)$. 
In particular, for the case with $m_b \ll m_{H_{2,3}} < 2 m_t$, which will be mainly considered in Sec.~\ref{sc:LHC}, 
$\zeta^2$ takes a significantly simple form as 
\begin{align}
\zeta^2 &\simeq \frac{1}{m_\tau^2}\sum_{f \neq t } m_f^2 N_f^c |\zeta_f|^2. 
\end{align}
We note that $H_2^0$ can only decay into a fermion pair at tree level, so that the $R$ value for $H_2^0$, i.e., $R|_{H_3^0 \to H_2^0}$ is unity.

Similar to the neutral Higgs bosons, we define the ratio parameters for $H^\pm$ as follows: 
\begin{align}
R^\pm &\equiv \frac{\sum_f\Gamma(H^\pm \to f\bar{f}')}{\sum_f\Gamma(H^\pm \to f\bar{f}') + \Gamma(H^\pm \to H_2^0 W^{\pm(*)})} \simeq \sum_f {\cal B}(H^\pm \to f\bar{f}'), \\
R_\tau^\pm & \equiv \frac{\Gamma(H^\pm \to \tau\nu)}{\sum_f \Gamma(H^\pm \to f\bar{f}')}.      \label{eq:Rtaucharged}
\end{align}
By introducing the quantities $\zeta_\pm^2$ and $r_\pm$, these $R$ parameters can be expressed as 
\begin{align}
R^\pm =  \frac{1}{1 + r_\pm/\zeta_\pm^2}, \quad 
R_\tau^\pm  = \frac{|\zeta_e|^2}{\zeta_\pm^2}, \label{eq:R3}
\end{align}
where 
\begin{align}
\zeta_\pm^2 &=  \frac{\sum_f\Gamma(H^\pm \to f\bar{f}')}{\Gamma_0|_{H_3^0 \to H^\pm}} \simeq |\zeta_e|^2 + 3\left(1 -\frac{m_t^2}{m_{H^\pm}^2}\right)^2 \left(\frac{m_t^2}{m_\tau^2}|\zeta_u|^2 + \frac{m_b^2}{m_\tau^2}|\zeta_d|^2\right), \\
r_\pm &= 
\begin{cases}
\frac{m_{H^\pm}^2}{2m_\tau^2} \lambda^{3/2}\left(\frac{m_{H_2^0}^2}{m_{H^\pm}^2}, \frac{m_W^2}{m_{H^\pm}^2}\right) \quad (m_{H^\pm} - m_{H_2^0} \ge  m_W)\\
\frac{9}{2\sqrt{2}\pi^2}\frac{G_F}{m_\tau^2}  m_W^4 G\left(\frac{m_{H_2^0}^2}{m_{H^\pm}^2}, \frac{m_W^2}{m_{H^\pm}^2}\right)\quad (m_{H^\pm} - m_{H_2^0} <  m_W)
\end{cases}, 
\end{align}
Unlike the $H_3^0$ decays, only the decay of $H^\pm \to H_2^0 W^{\pm(*)}$ is allowed for the bosonic decay mode. 

In terms of these $R$ parameters, the decay branching ratios into a tau lepton pair, which will be important in the discussion for the phenomenology at LHC, can simply be expressed as 
\begin{align}
{\cal B}(H_2^0\to \tau^+\tau^-) \simeq R_\tau,\quad 
{\cal B}(H_3^0 \to \tau^+\tau^-) \simeq RR_\tau,\quad 
{\cal B}(H^\pm \to \tau^\pm\nu) \simeq R^\pm R_\tau^\pm. \label{eq:br_r}
\end{align}

It would be important to discuss the critical values of $\zeta$ and $\zeta_\pm$ which give the sum of the fermionic decay branching ratios to be 50\%. 
Such critical values, denoted as $\zeta_{50}$ and $\zeta_{\pm 50}$, can be expressed as $\zeta_{50} = \sqrt{r}$ and $\zeta_{\pm 50} = \sqrt{r_\pm}$ from Eqs.~(\ref{eq:R2}) and (\ref{eq:R3}), respectively.
Since $r$ and $r_\pm$ depend only on the masses of additional Higgs bosons, $\zeta_{50}$ and $\zeta_{\pm 50}$ are determined as a function of these masses. 

\begin{figure}
\centering
\includegraphics[width=80 mm]{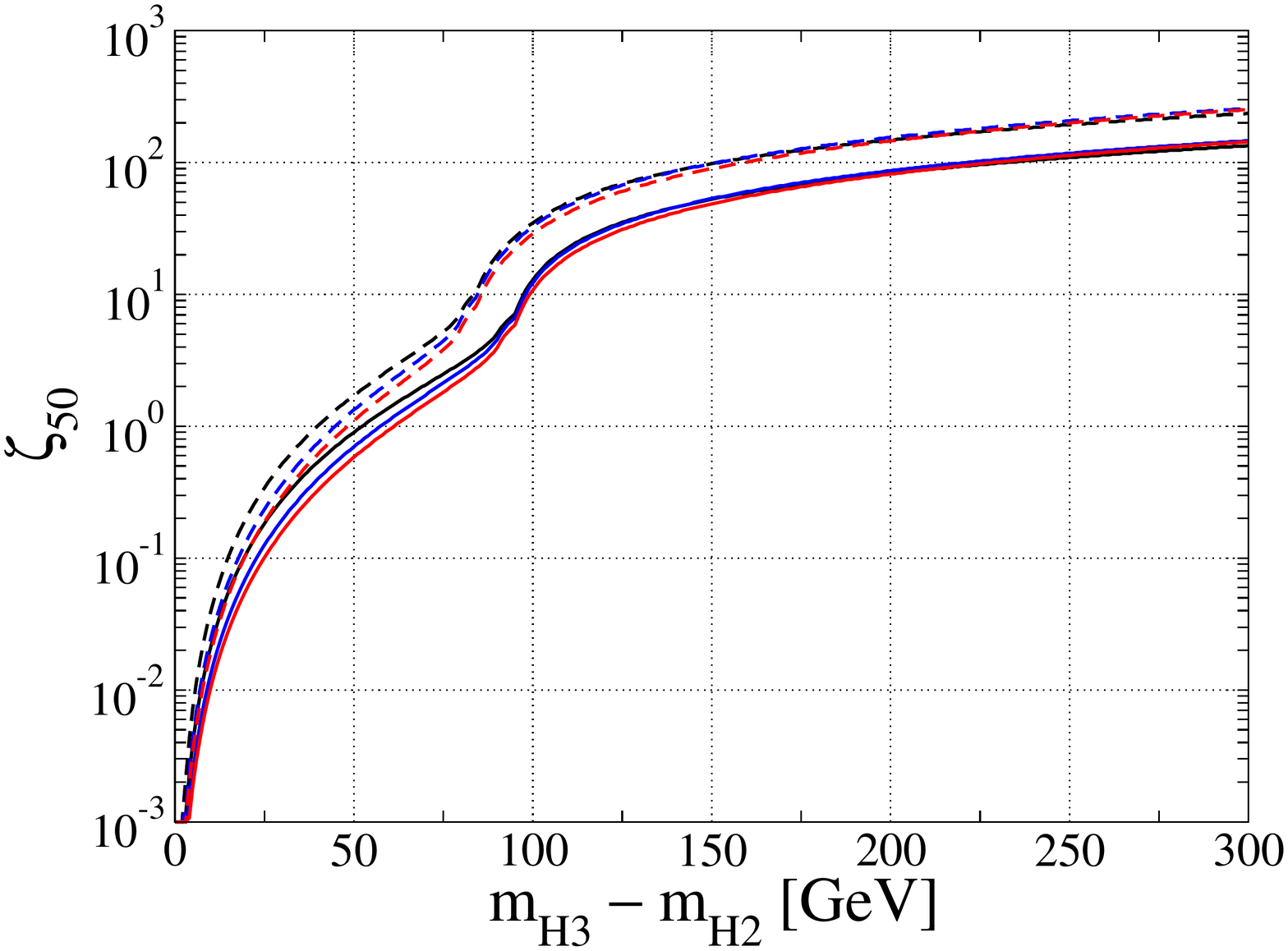}
\includegraphics[width=80 mm]{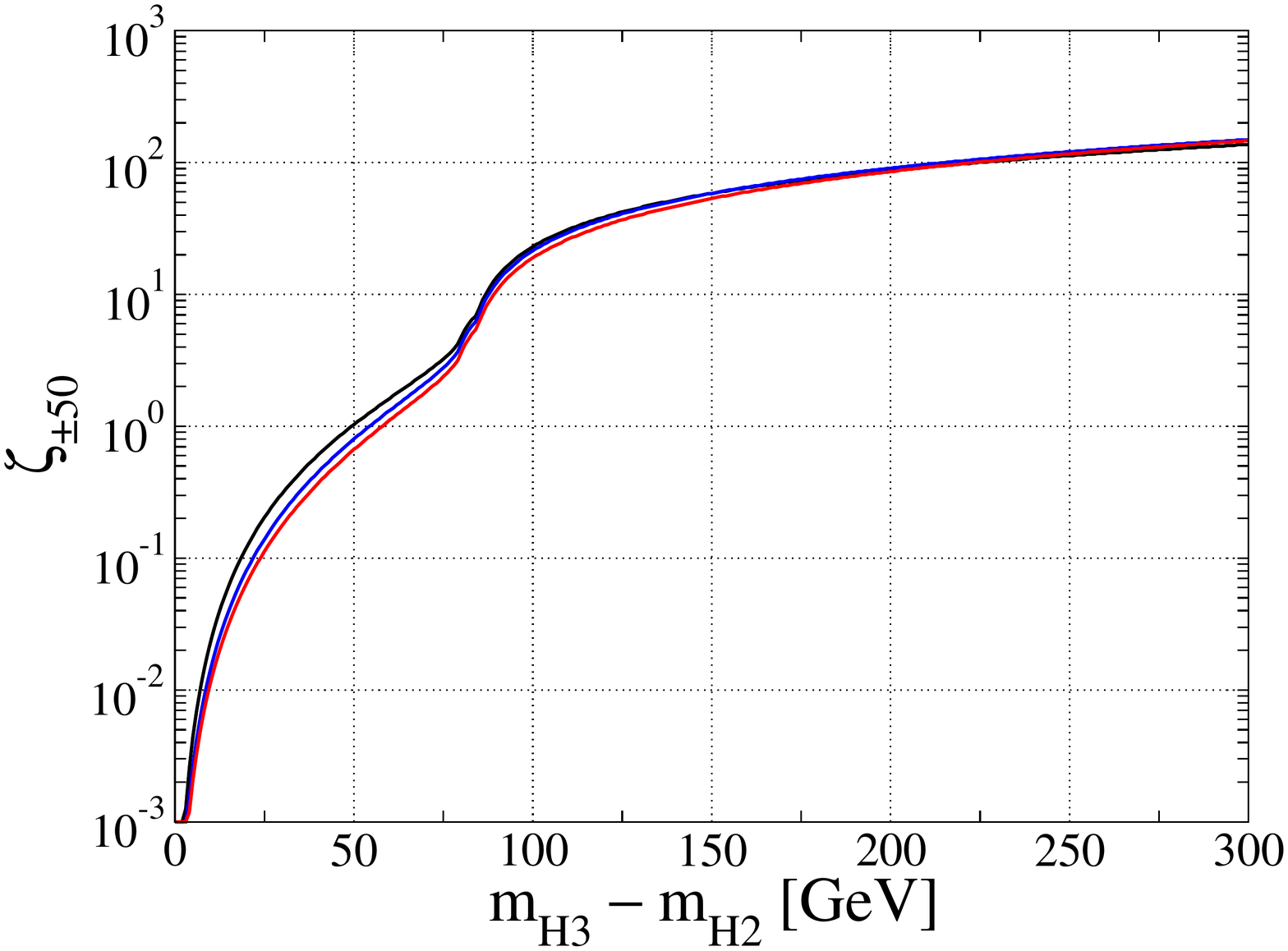}
\caption{Values of $\zeta_{50}$	(left) and $\zeta_{\pm 50}$ (right) as a function of the mass difference $m_{H_3^0} - m_{H_2^0}$ for $m_{H_2^0} = 100$ GeV (black), 300 GeV (blue) and 500 GeV (red). 
For the left panel, the solid and dashed curves show the case with the heavy $(m_{H^\pm} = m_{H_3^0})$ and the light ($m_{H^\pm} = m_{H_2^0}$) $H^\pm$ scenario, respectively.  
}
\label{fig:zeta50}
\end{figure}

Fig.~\ref{fig:zeta50} shows the values of $\zeta_{50}$ (left panel) and $\zeta_{\pm 50}$ (right panel) as a function of $m_{H_3^0} - m_{H_2^0}$. 
We see that for $m_{H_3^0} - m_{H_2^0} \ll m_W$ both $\zeta_{50}$ and $\zeta_{\pm 50}$ are much smaller than unity, because the bosonic decay modes are significantly suppressed by the phase space, 
while they can be of order one for $m_{H_3^0} - m_{H_2^0} \sim m_W$. 
When the on-shell decays open, $\zeta_{50}$ and $\zeta_{\pm 50}$ are of order 10 or larger. 
We also see that $\zeta_{50}$ and $\zeta_{\pm 50}$ almost do not depend on $m_{H_3^0}$, so that they are essentially determined by the mass difference $m_{H_3^0} - m_{H_2^0}$. 
It is clear that larger values of $\zeta_{50}$ are required for $m_{H^\pm} = m_{H_2^0}$ (dashed curves) as compared with that with $m_{H^\pm} = m_{H_3^0}$ (solid curves), because 
the $H_3^0 \to H^\pm W^{\pm(*)}$ mode also contribute to the $r$ value. 

\begin{figure}
\centering
\includegraphics[width=80 mm]{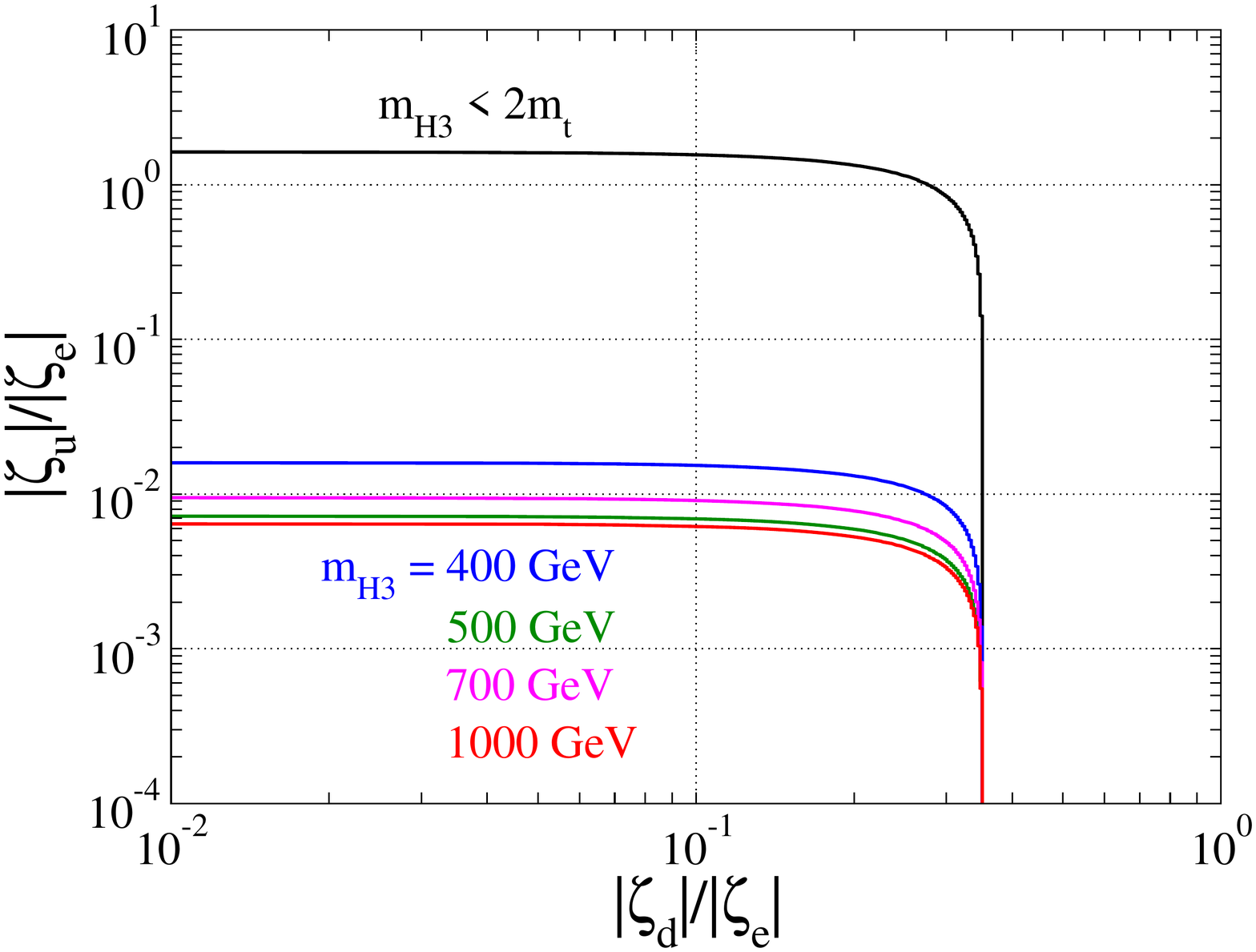}
\includegraphics[width=80 mm]{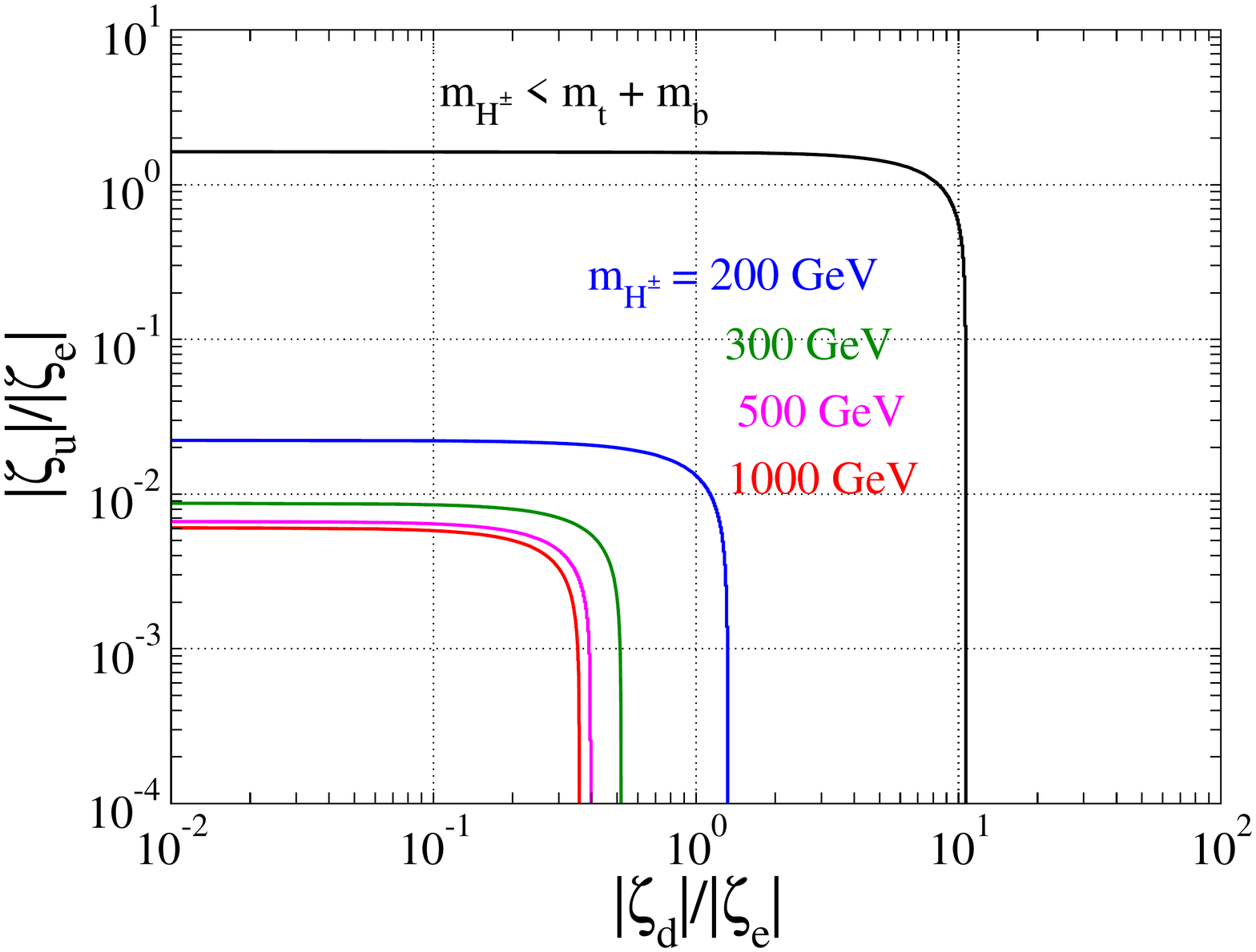}
\caption{Contour plots for $R_\tau = 50\%$ (left) and $R_\tau^\pm = 50\%$ (left) on the $|\zeta_d|/|\zeta_e|$ and $|\zeta_u|/|\zeta_e|$ plane for each fixed value of $m_{H_2^0}$ (left) and $m_{H^\pm}$ (right).
}
\label{fig:rtau50}
\end{figure}

Similarly, we discuss the critical values of $\zeta_f$ which lead to $R_\tau = 50\%$ or $R_\tau^\pm = 50\%$. 
From Eqs.~(\ref{eq:R2}) and (\ref{eq:R3}), $R_\tau$ and $R_\tau^\pm$ are determined by the ratios of the $\zeta_f$ parameters, i.e., $\zeta_u/\zeta_e$ and $\zeta_d/\zeta_e$. 
More concretely, the inverse of $R_\tau$ and $R_\tau^\pm$ can be expressed for $\theta_f = 0$ as
\begin{align}
R_\tau^{-1} &\simeq 1+\frac{3m_b^2}{m_\tau^2}\frac{|\zeta_d|^2}{|\zeta_e|^2} + \left[\frac{3m_c^2}{m_\tau^2} + \theta_{tt}\frac{3m_t^2}{m_\tau^2}\left(1 - \frac{4m_t^2}{m_{H_3^0}^2}\right)^{3/2}\right]\frac{|\zeta_u|^2}{|\zeta_e|^2}, 
\label{eq:ellipse1}\\
(R_\tau^{\pm})^{-1}&\simeq 1 + \left[\frac{3m_s^2}{m_\tau^2} + \theta_{tb}\frac{3m_b^2}{m_\tau^2}\left(1 - \frac{m_t^2}{m_{H^\pm}^2}\right)^2\right]\frac{|\zeta_d|^2}{|\zeta_e|^2}
+ \left[\frac{3m_c^2}{m_\tau^2} + \theta_{tb}\frac{3m_t^2}{m_\tau^2}\left(1 - \frac{m_t^2}{m_{H^\pm}^2}\right)^2\right]\frac{|\zeta_u|^2}{|\zeta_e|^2}, \label{eq:ellipse2}
\end{align}
where $\theta_{tt} \equiv \theta(m_{H_3^0} - 2m_t)$ and $\theta_{tb} \equiv \theta(m_{H^\pm} - m_t - m_b)$ with $\theta(x)$ being the Heaviside step function. 
These expressions show that the values of $(R_\tau)^{-1} - 1$ and $(R_\tau^\pm)^{-1} - 1$ obey the equation of an ellipse with 
the $x$ ($y$) axis corresponding to $|\zeta_d|/|\zeta_e|$ $(|\zeta_u|/|\zeta_e|)$. 
This is numerically shown in Fig.~\ref{fig:rtau50}. 
Here, we show the first quadrant of the ellipse. 
The points on each curve represent the required values of $|\zeta_d|/|\zeta_e|$ and $|\zeta_u|/|\zeta_e|$ to obtain $R_\tau = 50\%$ (left panel) and $R_\tau^\pm = 50\%$ (right panel).
For the case below the top threshold, 
the curve almost does not depend on $m_{H_2^0}$, and the interceptions of $x$ and $y$ axis are simply determined by the fermion mass ratio extracted from Eqs.~(\ref{eq:ellipse1}) and (\ref{eq:ellipse2}), i.e., 
$|\zeta_d|/|\zeta_e| \simeq m_\tau/(\sqrt{3}m_b) \simeq 0.35$ and $|\zeta_u|/|\zeta_e| \simeq m_\tau/(\sqrt{3}m_c) \simeq 1.6$.  
On the other hand, above the top threshold, 
the dependence of $m_{H_2^0}$ slightly appears due to the phase function as seen in Eqs.~(\ref{eq:ellipse1}) and (\ref{eq:ellipse2}). 
For $R_\tau^\pm$ (right panel), both the $|\zeta_d|/|\zeta_e|$ and  $|\zeta_u|/|\zeta_e|$ depend on $m_{H_2^0}$, because the phase space suppression is multiplied to both the $|\zeta_d|^2/|\zeta_e|^2$ and 
$|\zeta_u|^2/|\zeta_e|^2$ terms for $m_{H^\pm} > m_t + m_b$. 

By looking at Figs.~\ref{fig:zeta50} and \ref{fig:rtau50} and using Eq.~(\ref{eq:br_r}), 
the Higgs boson decays into a tau lepton pair can well be estimated.

\section{Flavor Constraints}\label{sc:flavor}

\begin{table}[!t]
\begin{center}
\begin{tabular}{cccccccccccccc}\hline\hline
  $m_t$ [GeV] & $m_b$ [GeV] & $m_c$ [GeV]  & $m_Z^{}$ [GeV] & $m_W^{}$ [GeV] & $m_{H_1}^{}$ [GeV] & $\alpha_{s}(m_Z)$  & $\alpha_{\rm em}^{-1}$\\\hline
  172.76~\cite{ParticleDataGroup:2020ssz}     & 4.561~\cite{Alberti:2014yda} & 1.092~\cite{Alberti:2014yda}
  & 91.1876~\cite{ParticleDataGroup:2020ssz} & 80.379~\cite{ParticleDataGroup:2020ssz} & 125.25~\cite{ParticleDataGroup:2020ssz}  &
  0.1182~\cite{FlavourLatticeAveragingGroup:2019iem}  & 137.036~\cite{ParticleDataGroup:2020ssz}\\ \hline\hline
  ${\cal B}(\bar{B} \to X_c \ell^-\nu)$ &
  $|V_{ts}^*V_{tb}/V_{cb}|^2$ &$|V_{cb}|$ & $\bar{\tau}(B_s^0)$ [ps] & $\tau(B_{sL}^0)$ [ps] & $\tau(B_{sH}^0)$ [ps] & $\Delta\Gamma_s$ [ps$^{-1}$] & $m_{B_s}$ [GeV] & $f_{B_s}$ [GeV]\\ \hline
  0.1065~\cite{HFLAV:2019otj} &
  0.9626~\cite{Charles:2015gya,Czakon:2015exa} &
  0.041~\cite{Charles:2015gya} & 
  1.510~\cite{HFLAV:2019otj} &
   1.414~\cite{HFLAV:2019otj}&
 1.619~\cite{HFLAV:2019otj}&
 0.090~\cite{HFLAV:2019otj} & 
 5.367~\cite{ParticleDataGroup:2020ssz} &
 0.2284~\cite{FlavourLatticeAveragingGroup:2019iem} \\\hline\hline
\end{tabular}
\caption{Input SM parameters. We take the central values of these parameters. }
\label{tab:inputs1}
\end{center}
\end{table}


In our scenario, the additional Higgs bosons can mediate various flavor processes
and change their predictions from those in the SM.
Such effects can be translated into constraints on the masses of Higgs bosons and the $\zeta_f$ parameters, defined in Sec.~\ref{sc:model}. 

Let us first discuss the constraint from $B \to X_s \gamma$ decay process.
The current experimental value of the branching ratio is given by~\cite{HFLAV:2019otj} 
\begin{align}
  {\cal B}(B \to X_s\gamma) = (3.32\pm  0.15)\times 10^{-4}. 
\end{align}
In THDMs, the charged Higgs boson can run in the one-loop diagram instead of the W boson, so that
there is a sensitivity to $m_{H^\pm}$ and quark Yukawa couplings which correspond to $\zeta_u$ and $\zeta_d$ in our scenario.
It has been known that QCD corrections can sizably change the prediction, e.g., about a few 10\% at NLO with respect to the LO prediction in the SM~\cite{Ciuchini:1997xe}. 
The NNLO QCD corrections have also been calculated in Ref.~\cite{Misiak:2017bgg} in the THDM.
We implement the NLO QCD and QED corrections to the decay rate according to Ref.~\cite{Borzumati:1998tg} and \cite{Kagan:1998ym}, respectively. \footnote{We set
  the renormalization scale of the $B$ meson decay $\mu_b$ to be the pole mass of the bottom quark $m_b$, while that
  of the matching to the full EW theory to be $m_W$. }
The constraint from $B \to X_s \gamma$ is particularly important for larger values of $\zeta_u$ which correspond to the case with a smaller $\tan\beta$ value in the THDMs with the softly-broken $Z_2$ symmetry, 
see e.g.,~\cite{Misiak:2017bgg}.  

In addition to $B \to X_s \gamma$, we take into account the constraint from $B_s \to \mu\mu$ decay.
The current experimental value of the branching ratio is given by~\cite{HFLAV:2019otj} 
\begin{align}
  {\cal B}(B_s \to \mu\mu) = (3.1\pm  0.6)\times 10^{-9}. 
\end{align}
In the aligned THDM, the charged Higgs boson can contribute to the process via box and penguin type diagrams.
The latter also contains neutral Higgs exchanges.
Unlike the $B \to X_s \gamma$ process,
the $B_s \to \mu\mu$ process has a sensitivity to the masses of neutral Higgs bosons and
the $\zeta_e$ parameters in addition to $m_{H^\pm}$, $\zeta_u$ and $\zeta_d$.
Therefore, even in a case with smaller $\zeta_u$ which is not excluded by $B \to X_s \gamma$, scenarios with larger $\zeta_d$ and/or $\zeta_e$ can be excluded.
We implement the decay rate of $B_s \to \mu\mu$ at one-loop level according to Ref.~\cite{Li:2014fea} by utilizing the Wilson coefficient
$C_{10} (\bar{s}\gamma_\mu P_L b)(\bar{\mu}\gamma^\mu \gamma_5 \mu)$ in the SM evaluated at NNLO of QCD as~\cite{Bobeth:2013uxa}
\begin{align}
 C_{10}^{\rm SM}  = -0.938\times \left(\frac{m_t}{173.1~\text{GeV}}\right)^{1.53}\times \left(\frac{\alpha_s(m_Z)}{0.1184}\right)^{-0.09}. 
\end{align}

Finally, we take into account the leptonic tau decays which are mediated by the charged Higgs boson in addition to the contribution from the $W$ boson at tree level.
Thus, the combination of $\zeta_e^2/m_{H^\pm}^2$ is constrained~\cite{Krawczyk:2004na}.
Because the coupling of $H^\pm$ with an electron is negligibly small, the contributions of $H^\pm$ to $\tau \to e\nu_\tau\bar{\nu}_e$
and $\mu \to e\nu_\mu\bar{\nu}_e$ are negligible. This causes the violation of lepton flavor universality, which has been stringently constrained by experiments.
The current world average is~\cite{HFLAV:2019otj}
\begin{align}
  (g_\tau/g_\mu)_{\rm ave} = 0.9999 \pm 0.0014. 
\end{align}
The left-hand side of the above expression can be compared with the theory prediction of $\Gamma(\tau \to \mu\nu_\tau\bar{\nu}_\mu)_{\rm THDM}/\Gamma(\tau \to \mu\nu_\tau\bar{\nu}_\mu)_{\rm SM}$. 

For numerical evaluations, we use the SM input parameters shown in Table~\ref{tab:inputs1}, and show 
the parameter region constrained by the flavor data in the next section.

\section{Direct searches for additional Higgs bosons at LHC}\label{sc:LHC}

We discuss constraints on the model parameters from direct searches for the additional Higgs bosons by using the current LHC data in the double-aligned THDM. 
We also study the parameter space expected to be explored at the HL-LHC. 
For simplicity, we here consider the CP-conserving case, i.e., $\theta_f = 0$, in which the constraint from LHC data almost does not depend on these phases while 
that from flavor experiments discussed in the previous section has a sensitivity to these phases. 
In addition, we concentrate on the case where the additional neutral Higgs bosons are lighter than $2 m_t$ as the phenomenologically interesting region. 

We particularly focus on the EW pair production of the additional Higgs bosons. 
The cross sections of these processes are simply determined by the gauge coupling for fixed Higgs boson masses~\cite{Kanemura:2001hz,Cao:2003tr,Belyaev:2006rf}, 
so that the number of signal events is determined by decays of the Higgs bosons. 
This is quite different from single Higgs productions such as 
the gluon fusion process and the bottom or top quark associated process
whose cross sections depend on Yukawa couplings. 
We note that the single additional Higgs boson production associated with a weak boson~\cite{Ellis:1975ap}, e.g., $pp \to H_{2,3}^0Z$, 
is absent at tree level in the Higgs alignment limit, so that we do not consider these production modes. 
Thus, we focus on the EW production, by which 
we obtain a robust constraint on the masses of the additional Higgs bosons for given decay branching ratios, because 
the production cross section cannot be tuned to be small by taking model parameters. 

\subsection{EW pair productions of the additional Higgs bosons}\label{production}

We study the pair productions of the additional Higgs bosons. 
There are the following 6 subprocesses~\cite{Eichten:1984eu, Gunion:1986nh, Djouadi:1999rca, Cao:2003tr}: 
\begin{align}
pp \to H_2^0 H_3^0,~~
pp \to H_2^0 H^\pm,~~
pp \to H_3^0 H^\pm,~~
pp \to H^+ H^-. 
\end{align}
%

\begin{figure}
\centering
\includegraphics[width=93 mm]{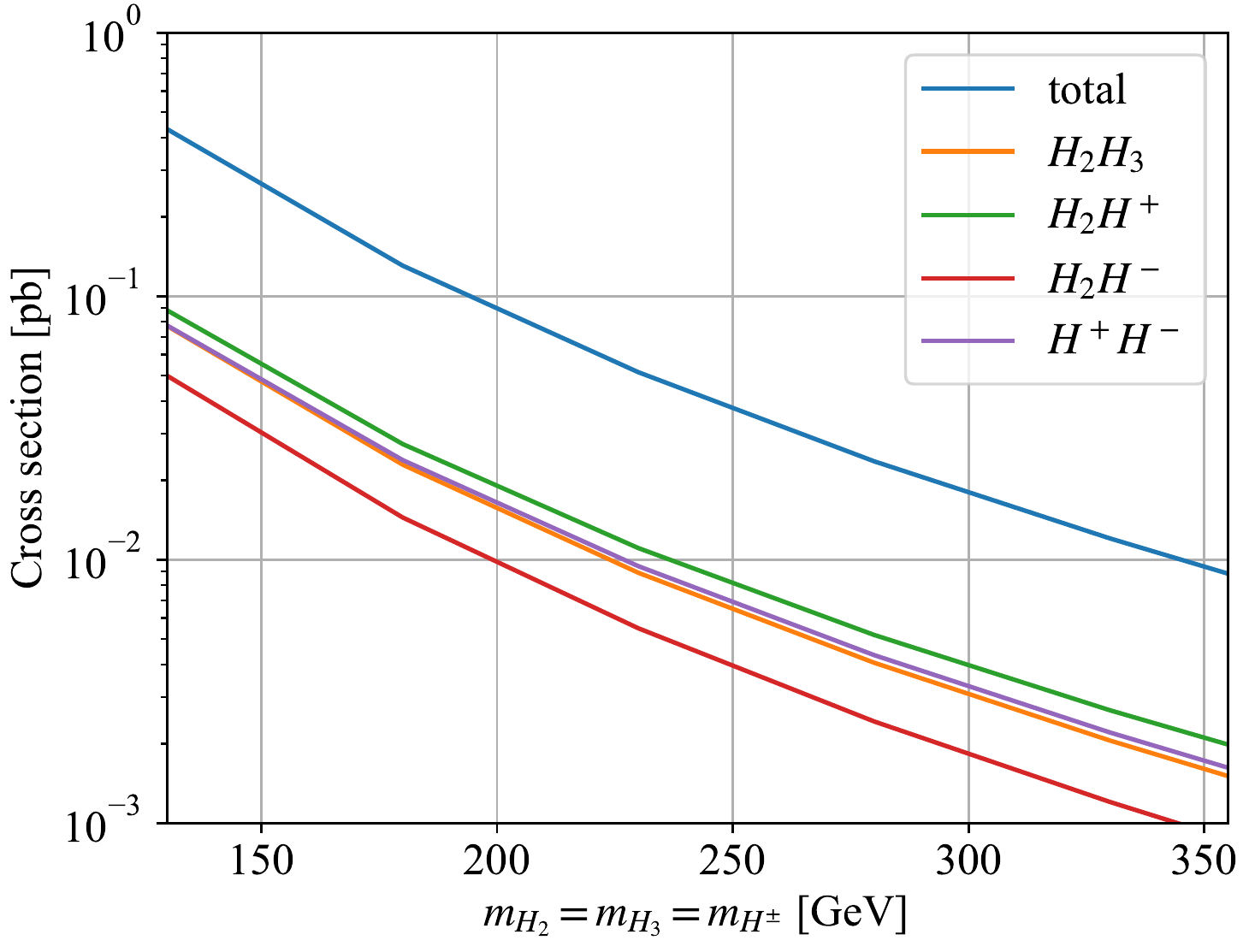}
\caption{Cross sections for the EW pair production of the additional Higgs bosons 
at LHC with the collision energy of 13~TeV as a function of the degenerate masses $m_{H_2^0} (= m_{H_3^0} = m_{H^\pm})$. }
\label{fg:xs}
\end{figure}

Fig.~\ref{fg:xs} shows the total cross section of the EW pair production 
at LO in the unit of pb as a function of the additional 
Higgs boson mass, where $m_{H_2^0}=m_{H_3^0} = m_{H^\pm}$ is assumed. 
We also show the cross sections of each subprocess. 
As we see in this plot, the cross section can be of ${\cal O}(10-500)$~fb at LHC with 13~TeV 
for $m_{H_2^0}, m_{H_3^0} \lesssim 350$~GeV. 
Thus, ${\cal O}(10^3 - 10^5)$ events are already expected even in the current LHC data with the integrated luminosity of 139~fb$^{-1}$, and 
we would expect ${\cal O}(10^4 - 10^6)$ events are produced at the HL-LHC with 3~ab$^{-1}$.~\footnote{In this paper, 
we show the results for 13~TeV although the ultimate planned center of mass energy 
at HL-LHC is 14~TeV, where about 10-15~\% more events are expected for our signals.}

As we discussed in Sec.~\ref{sc:decay}, the $\zeta_f$ parameters control the branching ratios of the additional Higgs bosons. 
Especially, the leptonic branching ratios are described through the four $R$ parameters; $R$, $R_\tau$, $R^{\pm}$ and $R^{\pm}_\tau$, see Eq.~(\ref{eq:br_r}).
The $\zeta_f$ dependence of the model then appears in the signal cross sections including the decay of the Higgs boson, which can be expressed as   
\begin{align}
\sigma(pp \to AB \to  i_A i_B) \sim \sigma(pp \to AB)\cdot {\cal B}(A \to i_A)\cdot {\cal B}(B \to i_B),
\end{align}
where $A$ and $B$ are $H_2^0$, $H_3^0$ or $H^\pm$. 
Their decay modes including the subsequent additional Higgs decay modes are indicated by $i_A$ and $i_B$.
For example, $i_{H_3^0}=\{\tau^+\tau^-,b\bar{b},Z^{(\ast)}\tau^+\tau^-,Z^{(\ast)}b\bar{b},W^{\pm(\ast)}\tau^\mp \nu,$
$W^{\pm(\ast)}tb, \cdots \}$ and $i_{H^\pm}=\{\tau^\pm \nu,tb,W^{\pm(\ast)}\tau^+\tau^-,W^{\pm(\ast)}b\bar{b}, \cdots \}$. 

When the branching ratios of the additional Higgs bosons into a charged lepton pair are large, 
multi-tau signatures from the EW pair production can be used to probe such models~\cite{Kanemura:2011kx,Chun:2015hsa}.  
We will show that the current multi-lepton searches at LHC
provide strong constraints on the model especially 
when the additional Higgs bosons dominantly decay into charged leptons.
We will also show the expected constraint on the parameter space at the HL-LHC by extrapolating these data.

\subsection{Constraints from multi-lepton searches at LHC}

We here consider the constraints on the model by the latest multi-lepton searches at LHC reported by the ATLAS collaboration~\cite{ATLAS:2021yyr}.\footnote{The corresponding analysis by the CMS collaboration utilizes the Neural Network approach~\cite{CMS:2021cox}, which is difficult to interpret in our model, 
and thus we do not include it.}
In this analysis, events with 4 or more leptons including tau leptons have been searched, which are classified into 14 signal regions with different combinations 
of the requirements on the lepton species, the number of $b$-tagged jets, and the existence of the lepton pair consistent with a $Z$-boson decay. 
For these signal regions, the minimum numbers of 
isolated electrons, muons, $\tau$-tagged jets and $b$-tagged jets are required.
The following kinematical cuts are imposed for signal events, i.e., 
$p_T > 7$~GeV, $|\eta|<2.47$ for electrons 
$p_T > 5$~GeV, $|\eta|<2.7$ for muons and 
$p_T > 20$~GeV, $|\eta|<2.5$ for jets. 

We find that the seven signal regions among them are relevant to constrain our model, and call the three 3L1T (L: $e$ or $\mu$ and T: $\tau$)
signal regions defined in the ATLAS analysis as SR1-SR3, the three 2L2T regions as SR4-SR6, 
and the 5L0T region as SR7. 
A summary of the definition of the relevant SRs is shown in Table~\ref{tab:SRs}, where 
the observed and the expected number of SM backgrounds and that of 95\% CL upper bounds for the signal in each SR are also provided.
We have checked that the other seven signal regions provide weaker constraints
than the seven signal regions considered here, and therefore, they are omitted in the following analysis.

As future prospects at the HL-LHC, 
we set the constraints on the model parameters based on the assumption with 
null contribution from new physics. 
The expected upper limit on the number of signal events is then 
estimated by using the Poisson distribution, in which 
we use the expected number of SM events obtained by extrapolating $N_{\rm exp}^{\rm BG}$ (expected number of SM events for 139 fb$^{-1}$) 
to that given at the HL-LHC, i.e.,  $N_{\rm exp}^{\rm BG} \times 3000/139$. 
For instance for SR6, the upper limit on the signal event number at 95\% CL is given to be about 9.  

\begin{table}[t]
    \centering
    \begin{tabular}{p{1cm}|c|cccc||rr|rr}    \hline    \hline
    SR & SR in Ref.~\cite{ATLAS:2021yyr} & $N(e,\mu)$ &  $N(\tau_{\rm had})$ & $N$($b$-jets) & $m_{\rm eff}^{\rm th}$~[{\rm GeV}] &
    $N_{\rm obs}^{\rm BG}$ & $N_{\rm exp}^{\rm BG}$ &
    $S_{\rm obs}^{95}$ & $S_{\rm exp}^{95}$\\
    \hline
    SR1 & ${\tt SR1}^{\rm loose}_{\rm bveto}$ 
    & =3 & $\ge 1$ & $0$ & 600 & 7 & 7.7 & 6.37 & 7.46 \\
    SR2 & ${\tt SR1}^{\rm tight}_{\rm bveto}$
    & =3 & $\ge 1$ & $0$ & 1000& 2 & 1.6 & 4.47 & 4.22 \\
    SR3 & ${\tt SR1}_{\rm breq}$
    & =3 & $\ge 1$ & $\ge 1$ & 1300& 2 & 2.2 &  4.56 & 4.59 \\
    \hline
    SR4 & ${\tt SR2}^{\rm loose}_{\rm bveto}$ 
    & =2 & $\ge 2$ & $0$ & 600 & 5 & 3.4 & 8.45 & 7.45\\
    SR5 & ${\tt SR2}^{\rm tight}_{\rm bveto}$ 
    & =2 & $\ge 2$ & $0$ & 1000 & 2 & 0.35 & 5.63 & 3.53\\
    SR6 & ${\tt SR2}_{\rm breq}$
    & =2 & $\ge 2$ & $\ge 1$ & 1100 & 1 & 0.52 & 4.17 & 3.16 \\
    \hline
    SR7 & ${\tt SR5L}$
    & =5 & $\ge 0$ & $\ge 0$ & -- & 21 & 12.4 & 17.88 & 9.88 \\
    \hline    \hline
    \end{tabular}
    \caption{Summary of the relevant SRs and corresponding limits for 139 fb$^{-1}$ quoted from Ref.~\cite{ATLAS:2021yyr}, 
where $N_{\rm obs}^{\rm BG}$ ($N_{\rm exp}^{\rm BG}$) is the observed (expected) number of background events and 
$S_{\rm obs}^{95}$ ($S_{\rm exp}^{95}$) is the observed (expected) upper limits on the number of signal events at 95\% CL. }
    \label{tab:SRs}
\end{table}

We perform the simulation study by 
using {\tt MadGraph5}~\cite{Alwall:2011uj} and {\tt Pythia 8}~\cite{Sjostrand:2007gs}
with the detector simulation using {\tt Delphes3}~\cite{deFavereau:2013fsa}.
We consider the $16$ model points with different additional Higgs masses labeled with $0 \le n_2 \le n_3 \le 3$, 
corresponding to $180$~GeV~$\le m_{H_2^0} \le m_{H_3^0}\le 330$~GeV, as follows:
\begin{align}
m_{H_2^0}=180 + 50 n_2~{\rm GeV}, \ m_{H_3^0}=180 + 50 n_3~{\rm GeV}, \ 
m_{H^\pm} = 
\begin{cases}
m_{H_3^0} & {\rm (heavy}\  H^\pm \ {\rm scenario)}\cr
m_{H_2^0} & {\rm (light}\  H^\pm \ {\rm scenario)}
\end{cases}.  \label{eq:spectra}
\end{align}

In our simulation, 
we fix $\zeta_u=0.01$, $\zeta_d=0.1$ and $\zeta_e=0.5$ for the model points with $|\Delta m| = m_{H_3^0} - m_{H_2^0} \le 50$~GeV, where we define $\Delta m = m_{H^\pm} - m_\text{no-deg}$ with 
$m_\text{no-deg}$ being the mass of the additional neutral Higgs which is not degenerate with $m_{H^\pm}$. 
For the model points with $|\Delta m| \ge 100$~GeV,
larger $\zeta_e$ and $\zeta_d$ values are adopted to achieve
$R\sim R_\tau \sim 50~\%$  to keep the several decay modes contribute in a similar size.

For different values of  $\zeta_f$, we estimate the number of events
falling down in a certain SR by multiplying the scaling factors defined as the ratio of the product of the branching ratios, i.e.,  
${\cal B}(A \to i_A){\cal B}(B \to i_B)|_{\zeta_f}/{\cal B}(A \to i_A){\cal B}(B \to i_B)|_{\rm fixed}$ with the numerator (denominator) being the value at arbitrary $\zeta_f$ ($\zeta_f$ fixed to be the above values). 
We note that the production cross section does not depend on the $\zeta_f$ parameters, so that the scaling factor is needed to be multiplied only to the branching ratios as explained in the above. 

\begin{figure}[t]
 \includegraphics[width=8cm]{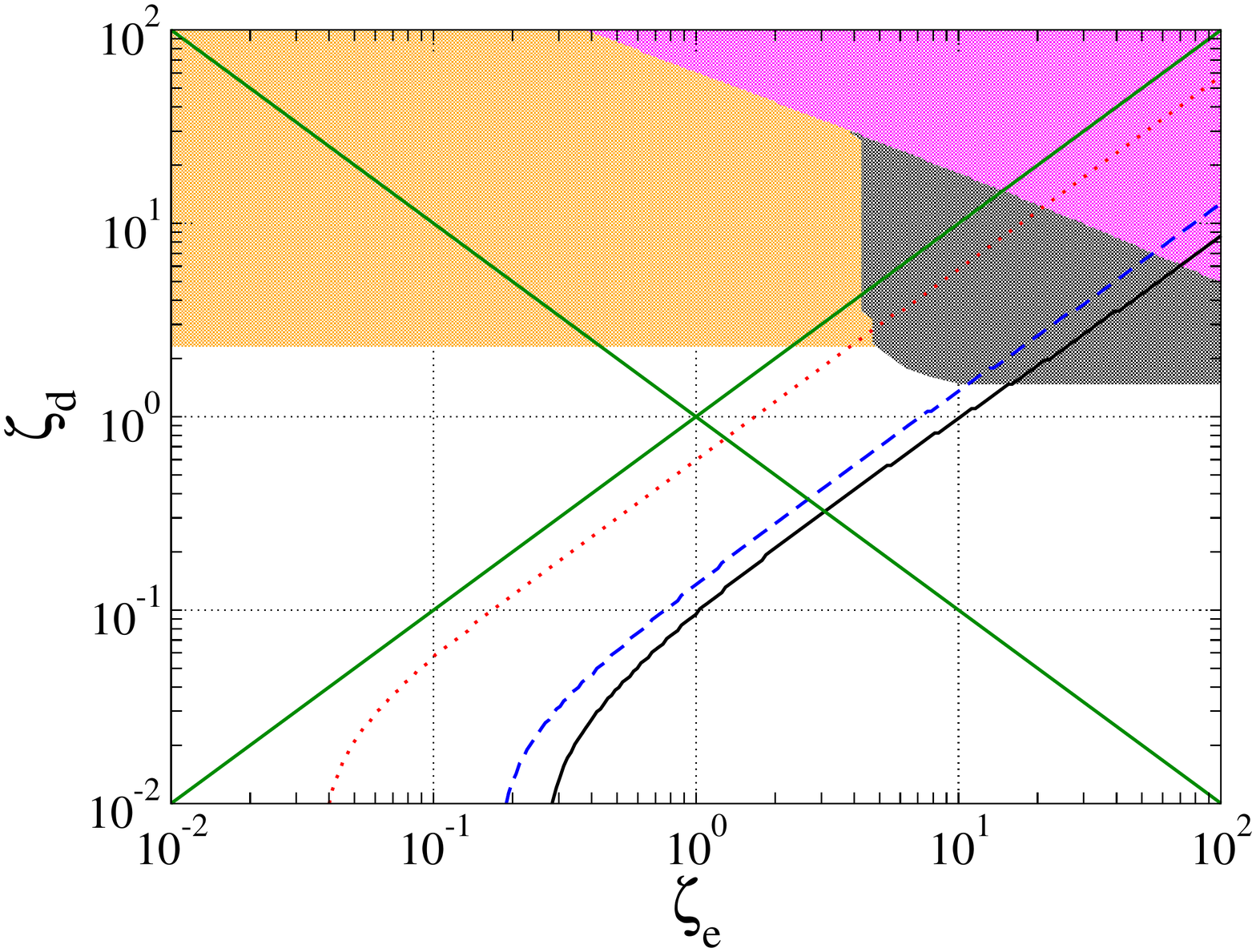}
 \includegraphics[width=8cm]{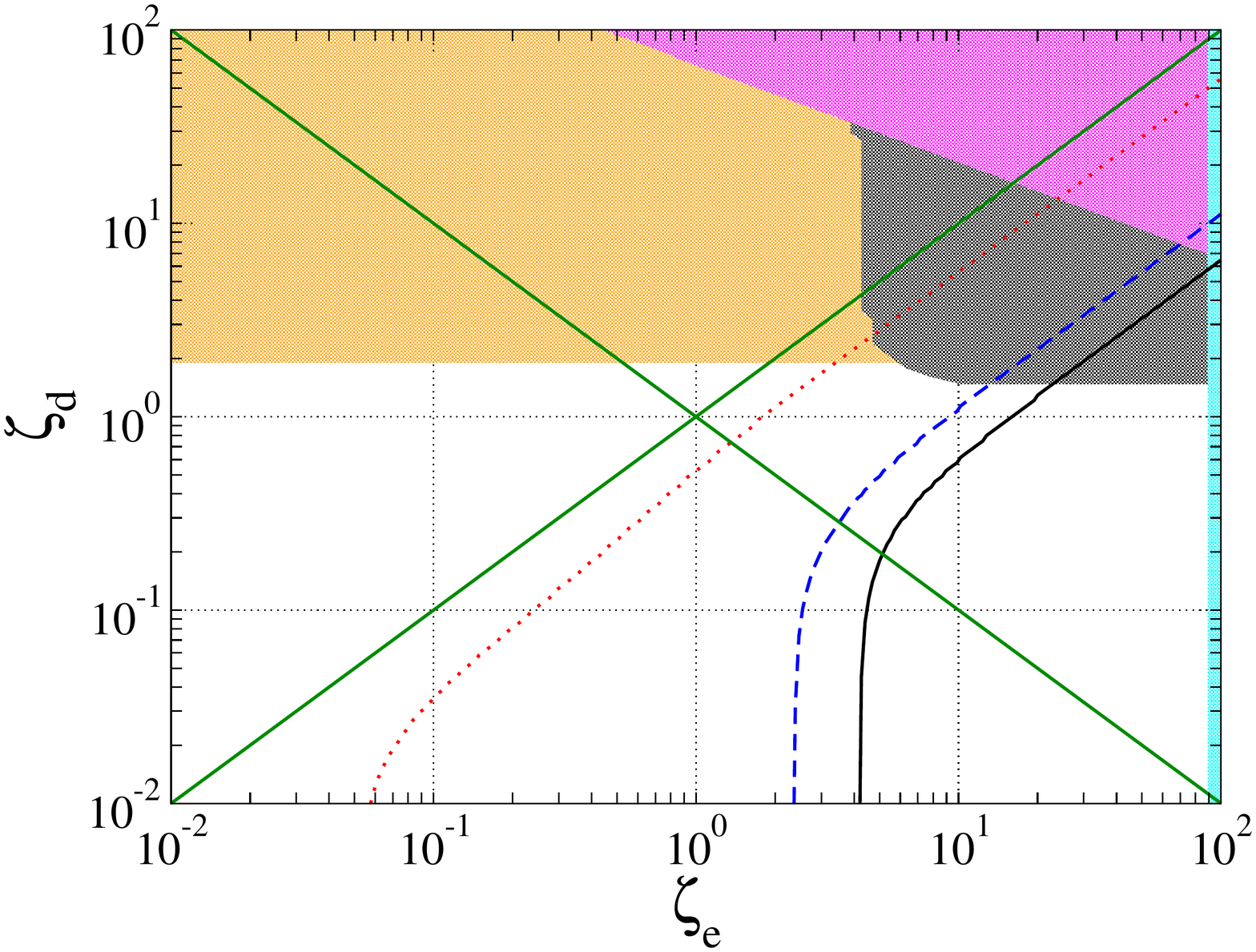}
 \caption{Constraint on the parameter space on the $\zeta_e$--$\zeta_d$ plane in the case with $\zeta_u = 0.1$ for 
 $(m_{H_2^0}, m_{H_3^0}, m_{H^\pm})= (230, 280, 280)$~GeV in the left panel and $(230, 280, 230)$~GeV in the right panel.
 The regions below the black-solid, blue-dashed and red-dotted curves are excluded by the observed LHC Run-II data, 
 the expected LHC Run-II data and data expected at the HL-LHC, respectively.
 The regions shaded by orange, magenta and cyan are excluded by the constraints from $B \to X_s\gamma$, $B_s \to \mu\mu$ and the leptonic tau decay, respectively, 
  while those shaded by black are excluded by the searches for $gg \to H_{2,3}^0 \to \tau^+\tau^-$ and $gg \to b\bar{b}H_{2,3}^0 \to b\bar{b}\tau^+\tau^-$ at the LHC Run-II experiments. 
The green lines indicate the case with $\zeta_d = \zeta_e$ and $\zeta_d = \zeta_e^{-1}$. }
 \label{fig:bp1}
\end{figure}

In Fig.~\ref{fig:bp1}, we show the region of the parameter space excluded by the LHC data and flavor experiments at 95~\% CL in the $\zeta_e$--$\zeta_d$ plane 
for the case with $(m_{H_2^0}, m_{H_3^0}, m_{H^\pm})= (230, 280, 280)$~GeV (left panel) and $(230, 280, 230)$~GeV (right panel). 
These two mass spectra are the representative choices for the heavy ${H^\pm}$ and the light ${H^\pm}$ scenarios.
The $\zeta_u$ parameter is fixed to be $0.1$ in these plots. 
The region below the black-solid (blue-dashed) curves are excluded by using $S_{\rm obs}^{95}$ ($S_{\rm exp}^{95}$) given in Table~\ref{tab:SRs}. 
The red-dotted curve shows the expected exclusion of the parameter space at the HL-LHC (3 ab$^{-1}$). 
We note that these excluded regions are obtained by taking into account all the seven SRs defined in Table~\ref{tab:SRs}. 
Namely, we draw these curves so as to maximize the area of the excluded region. 
We find that these constraints are dominantly determined from SR4, where two isolated leptons 
and at least two $\tau$-tagged jets are required with the lowest $m_{\rm eff}$ threshold $m_{\rm eff}^{\rm th}=600$~GeV. 
The sensitivity tends to increase as the leptonic (mainly tau) branching ratio increases, 
so that it is highly correlated to $R_\tau$ which is the function of $|\zeta_e/\zeta_d|$ and $|\zeta_e/\zeta_u|$ as shown in Fig.~\ref{fig:rtau50}.
Interestingly, even though the light ${H^{\pm}}$ scenario (shown in the right panel) predicts a larger production cross section than the heavy $H^\pm$ case (shown in the left panel), 
the constraint from the LHC data in the former case is weaker. This is because the heavier $H^\pm$ provide more $H_2^0$ via the decay $H^\pm \to W^{\pm(*)}H_2^0$, which give more tau leptons in the final states.
More concretely, we find that the current LHC data excludes the region with $\zeta_e/\zeta_d \gtrsim 10$ $(\zeta_e/\zeta_d \gtrsim 20)$ for the heavy (light) $H^\pm$ case as long as 
$\zeta_e \gtrsim 0.1$ ($\zeta_e \gtrsim 3$). 
These bounds are expected to be significantly improved at the HL-LHC, where the region with $\zeta_e/\zeta_d \gtrsim 2$ can be excluded, shown as the red-dotted curves. 

The constraints from the precision measurements of the flavor observables discussed in Sec.~\ref{sc:flavor} are also overlaid on these panels.
The regions excluded by the measurements of $B\to X_s \gamma$, $B_s \to \mu\mu$ and the leptonic $\tau$ decay are depicted in the orange, magenta and cyan regions, respectively.
In addition, those excluded by the current LHC searches for additional neutral Higgs bosons via the gluon fusion production ($ggH_{2,3}^0$) and the bottom quark associated production $b\bar{b}H_{2,3}^0$ with the subsequent $H_{2,3}\to \tau^+\tau^-$ decay~\cite{ATLAS:2020zms} are depicted by the black regions. 
We note that the flavor constraints are sensitive to the signs (complex phases) of the $\zeta_f$ parameters while
the constraints from the LHC multi-lepton searches are essentially insensitive to those.
We see that $B \to X_s \gamma$ sets the upper limit on the value of $\zeta_d$ to be about 2, which does not depend on $\zeta_e$. 
On the other hand, the constraint from the single Higgs productions (black shaded region) excludes the region with larger values of $\zeta_e$ and $\zeta_d$, in which the exclusion is almost given only by the 
$pp \to b\bar{b}H_{2,3}^0 \to b\bar{b}\tau^+\tau^-$ channel. 
The region with $\zeta_e \gtrsim {\cal O}(100)$ is excluded by the leptonic tau decay. 
No further region is excluded by the $B_s \to \mu\mu$ data in this setup. 
As a summary of this figure, we clarify that except for the bottom-left region i.e., $\zeta_e/\zeta_d \lesssim {\cal O}(1)$ with $\zeta_d\lesssim {\cal O}(1)$, 
most of the parameter region can be excluded or explored after combining all the constraints (including the expected bound from the HL-LHC) considered in this section.

\subsection{Dependence on the masses of the additional Higgs bosons}

\begin{figure}[t]
\hspace{11.8cm} {\includegraphics[width=4.55cm]{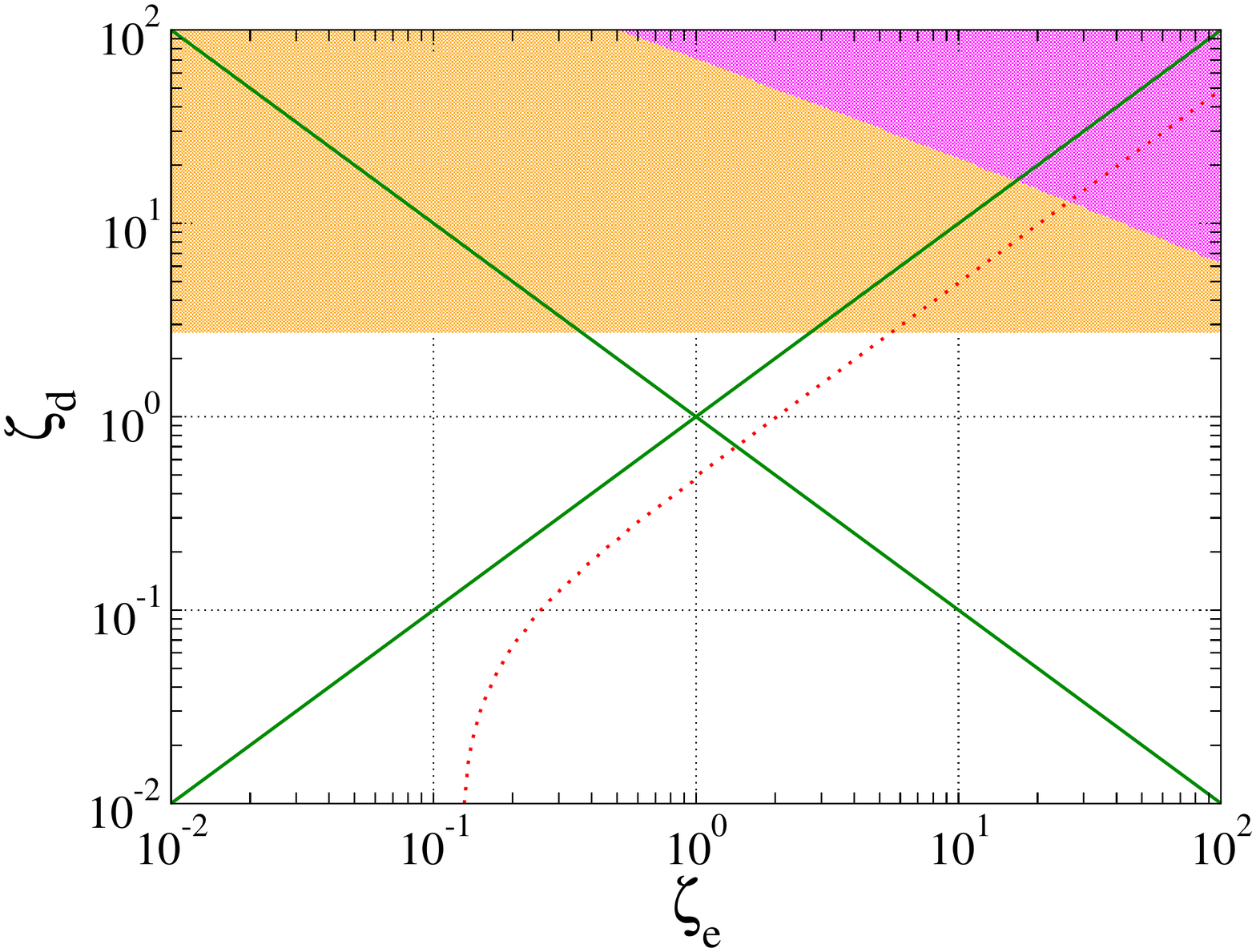}~\hspace{-0.5cm}~\ }\\
\hspace{7.5cm} 
{
\includegraphics[width=4.55cm]{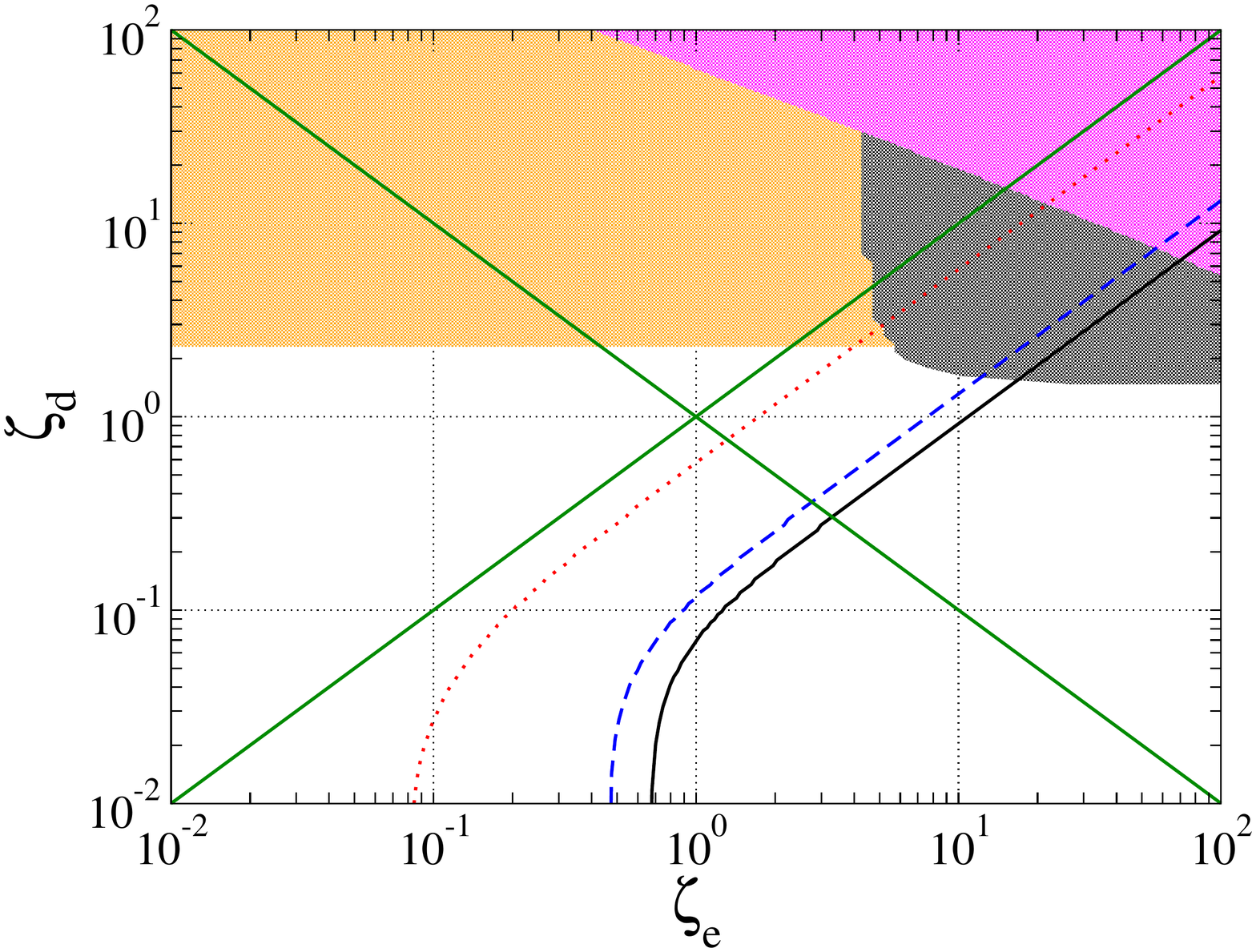}\hspace{-0.5cm}
\includegraphics[width=4.55cm]{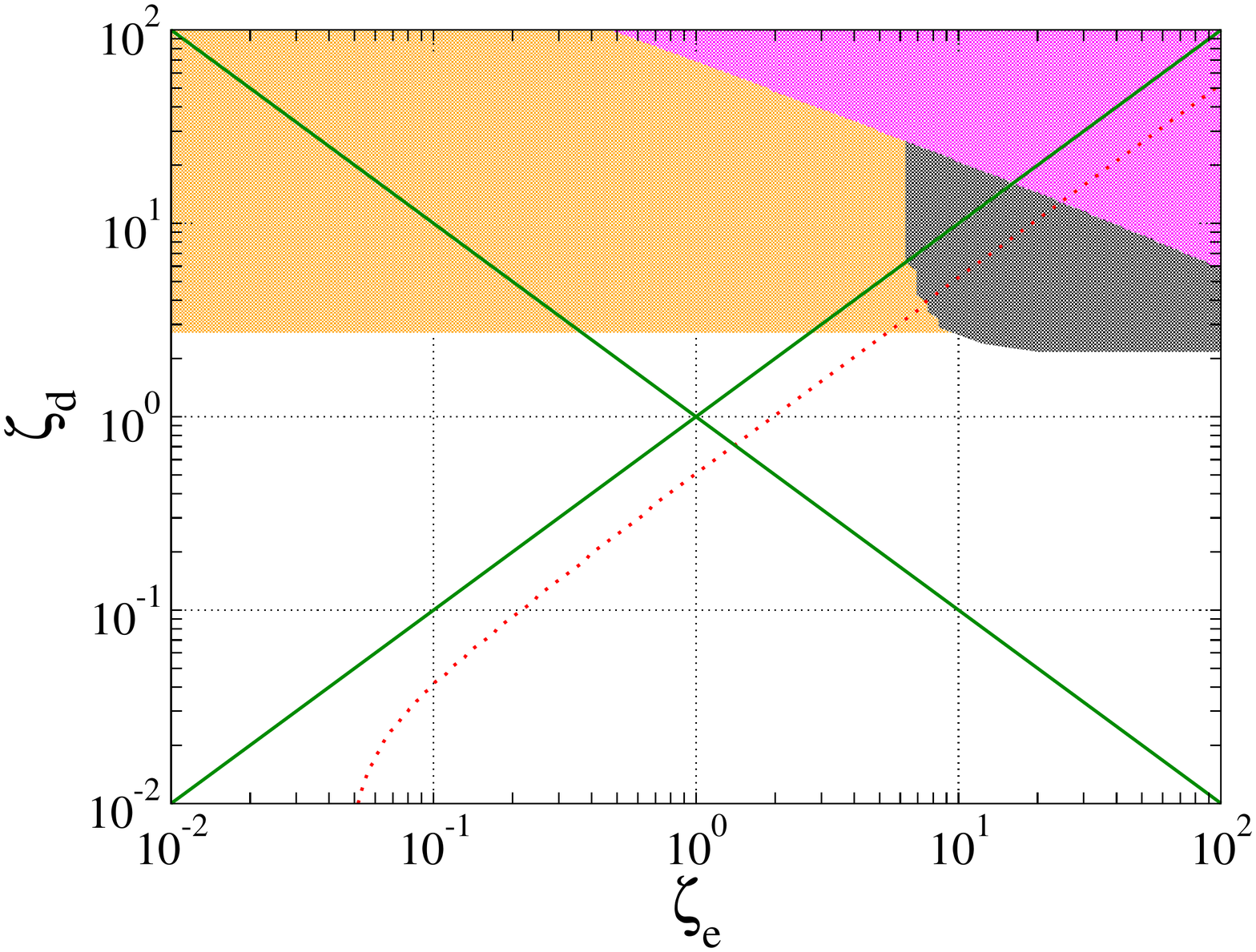}\hspace{-0.5cm}
}
\\
 \hspace{3.35cm}
{
\includegraphics[width=4.55cm]{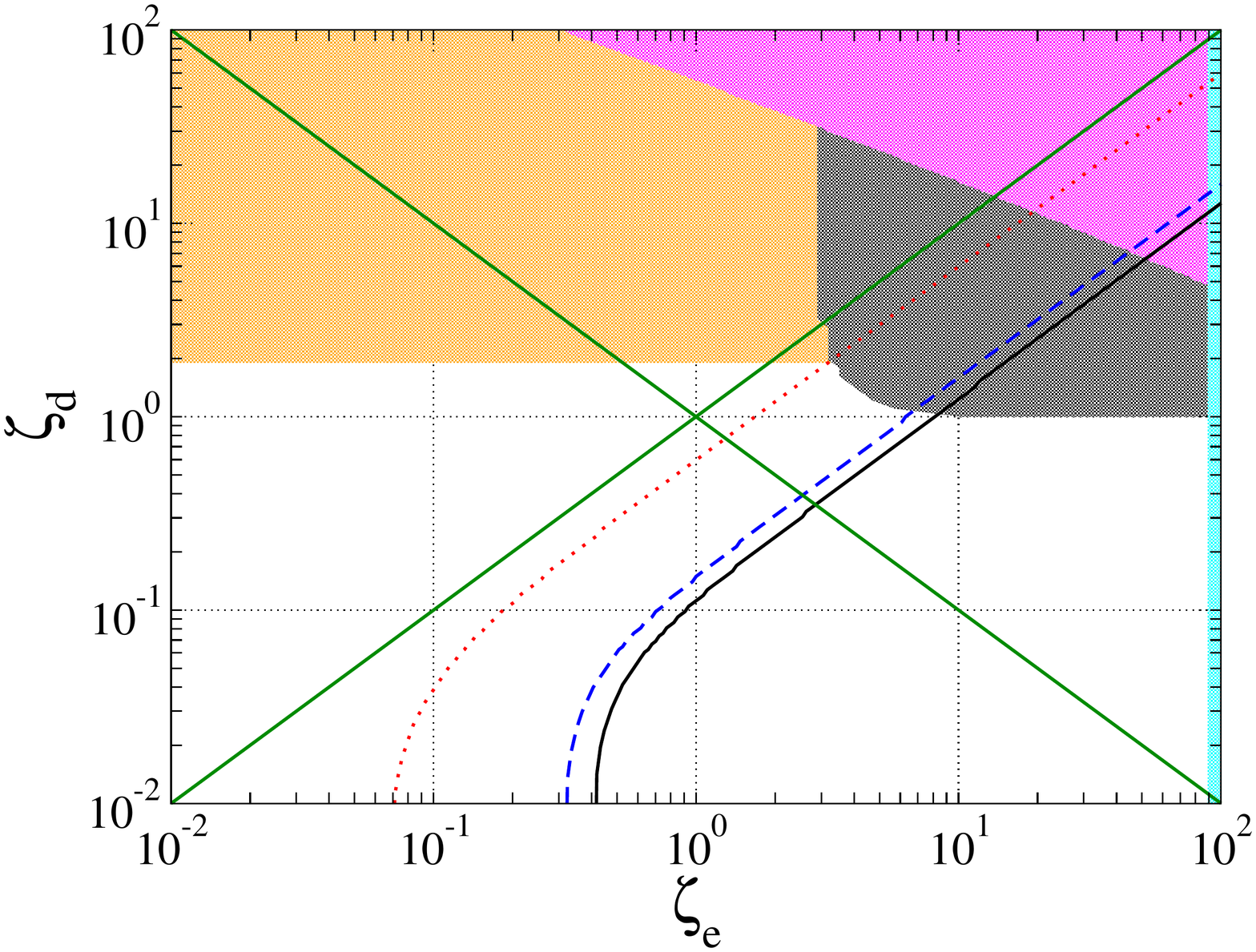}\hspace{-0.5cm}
\includegraphics[width=4.55cm]{./fig/280_230_280_combined_zetau01.pdf}\hspace{-0.5cm}
\includegraphics[width=4.55cm]{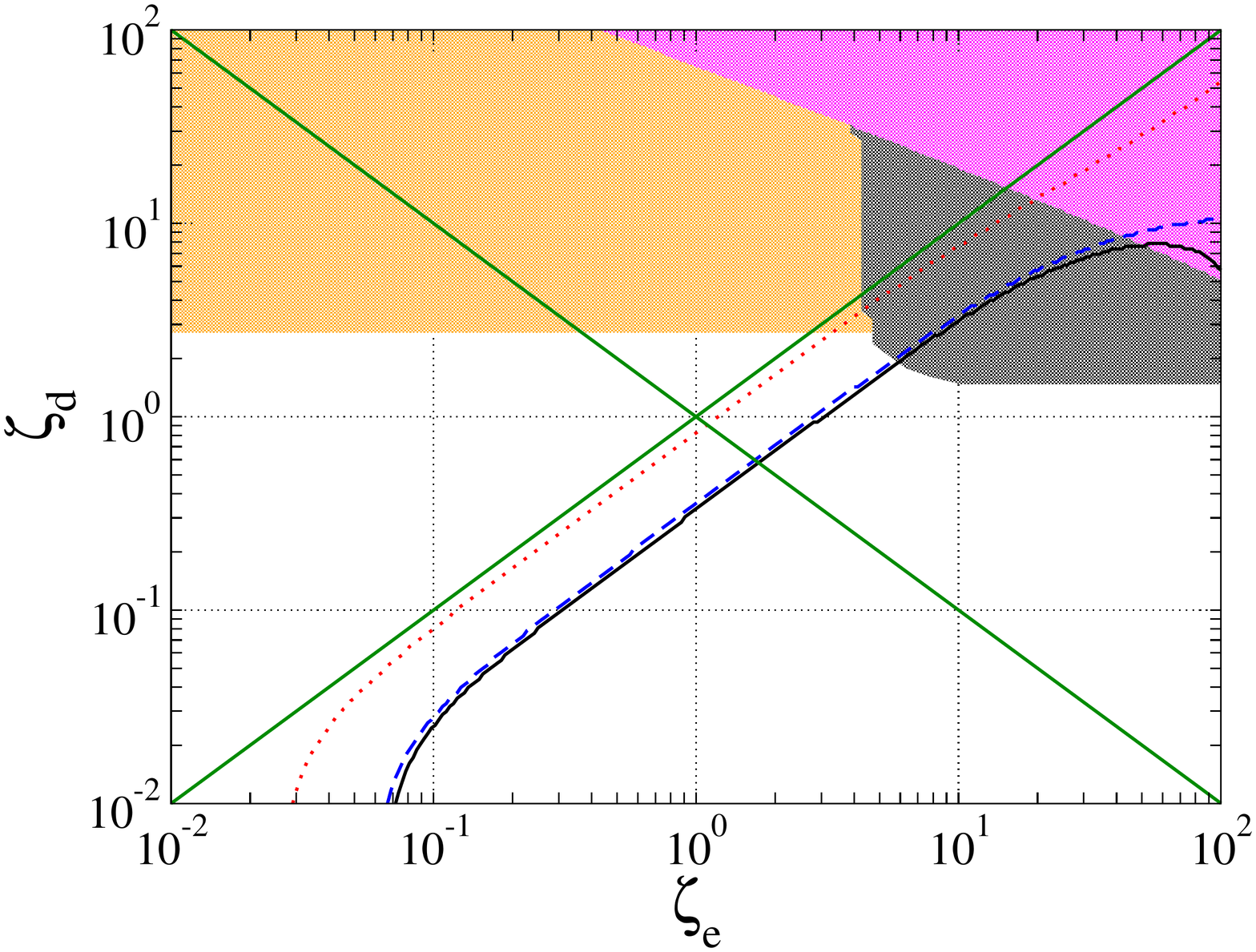}\hspace{-0.5cm}
}
\\
 \hspace{-0.7cm}
 {
 \includegraphics[width=4.55cm]{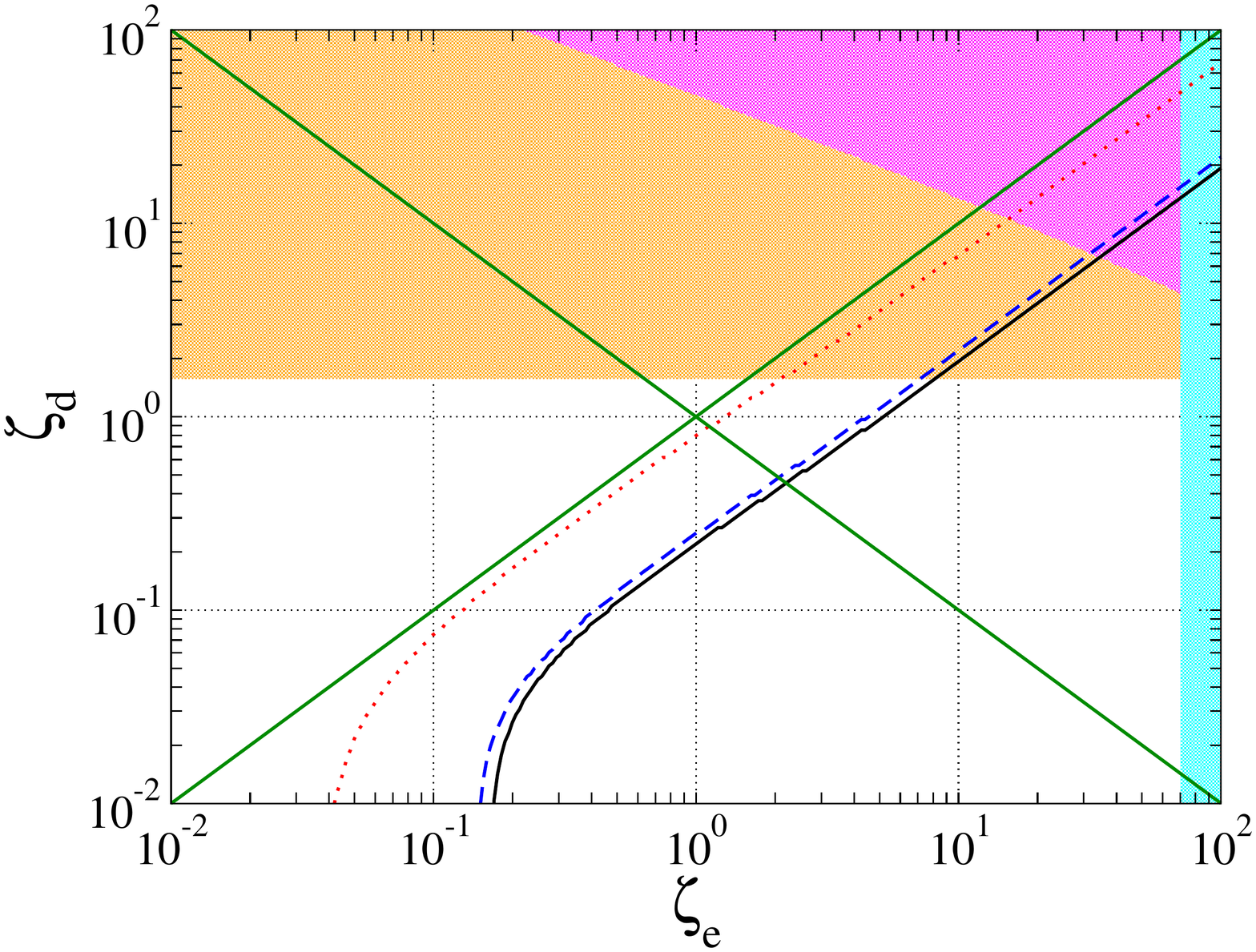}\hspace{-0.5cm}
 \includegraphics[width=4.55cm]{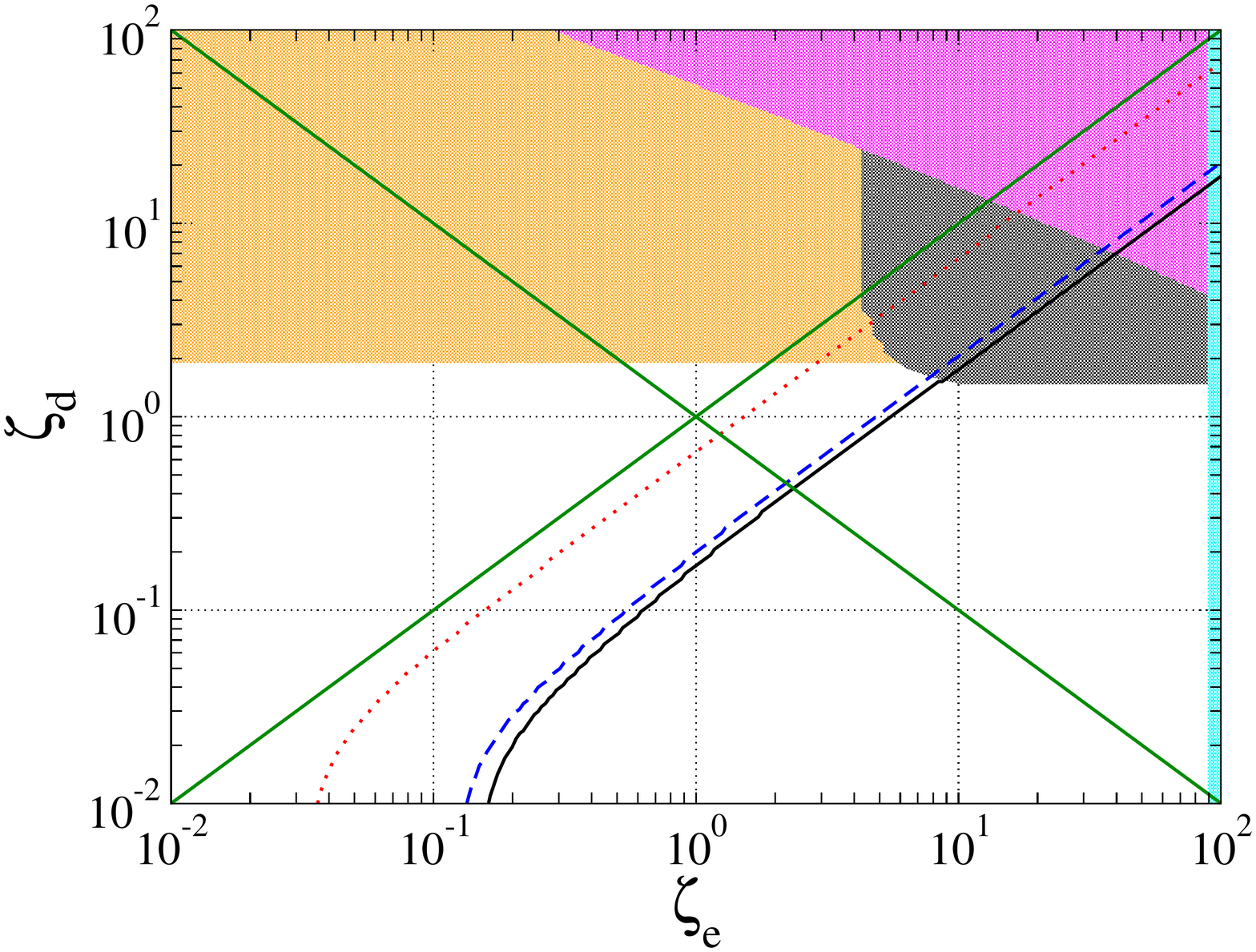}\hspace{-0.5cm}
 \includegraphics[width=4.55cm]{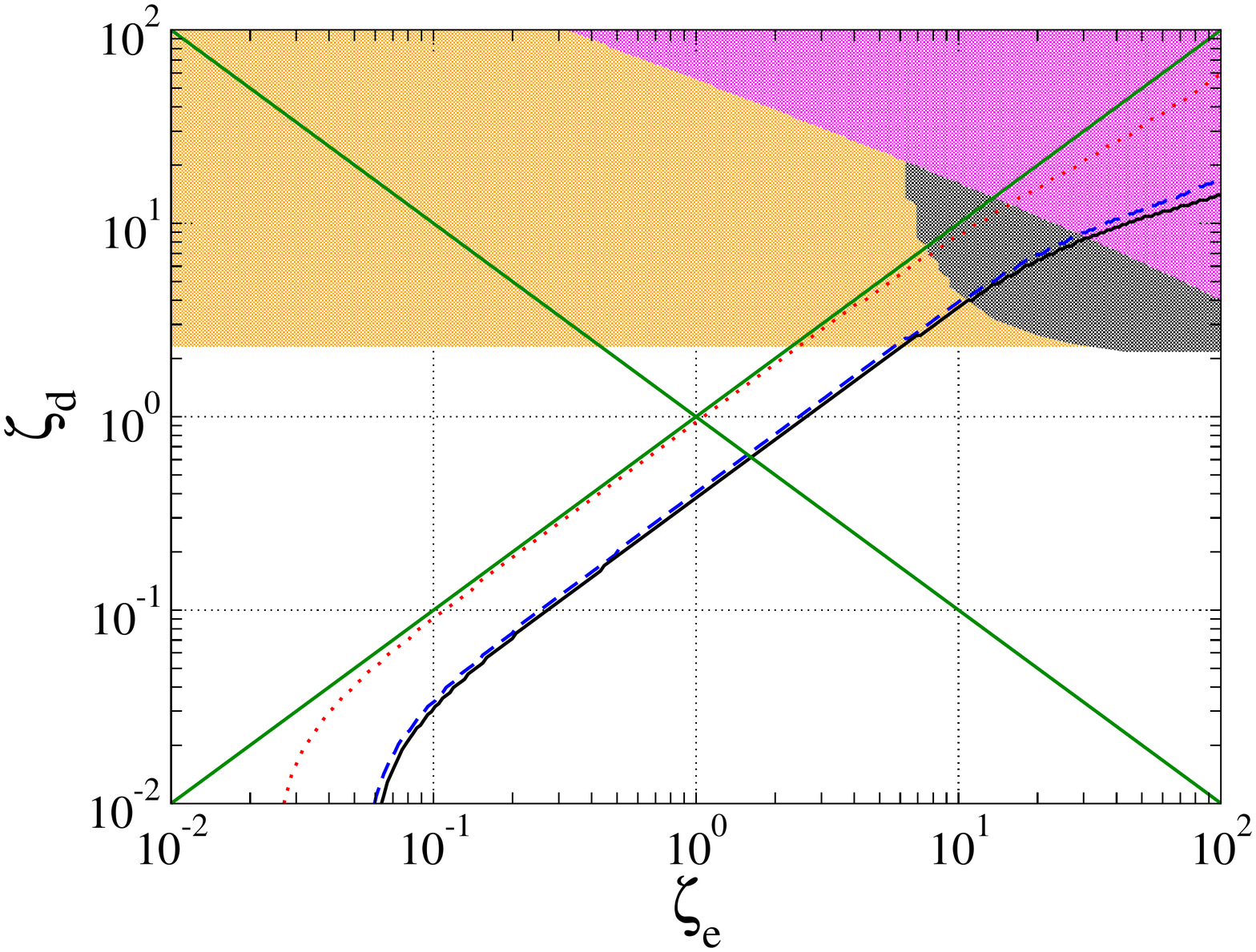}\hspace{-0.5cm}
 \includegraphics[width=4.55cm]{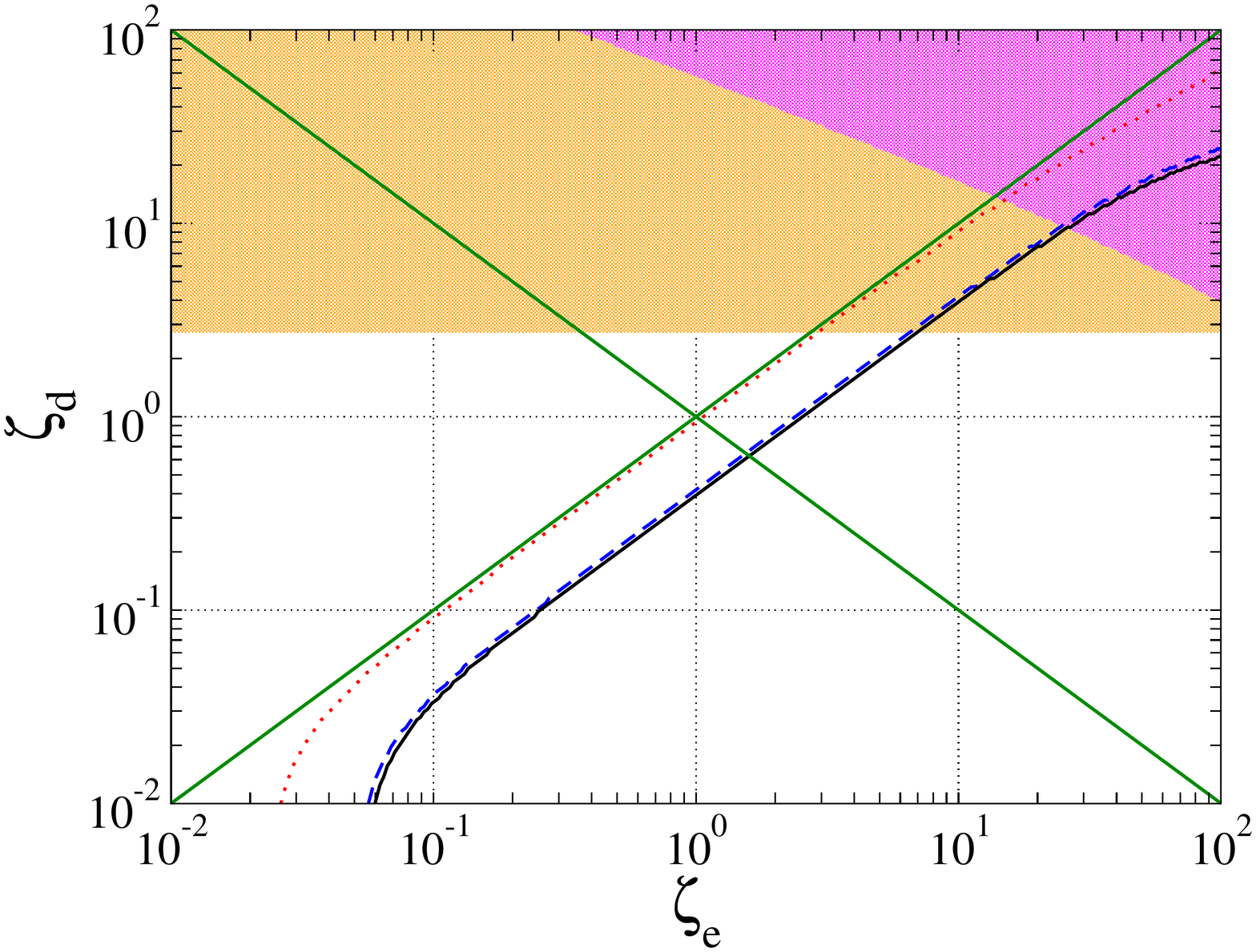}\hspace{-0.5cm}
}
\caption{Constraint on the parameter space on the $\zeta_e$--$\zeta_d$ plane for $\zeta_u = 0.1$ in
the heavy $H^\pm$ scenario ($m_{H^\pm} = m_{H_3^0} \geq m_{H_2^0}$).
The bottom to top panels show the case with 
$m_{H_2^0} = 180$, 230,  280 and 330 GeV
while the most left panels show the case with $|\Delta m| (=m_{H_3^0} - m_{H_2^0}) = 0$, 
and next panels show the case incrementing $\Delta m$ with a step of 50 GeV (up to $m_{H_3^0}=330$ GeV). 
In each panel, the descriptions of curves and shaded regions are the same as those given in Fig.~\ref{fig:bp1}. 
}
   \label{fig:comb1}
\end{figure}

\begin{figure}[t]
\hspace{11.8cm} {\includegraphics[width=4.55cm]{./fig/330_330_330_combined_zetau01.pdf}~\hspace{-0.5cm}~\ }\\
\hspace{7.5cm} 
{
\includegraphics[width=4.55cm]{./fig/280_280_280_combined_zetau01.pdf}\hspace{-0.5cm}
\includegraphics[width=4.55cm]{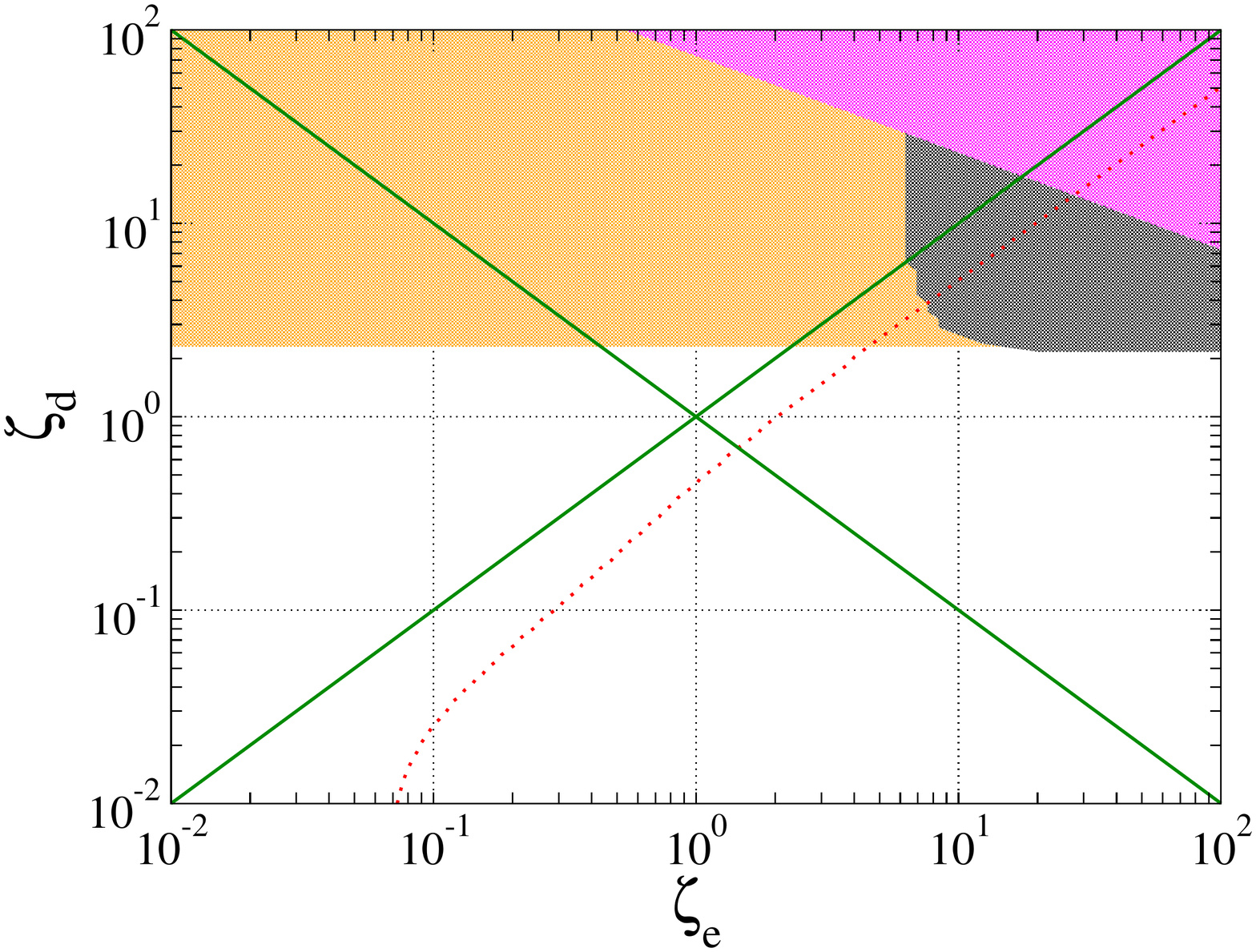}\hspace{-0.5cm}
}
\\
 \hspace{3.35cm}
{
\includegraphics[width=4.55cm]{./fig/230_230_230_combined_zetau01.pdf}\hspace{-0.5cm}
\includegraphics[width=4.55cm]{./fig/280_230_230_combined_zetau01.pdf}\hspace{-0.5cm}
\includegraphics[width=4.55cm]{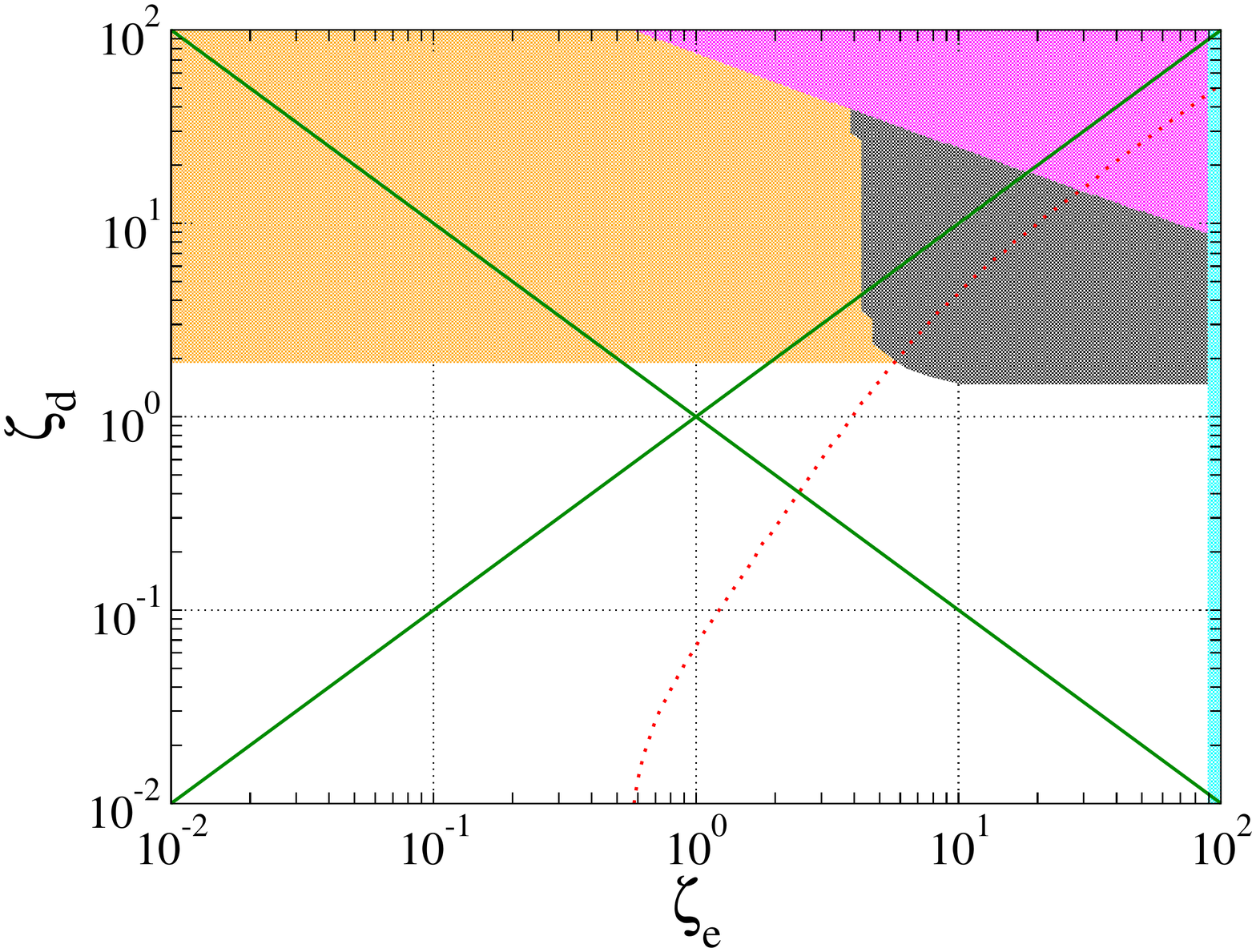}\hspace{-0.5cm}
}
\\
 \hspace{-0.7cm}
 {
 \includegraphics[width=4.55cm]{./fig/180_180_180_combined_zetau01.pdf}\hspace{-0.5cm}
 \includegraphics[width=4.55cm]{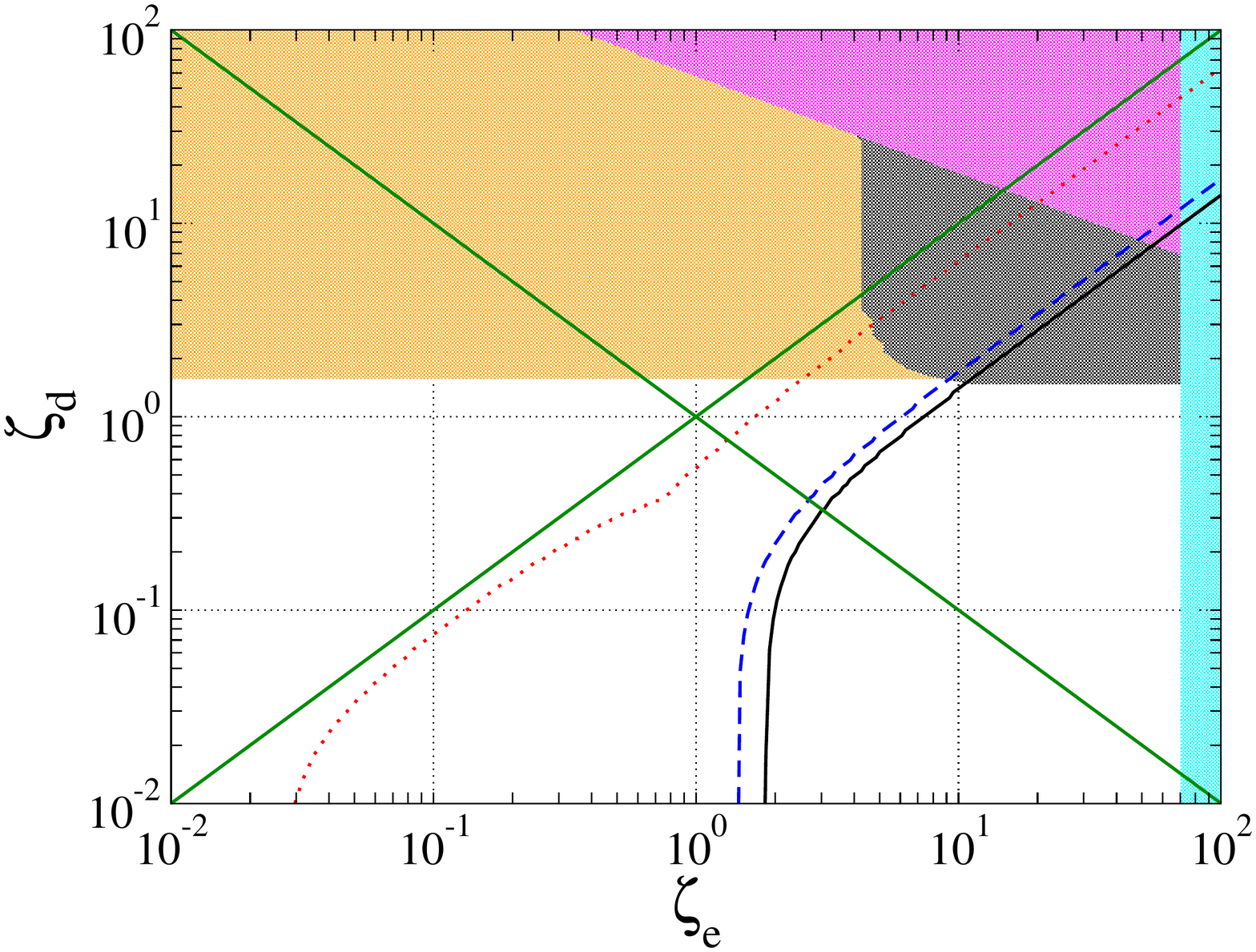}\hspace{-0.5cm}
 \includegraphics[width=4.55cm]{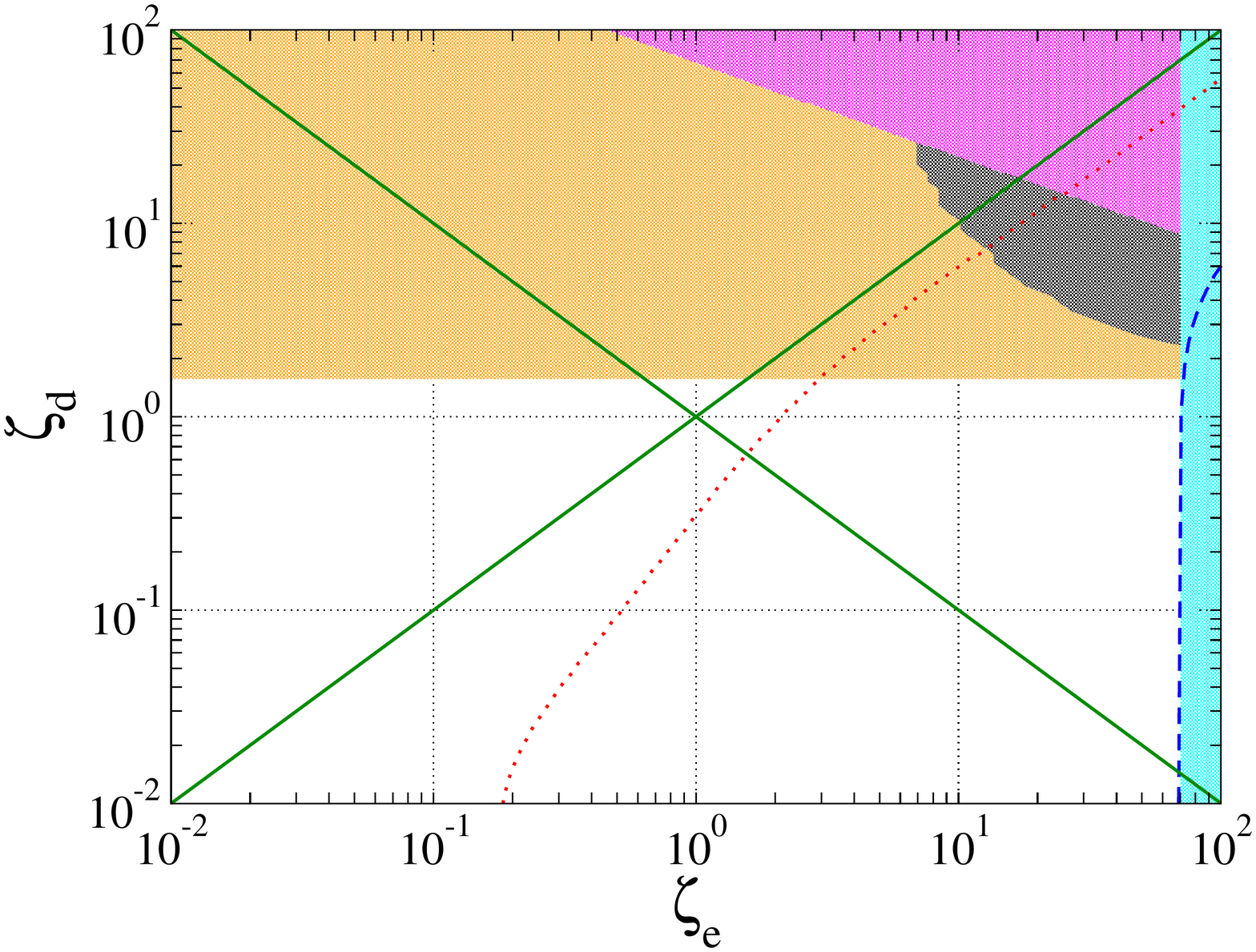}\hspace{-0.5cm}
 \includegraphics[width=4.55cm]{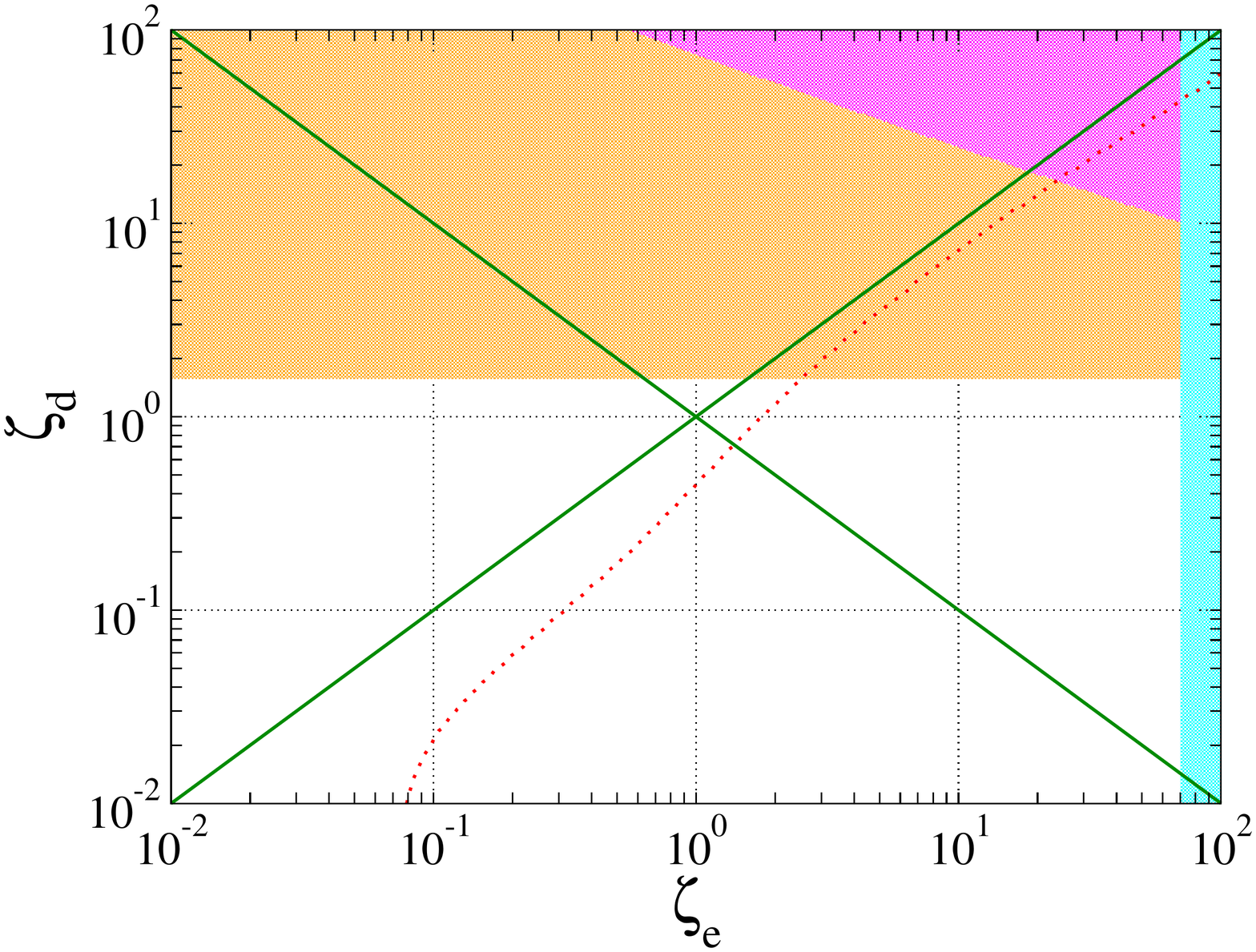}\hspace{-0.5cm}
}
\caption{Same as Fig.~\ref{fig:comb1}  but for the case with the light $H^\pm$ scenario ($m_{H^\pm} = m_{H_2^0} \leq m_{H_3^0}$). 
}
\label{fig:comb2}
\end{figure}

We discuss the constraints on the parameter space for the different 16 mass spectra defined in Eq.~(\ref{eq:spectra}) in order to see the mass dependence of the constraints. 
Figs.~\ref{fig:comb1} and \ref{fig:comb2} respectively show the results for the heavy $H^\pm$ scenario and the light $H^\pm$ scenario.
The descriptions for each curve and shaded region are the same as those explained in Fig.~\ref{fig:bp1}. 


In Fig.~\ref{fig:comb1}, we see that slightly larger regions are excluded for the case with $|\Delta m| \neq 0$ as compared to that with $|\Delta m| = 0$
for $m_{H_2^0} = 180$ GeV (bottom panels) and 230 GeV (next to bottom panels). 
This can be understood by the effect of the $H^\pm \to W^{\pm(*)} H_2^0$ decay with $H_2^0 \to \tau^+\tau^-$, providing more tau leptons at final states. 
We note that the decay of $H^\pm \to W^{\pm(*)} H_2^0$ is suppressed for larger values of $\zeta_e$ and/or $\zeta_d$ due to the enhancement of the $H^\pm \to \tau\nu$ and $H^\pm \to tb$ modes, 
so that in such a region the limit is slightly weaker.
For $m_{H_2^0} = 280$ GeV with $|\Delta m| =50$ GeV (right panel in the second row) and $m_{H_2^0} = 280$ GeV (top panel), 
no bound is obtained from the current LHC data, which is simply because of the smaller production cross section as compared with 
the case with $|\Delta m| =0$. On the other hand, the HL-LHC sets the limit on $\zeta_e$ and $\zeta_d$ for these cases. 
The behavior of the flavor constraints is similar to that shown in Fig.~\ref{fig:bp1}, while the larger region can be excluded for the case with smaller $m_{H^\pm}$. 


In Fig.~\ref{fig:comb2}, we see that the smaller area of the parameter region is excluded by the LHC data as compared with the heavy $H^\pm$ scenario shown in Fig.~\ref{fig:comb1}. 
This is because $H_3^0$ can decay into not only $Z^{(*)} H_2^0$ but also $W^{\pm(*)} H^\pm$, where the former produces a tau lepton pair in its subsequent decay while
the latter provides $\tau^\pm \nu$.  
In fact, it is seen that no bound is obtained for $|\Delta m|\geq 100$ GeV ($|\Delta m|\geq 50$ GeV) for $m_{H_2^0} = 180$ and 230 GeV (280 GeV) from the current LHC data, while 
the HL-LHC can exclude a portion of the parameter space in such cases. 




In Appendix~\ref{sec:zetau0}, we show the similar plots for $\zeta_u = 0$ in  Figs.~\ref{fig:comb1_zetau0} and \ref{fig:comb2_zetau0}. 
The main difference in the collider study from the case with $\zeta_u=0.1$ is 
the absence of $H_{2,3}^0 \to c\bar{c}$ and the $H^\pm \to tb$ decay being more suppressed.
We can see that the above effects slightly make the LHC bound stronger particularly in the region with smaller $\zeta_e$ and $\zeta_d$ values as compared with the case with $\zeta_u = 0.1$. 
In addition, the constraint from $B \to X_s \gamma$ disappears in this case, so that the case with smaller $\zeta_e/\zeta_d$ values is allowed. 
%

Since the LHC bound is almost described by a line with a constant $\zeta_e/\zeta_d$ value, 
we can extract the value of $\zeta_e$ along the line $\zeta_e=\zeta_d^{-1}$ as a representative point. 
In this way, we can set the upper limit on $\zeta_e$ as a function of the mass of the additional Higgs boson. 
We here particularly discuss the upper limit on $\zeta_e$ in the Type-X THDM which can be regarded as a special case of the aligned THDM, i.e.,  
$\zeta_e = \zeta_d^{-1} = \zeta_u^{-1}$, see Eq.~(\ref{eq:types}). 

\begin{figure}[t]
 \includegraphics[width=12cm]{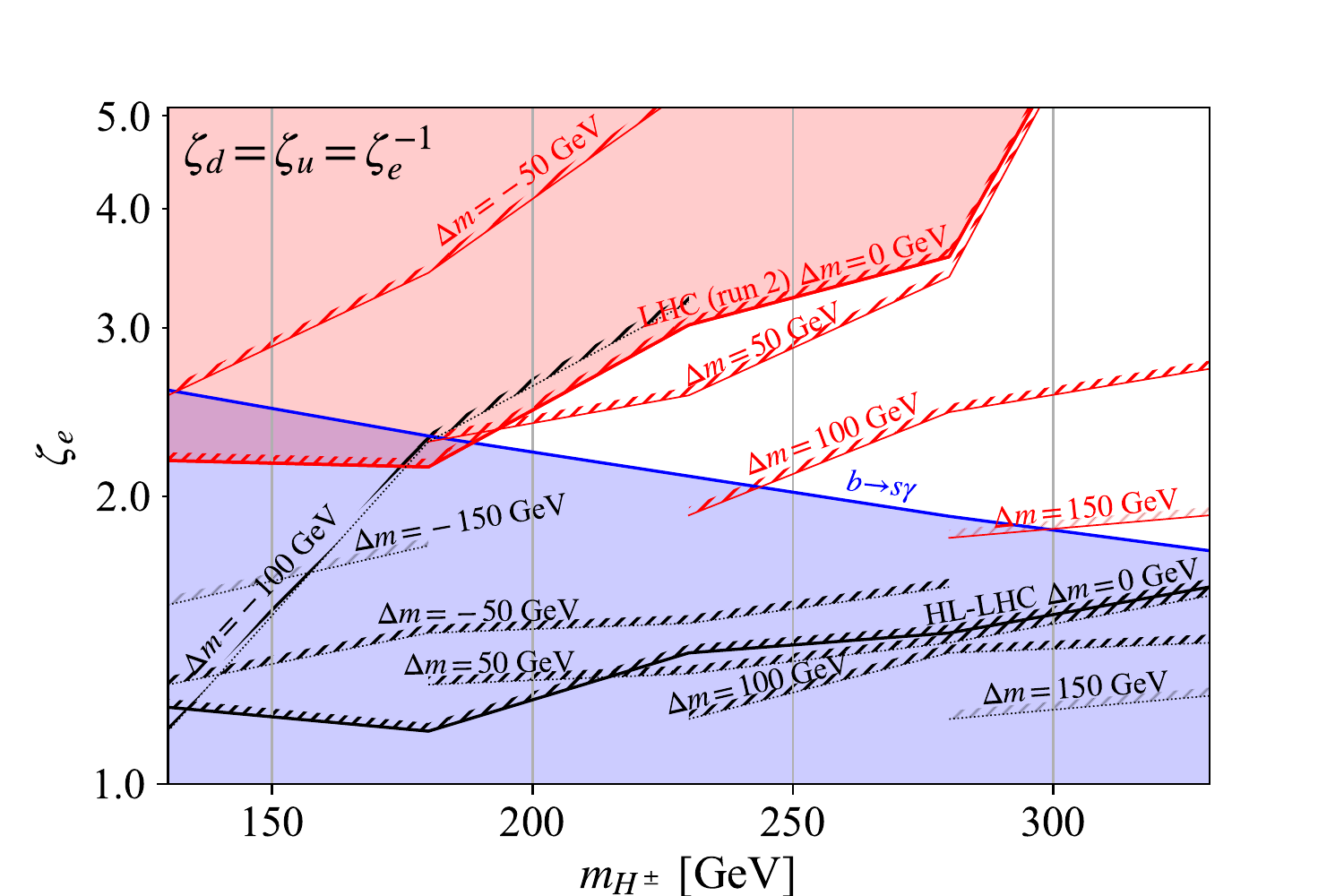}
 \caption{Constraints on the parameter $\zeta_e$ in the Type-X THDM ($\zeta_u =\zeta_d = \zeta_e^{-1}$) as a function of the mass of $H^\pm$.
The shaded regions are excluded by $B \to X_s \gamma$ (blue) and by the current LHC multi-lepton searches (red) with $\Delta m = 0$.
The region above each red (black) curve is excluded by the multi-lepton searches at LHC (the HL-LHC) with several fixed value of $\Delta m$. }
 \label{fig:typeXlike}
\end{figure}

In Fig.~\ref{fig:typeXlike}, we show the constraints on the $\zeta_e$ value by the LHC multi-lepton searches 
as a function of $m_{H^{\pm}}$ in the Type-X THDM $(\zeta_e=\zeta_d^{-1}=\zeta_u^{-1})$.
The region above the red line is excluded, where the different lines correspond to the cases with different $\Delta m$.
In particular, we fill out the excluded region by a red color for $\Delta m = 0$. 
%
%
For the same $m_{H^\pm}$ value, a larger $\Delta m$ would set a stronger constraint. 
To see the difference between the heavy/light $H^\pm$ scenarios discussed above, for example, 
comparing the sensitivity at $m_{H^\pm}=200$~GeV for $\Delta m=50$~GeV  and  
that at $m_{H^\pm}=250$~GeV for $\Delta m=-50$~GeV would be helpful, which share the 
same $m_{H_2^0}$ and $m_{H_3^0}$ values but with different $m_{H^\pm}$ values. One can see 
the stronger constraint on $\zeta_e$ for the larger $m_{H^\pm}$ case.
In this figure, the blue shaded region describes the excluded region from the $B \to X_s\gamma$ data. 
One can see that most of the region is highly constrained for $\Delta m > 0$, and especially most of the parameter space is already excluded for $\Delta m = 150$~GeV when we restrict the masses of the additional Higgs bosons to be lighter than $2m_t$.
The future HL-LHC reaches are shown by the black lines. 
There are almost no allowed region by combing the constraints from the HL-LHC and $B \to X_s\gamma$. 
Only exception is the case with $\Delta m = -100$~GeV, where the $H_3^0 \to H^\pm W^\mp$ and $H_3^0 \to H_2^0 Z$ modes dominate 
for such a small $\zeta_e$ case.
The reason why the constraint on $\zeta_e$ for $\Delta m = -100$~GeV is weak 
compared with the case for $\Delta m = -150$~GeV at $m_{H^\pm} = 180$~GeV
is that the $H_3^0 \to H_2^0 Z$ mode is relatively small compared with the $H_3^0 \to H^\pm W^\mp$ mode 
by the phase space suppression in the former case, where the $H_2^0 \to \tau^+\tau^-$ mode provides 
an important contribution to the multi-lepton events. On the other hand, at $m_{H^\pm} = 130$~GeV, 
that for $\Delta m = -100$~GeV is strong compared with that for $\Delta m = -150$~GeV. 
This is because the $H_3^0 \to H^\pm W^\mp \to (\tau^+\nu) (\tau^-\nu)$ mode provides an important contribution instead, where no competition between $\tau\nu$ and $tb$ modes exists.
We conclude that the Type-X THDM can be almost completely probed 
by the searches for the additional Higgs bosons at the HL-LHC when their masses are smaller than the $2m_t$.
We note again that the constraints from $B \to X_s\gamma$  are sensitive to the complex phases of the $\zeta_f$ parameters.


\subsection{Prospect for mass measurements of additional Higgs bosons}\label{sc:massmeasurement}
We consider here briefly the prospects of the mass measurements for the additional Higgs bosons at the HL-LHC.
It is important as a basis of more dedicated study for the property measurements of additional Higgs bosons 
at LHC as well as the other future collider experiments including ILC. 
For example, the measurements of the CP property of the additional Higgs bosons at ILC has 
been discussed assuming the mass measurements are available at LHC~\cite{Kanemura:2020ibp}.
We take the several benchmark points which are currently not excluded but in future possibly probed at LHC, and demonstrate the 
strategy to reconstruct the masses of the additional Higgs bosons.
A summary of the benchmark points taken are shown in Table~\ref{tab:bps}.

\begin{table}[t]
\begin{tabular}{c|c|ccc}
\hline\hline
\ \ scenario \ \ & \ \ \ BPs\ \ \  & $m_{H_2^0}$ & $m_{H_3^0}$ & $m_{H^\pm}$ \cr
\hline
&BP1 & 280~GeV & 180~GeV  & 280~GeV  \cr
heavy $H^\pm$ &BP2& 280~GeV  & 230~GeV  & 280~GeV  \cr
&BP3 & 230~GeV  & 180~GeV  & 230~GeV  \cr
\hline
light $H^\pm$  & BP4 & 280~GeV  & 180~GeV  & 180~GeV  \cr
\hline\hline
\end{tabular}
\caption{Summary of the benchmark points.
}
   \label{tab:bps}
\end{table}

\begin{figure}[]
\centering
\includegraphics[width=43mm]{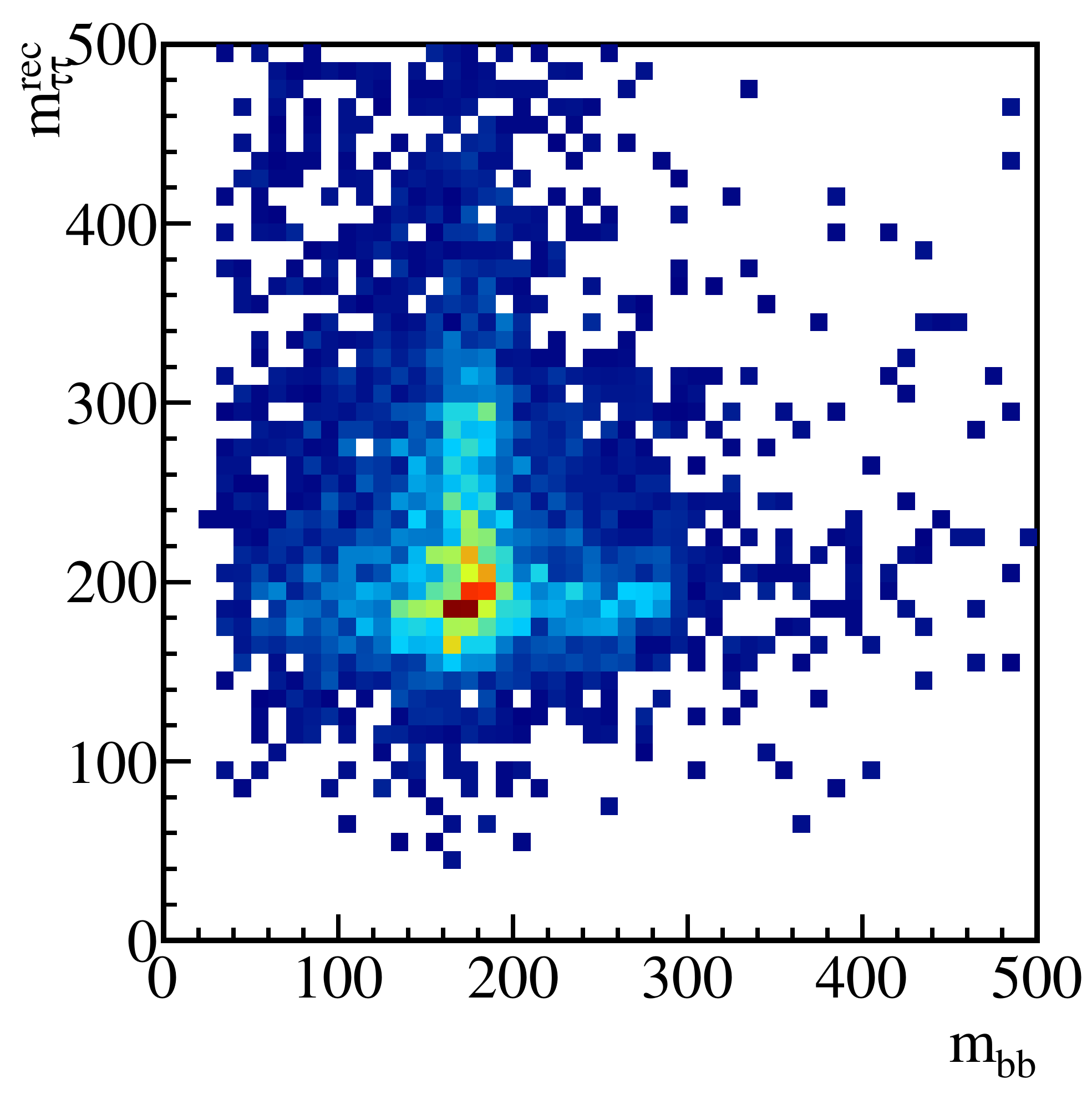}
\includegraphics[width=43mm]{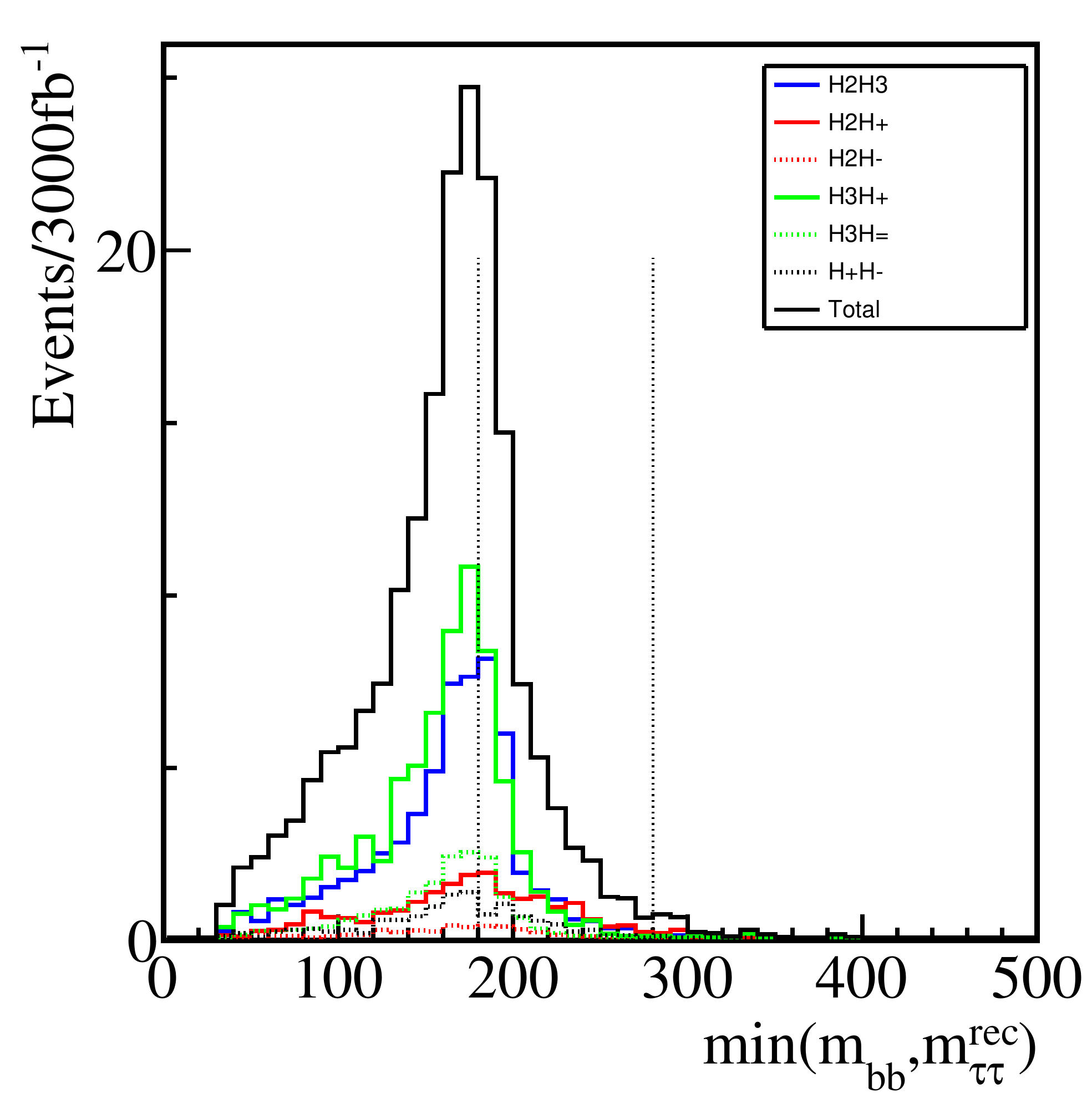}
\includegraphics[width=43mm]{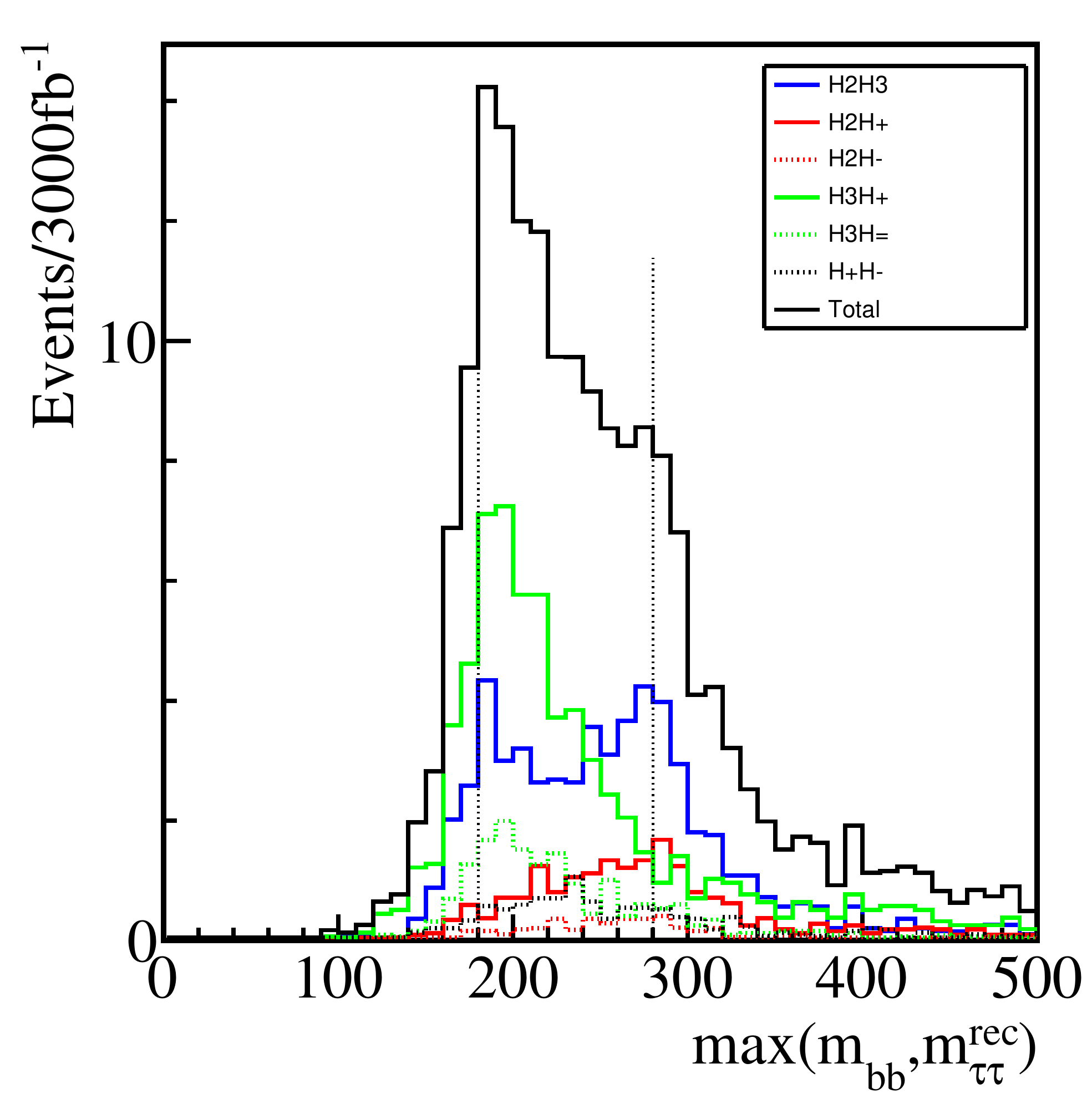}

\includegraphics[width=43mm]{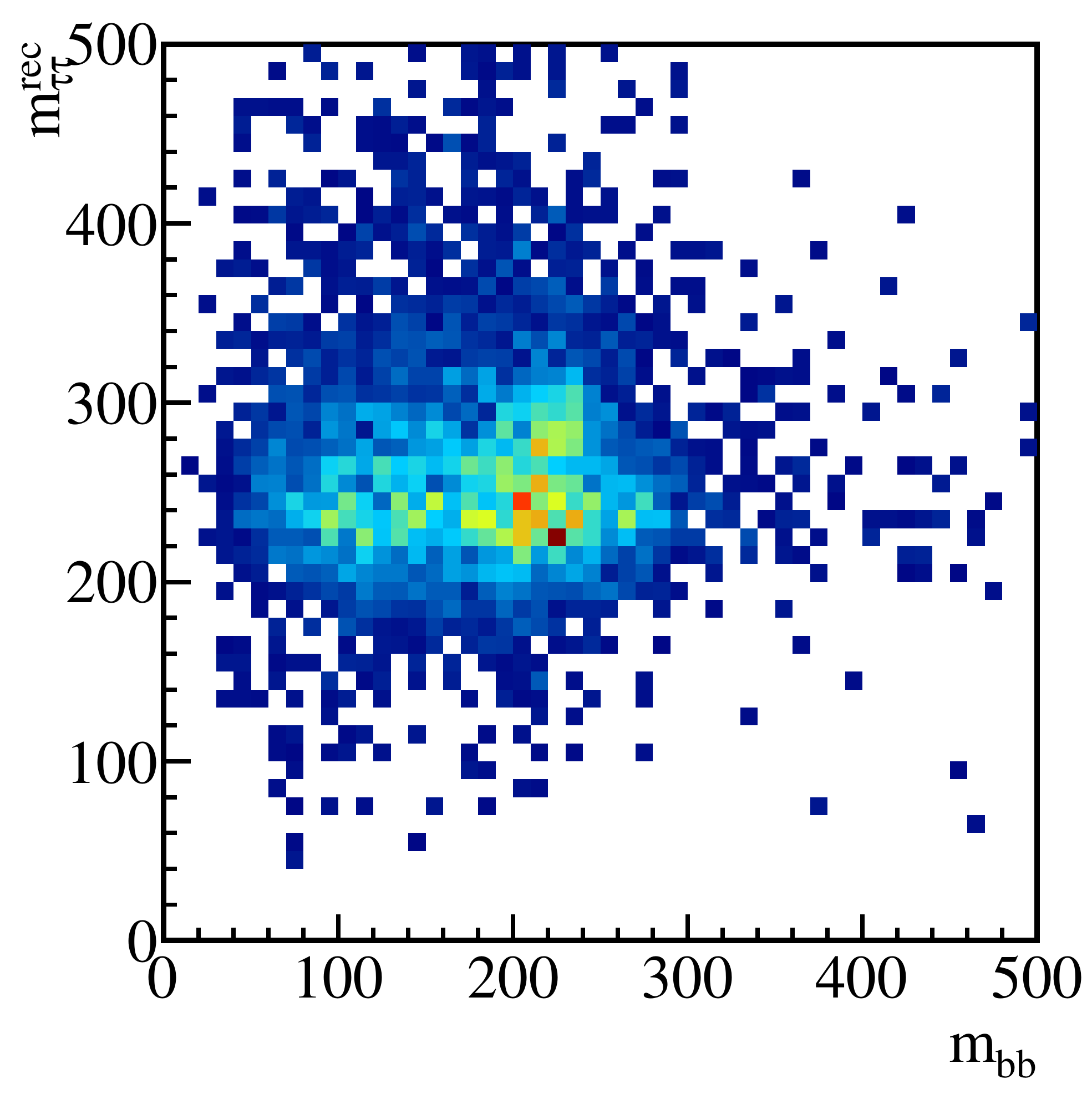}
\includegraphics[width=43mm]{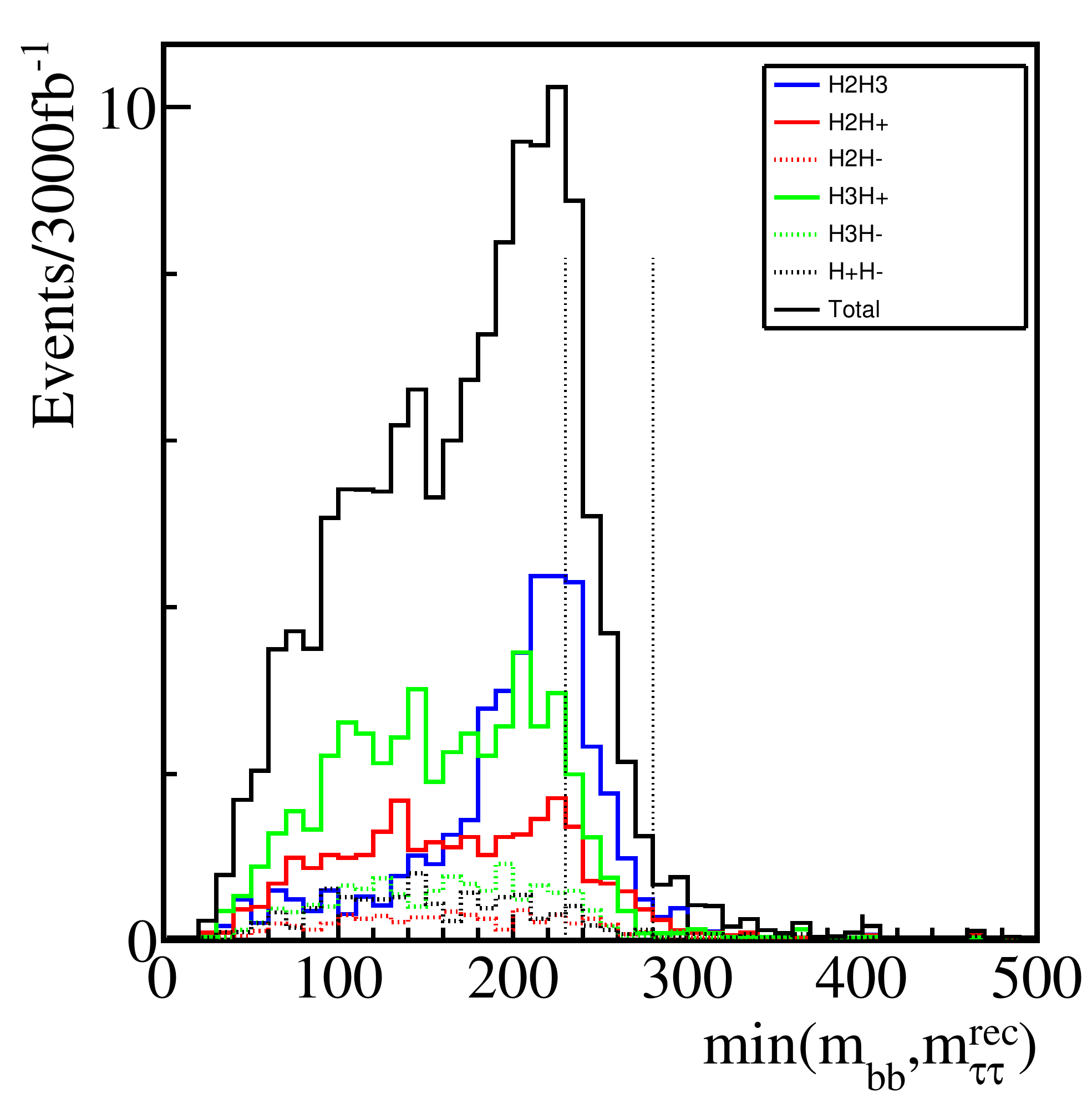}
\includegraphics[width=43mm]{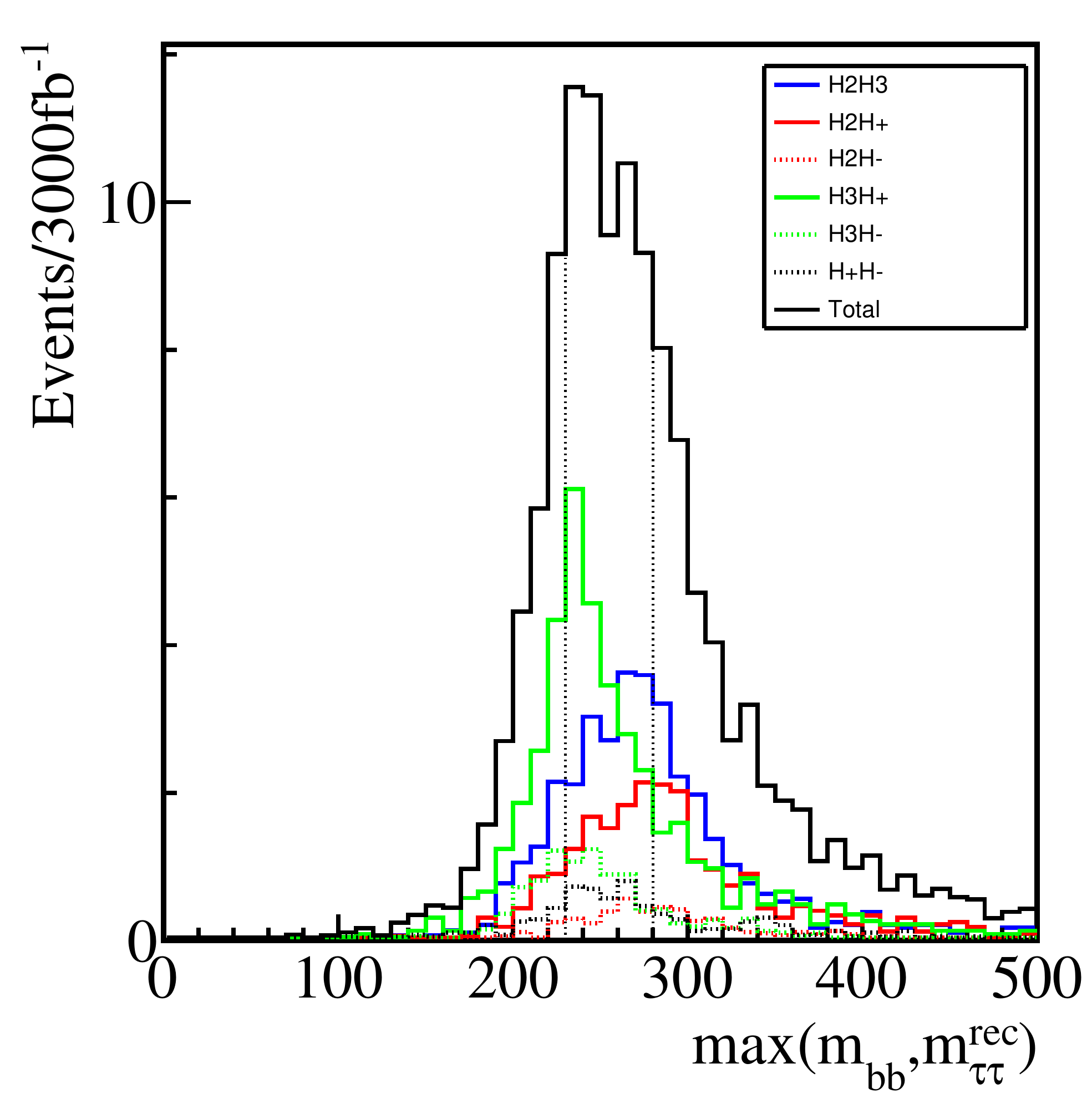}

\includegraphics[width=43mm]{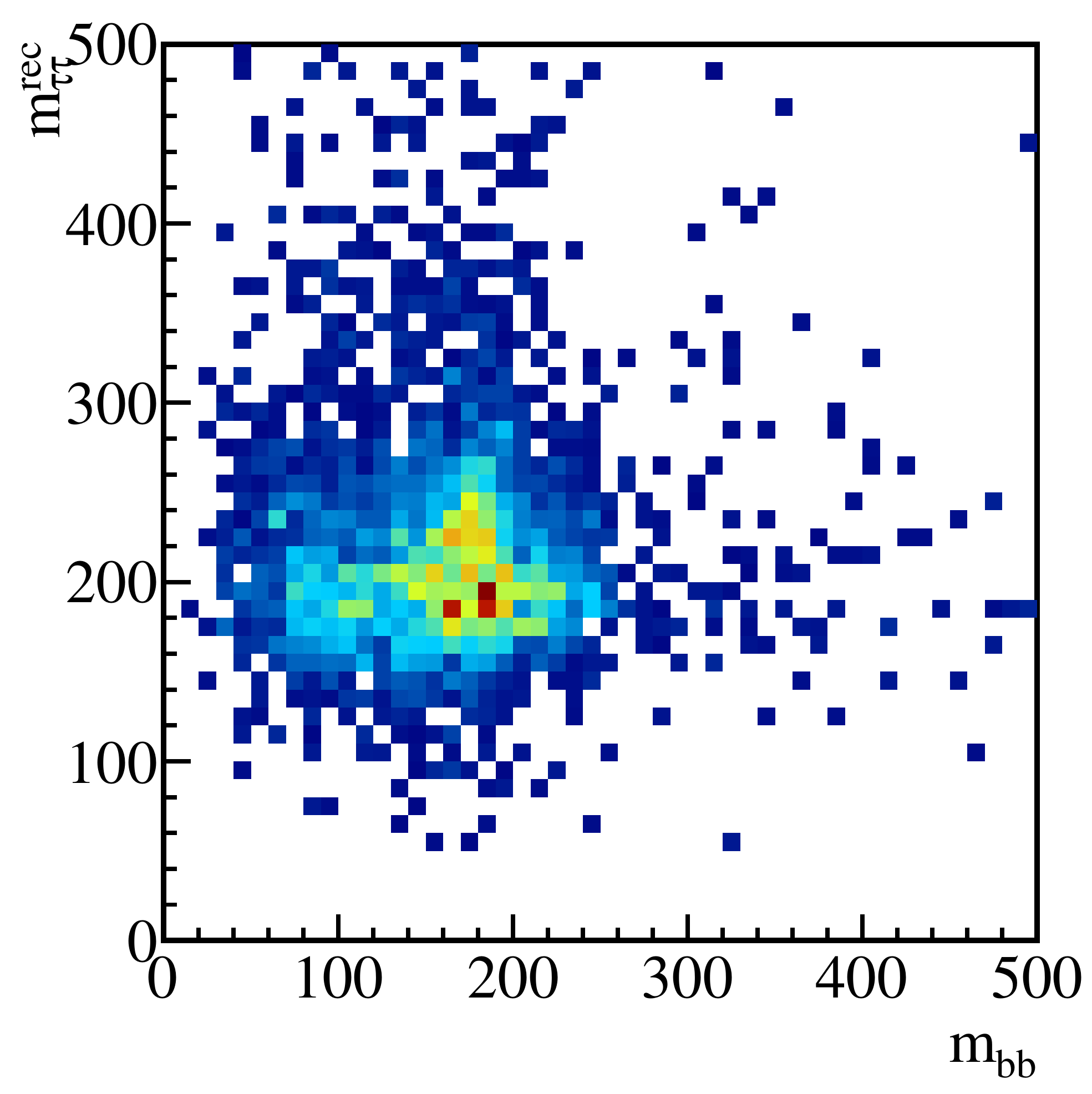}
\includegraphics[width=43mm]{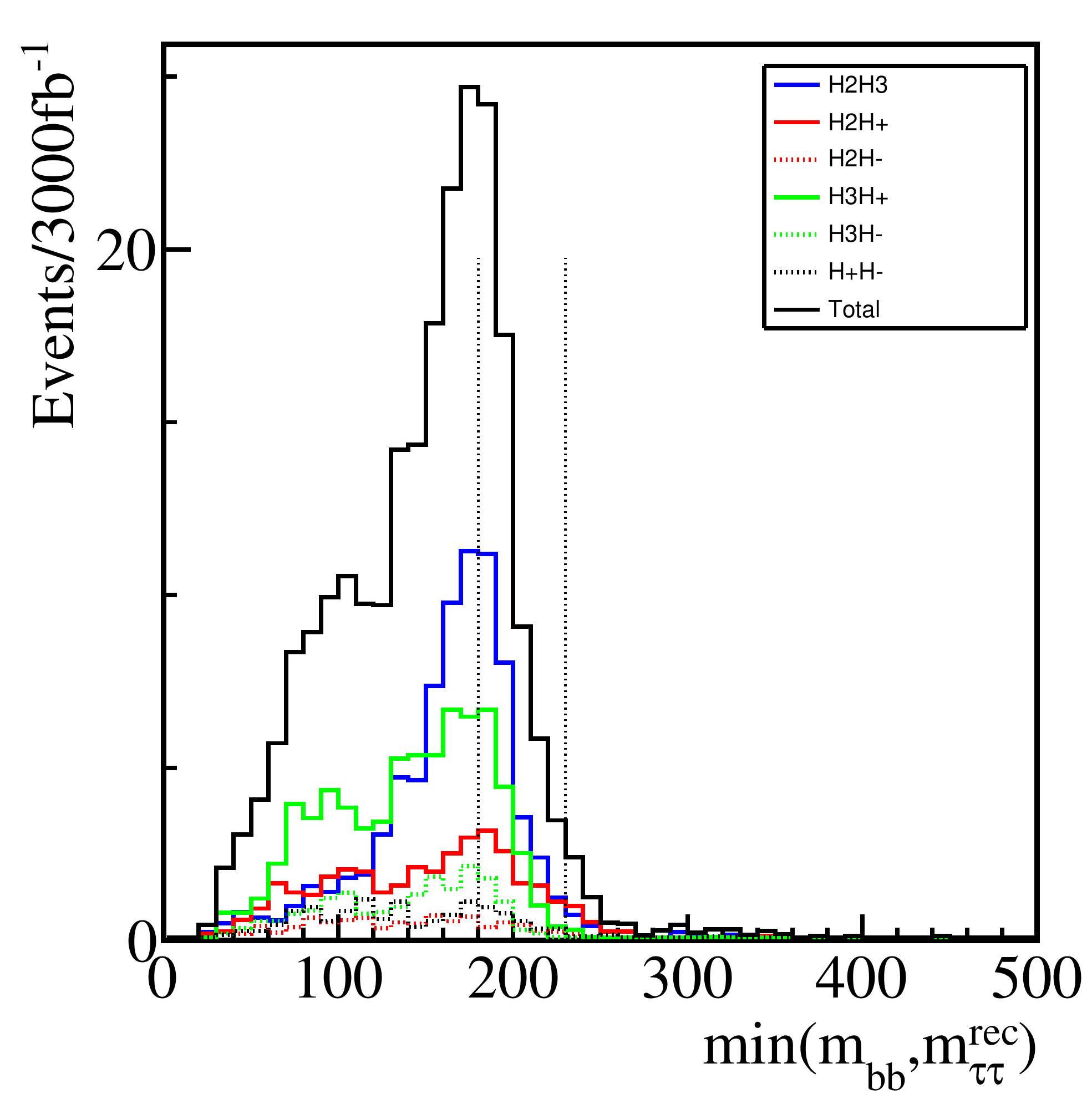}
\includegraphics[width=43mm]{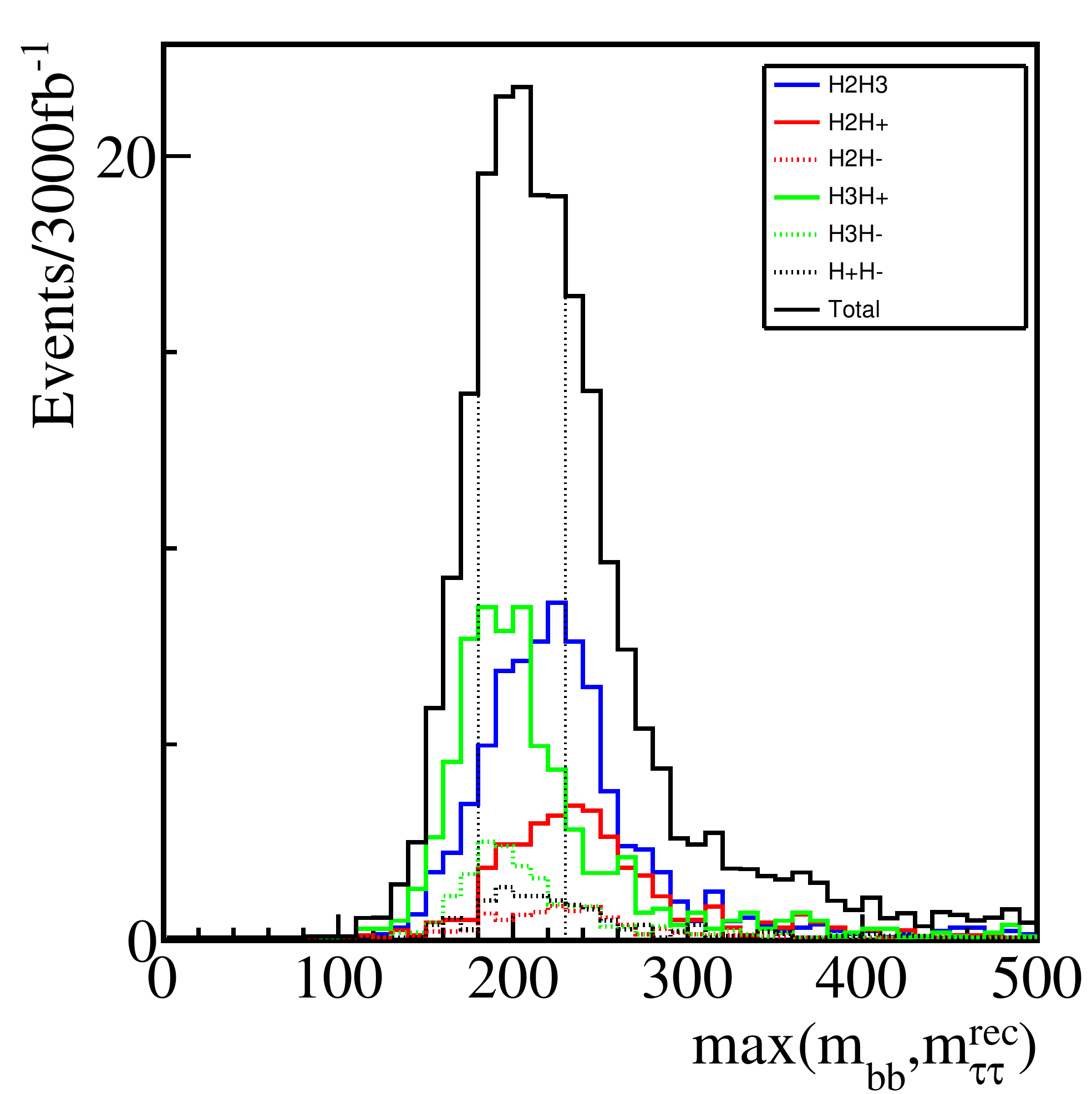}

\includegraphics[width=43mm]{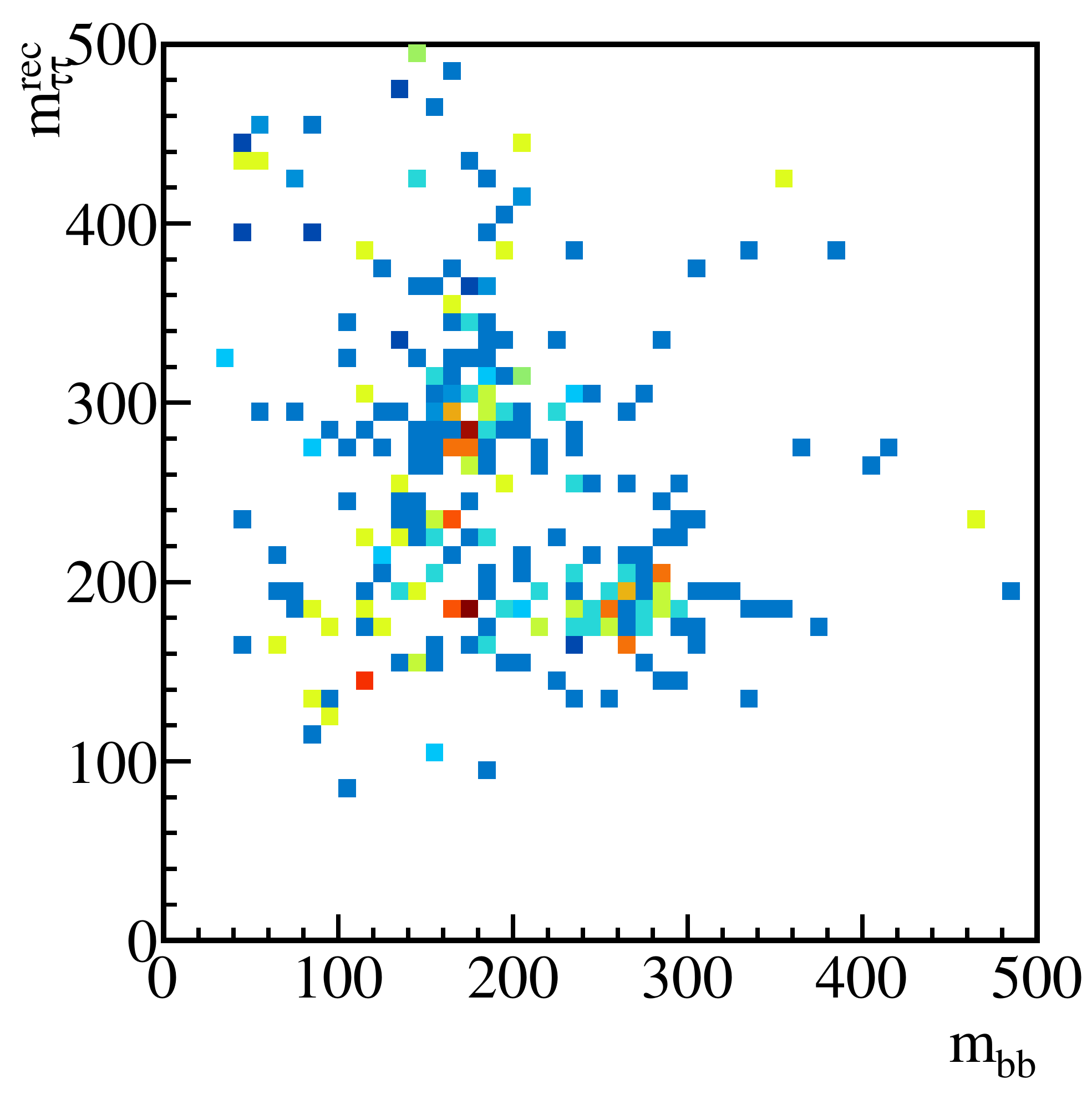}
\includegraphics[width=43mm]{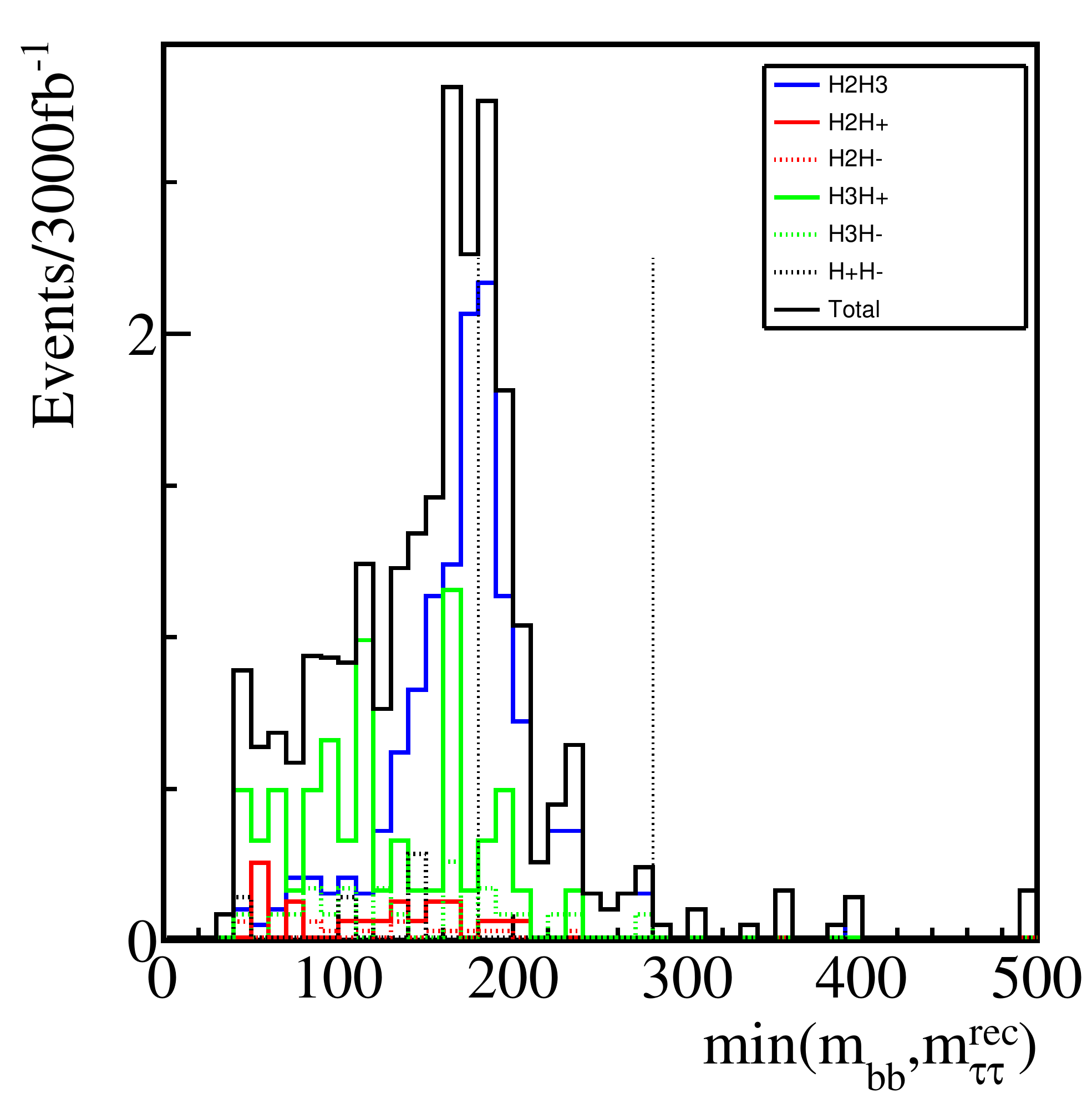}
\includegraphics[width=43mm]{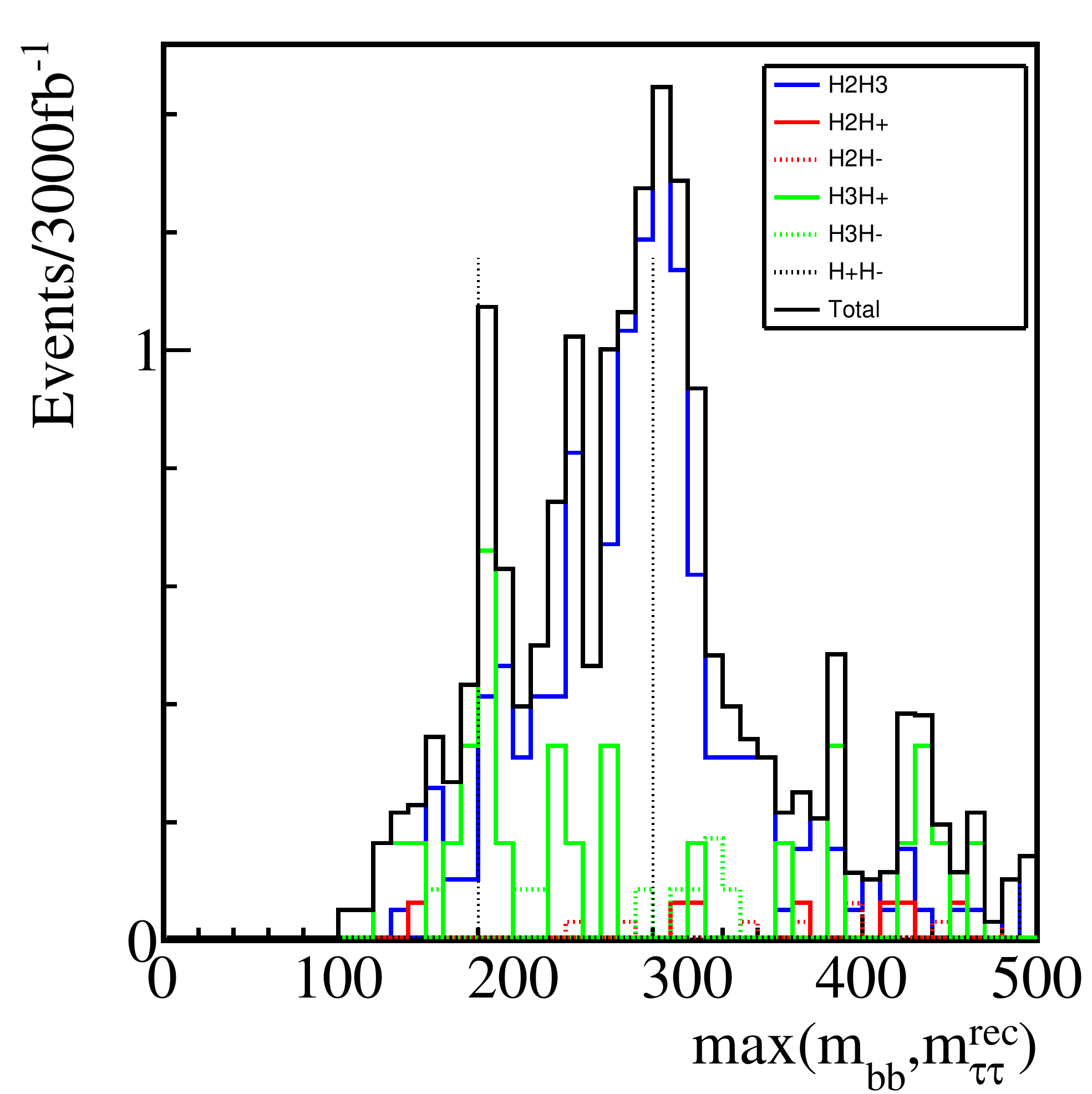}
\vspace{-0.2cm}
\caption{2-dimensional $m_{bb}$ vs. $m_{\tau\tau}^{\rm rec}$ distributions (left panels),
$\min(m_{bb}, m_{\tau\tau}^{\rm rec})$ distributions (central panels),  and
$\max(m_{bb}, m_{\tau\tau}^{\rm rec})$ distributions (right panels) in the $2b2\tau$ events at
BP1 - BP4 from top to bottom.
\vspace{-0.2cm}
}
\label{fig:mtautaumbb}
\end{figure}

We try to access the neutral Higgs boson masses using the $b\bar{b}\tau^+\tau^-$ mode, which is expected to be the dominant decay mode 
in the available parameter region. It is because the region with too large ratio of $\zeta_e/\zeta_d$ is already very constrained by the current LHC data as shown in the previous sections, which indicates that too large $R_\tau$ is not allowed unless the additional Higgs bosons are decoupled.
As a selection cut we require a event should contain 
at least two $b$-tagged jets and at least two tau-tagged jets with $p_T> 20$ GeV, and $|\eta|<2.5$.
We also require $E\!\!\!/_T > 50$ GeV. 
Since a visible hadronic tau jet carries only a part of the original tau lepton momentum due to the escaping neutrino momentum,
we adopt the collinear approximation~\cite{Rainwater:1998kj} to reconstruct the tau lepton momentum,
assuming the transverse missing momentum $\mathbf{p}\!\!\!/_T$ are generated solely by these neutrino momenta.
The explicit procedure is to obtain $c_1$ and $c_2$ by solving the following relations,
\begin{align}
&\mathbf{p}_{\tau_1} = (1 + c_1)  \mathbf{p}^{\rm vis}_{\tau_1}, \  \ \ 
\mathbf{p}_{\tau_2} = (1 + c_2)  \mathbf{p}^{\rm vis}_{\tau_2}\ \ \ \  (c_1, c_2> 0),
\cr
&\mathbf{p}\!\!\!/_T = c_1 \mathbf{p}^{\rm vis}_{T, \tau_1} + c_2  \mathbf{p}^{\rm vis}_{T,\tau_2}.
\label{coll}
\end{align}
We 
only accept the events where the above equation has a solution.
The larger the mass of the resonance is, the better 
this approximation provides the reconstructed tau momenta, since the momentum carried by the neutrino is aligned to the visible momentum.
With the reconstructed tau momenta and bottom momenta we can compute
$m_{\tau\tau}^{\rm rec}$ and $m_{bb}$ each event.
In Fig.~\ref{fig:mtautaumbb}, the 2-dimensional $m_{bb}$  vs. $m_{\tau\tau}$ distributions are shown as a scattering plot 
in the left panels for BP1 to BP4 from top to bottom.
The dense regions are depicted in red points.
In all cases, dense regions are observed either at the corresponding 
$(m_{H_2^0}, m_{H_3^0})$,  $(m_{H_3^0}, m_{H_2^0})$, and  $(m_{H_2^0}, m_{H_2^0})$ in each BP.

The central and right panels show the corresponding $\min(m_{bb}, m_{\tau\tau}^{\rm rec})$ and $\max(m_{bb}, m_{\tau\tau}^{\rm rec})$ projected distributions, respectively.
The expected number of events for 3~ab$^{-1}$ at HL-LHC including the effects of the selection cut and the efficiencies are shown in the plots.
The total and the breakdown of the contributions from the six production modes are plotted.
One can see that the $\min(m_{bb}, m_{\tau\tau}^{\rm rec})$ distribution exhibits the peak at the corresponding $m_{H_2^0}$ and 
the $\max(m_{bb}, m_{\tau\tau}^{\rm rec})$ distribution does at the $m_{H_3^0}$ in each BP.
For the benchmark points in the heavy $H^\pm$ scenarios~(BP1, BP2, BP3), all production modes contribute to the peak at $m_{H_2^0}$ in the 
$\min(m_{bb}, m_{\tau\tau})$ distributions. 
In the $\max(m_{bb}, m_{\tau\tau})$ distributions
only the $H_2 H_3$ and $H_2 H^\pm$ production modes 
contribute to the peak at $m_{H_3^0}$ as expected,
while all modes contribute to the peak at $m_{H_2^0}$.
The corresponding $m_{H_2^0}$ and $m_{H_3^0}$ values are indicated by the vertical dashed lines in each plot.

For the light $H^\pm$ scenario (BP4), only the $H_2 H_3$ production mode essentially contributes to these distributions since the decays involving charged Higgs bosons do not produce events to fulfill the $2b2\tau$ selection criteria. Therefore, 
the expected number of the signals is much smaller than the cases in the heavy $H^\pm$ scenario, 
although the BP4 has the lightest mass spectrum and the total production cross section is the largest among those of the four benchmark points.
Accessing the charged Higgs mass would be also possible, for example, 
following a similar procedure proposed in Ref.~\cite{Iguro:2019sly}, although we leave it for a future study.

As we have seen in this section, the masses of the additional Higgs bosons can be in principle determined 
using these characteristic peaks in the proposed distributions in the mass range we consider, whereas 
it would be rather difficult when the charged Higgs boson is the lightest 
among the three additional Higgs bosons.
For quantitative estimation of the precision of the mass measurements, clearly more dedicated signal and background analyses are necessary for our benchmark scenarios, as performed in the different occasion in Ref.~\cite{Cao:2003tr}.
It is beyond the scope of this paper, but we plan to do it as a future work.

\section{Conclusions and Dicussions } \label{sc:summary}

We have discussed the double-aligned THDM, where 
the coupling constants of the discovered Higgs boson to the SM particles are identical to 
those of the SM Higgs boson, and the FCNCs do not appear at tree level.
This scenario is well motivated by the scenario of EW baryogenesis, 
which is compatible with the current experimental data. 
In the double-aligned scenario, the additional Higgs bosons can mainly decay into a fermion pair or a lighter Higgs boson associated with a gauge boson depending 
on the $\zeta_f$ parameters and the masses of the additional Higgs bosons. 
We have discussed the behavior of the branching ratios in details
especially for the CP conserving case.
We then have explicitly shown the critical values of the $\zeta_f$ parameters, 
at which the sum of the branching ratios to the fermion pairs is 50\%. 
In addition, we have investigated the current constraints on the parameter space from the various flavor experiments, 
i.e., $B \to X_s \gamma$, $B_s \to \mu\mu$ and the tau decay, 
and from the searches in the multi-lepton final states at the LHC Run-II experiment. 
For the collider constraints, we particularly focused on the EW pair production of the additional Higgs bosons with their masses below $2m_t$, whose 
production cross sections are simply determined by the masses of the Higgs bosons. 
It has been found that a large portion of the parameter space is already excluded by the current LHC data  
when the leptonic decays of the additional Higgs bosons are dominant.
As a result, we have found that the remaining parameter space should satisfy $|\zeta_d/\zeta_e| \gtrsim {\cal O}(1)$.
Our results can be interpreted to the scenario in the Type-X THDM as a special case. 
In the case where all the additional Higgs bosons are degenerate in mass, we have found that 
the mass below 190 GeV has 
been already excluded by the data from the LHC Run-II in the Type-X THDM. If there is the mass difference the constraint is 
stronger. Namely, the charged Higgs mass below 250~GeV (300~GeV) has been excluded for $\Delta m = 100$~GeV 
(150~GeV).
Furthermore, at the HL-LHC we have shown that most of the parameter region in the Type-X THDM would be explored.
Since the available parameter region tends to predict 
a significant amount of the $b\bar{b}$ branching ratios for the additional Higgs bosons,
we have demonstrated the reconstruction of the masses of additional 
Higgs bosons from the $b\bar{b}\tau^+\tau^-$ final state in a few bench mark points.
Extension to the analysis with CP violating phases will be performed elsewhere.

\begin{acknowledgments}

This work was supported, in part, by the Grant-in-Aid on Innovative Areas, the Ministry of Education, Culture, Sports, Science and Technology, No. 16H06492, and by the JSPS KAKENHI Grant No.~20H00160 [SK, MT], 
the Grant-in-Aid for Scientific Research\,C, No.~18K03611~[MT] 
and the Grant-in-Aid for Early-Career Scientists, No.~19K14714 [KY]. 

\end{acknowledgments}

\clearpage

\appendix
\section{The results for $\zeta_u=0$}\label{sec:zetau0}
In this appendix, we show the plots as Fig.~\ref{fig:comb1} and Fig.~\ref{fig:comb2} in the heavy charged Higgs scenario and in 
the light charged Higgs scenario for $\zeta_u = 0$. Due to the suppression of the $H_{2,3} \to c\bar{c}$ and $H^\pm \to tb$ decay modes, 
stronger constraints are in general obtained from the LHC multi-lepton searches.

\begin{figure}[h!]
\hspace{11.8cm} {\includegraphics[width=4.55cm]{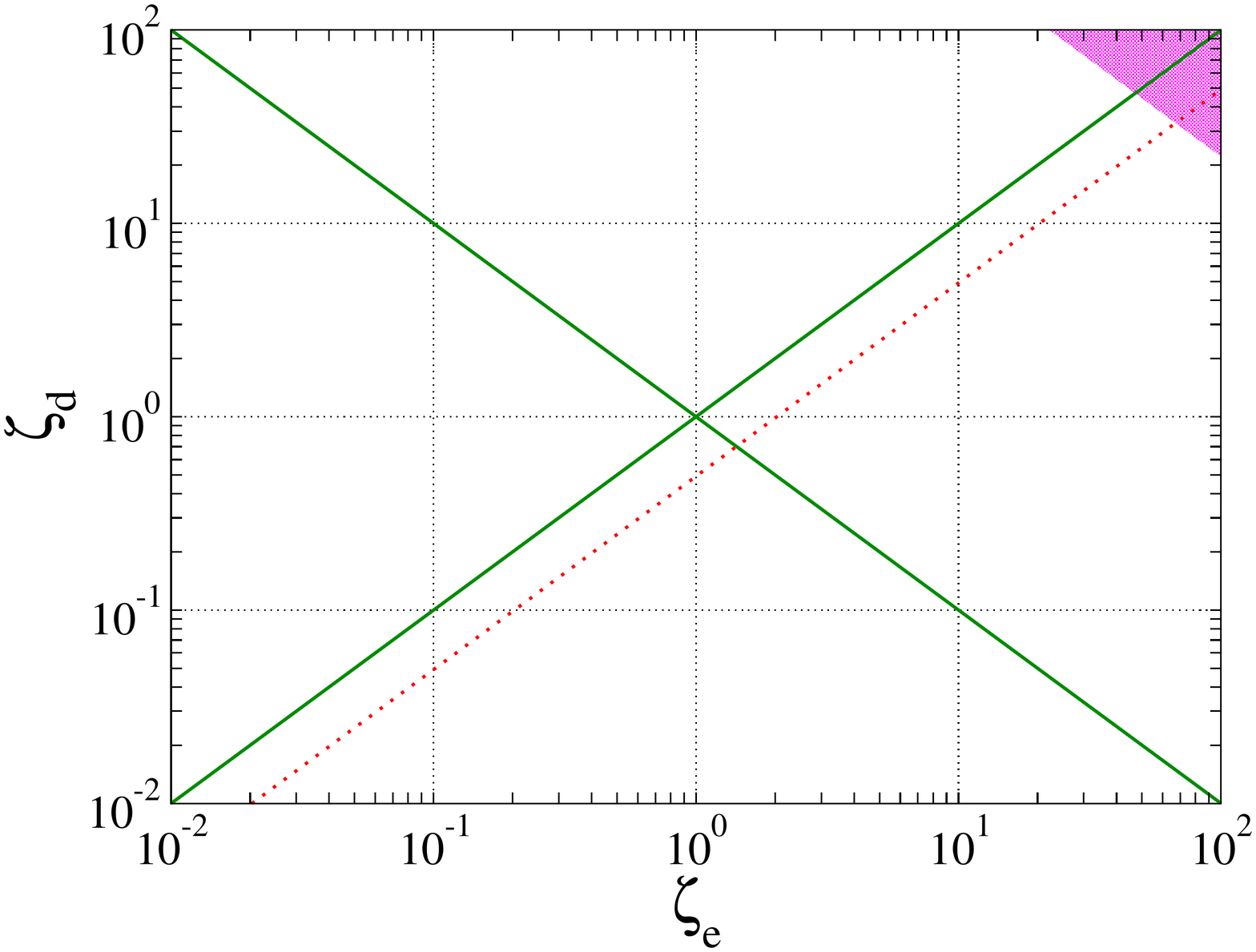}~\hspace{-0.5cm}~\ }\\
\hspace{7.5cm} 
{
\includegraphics[width=4.55cm]{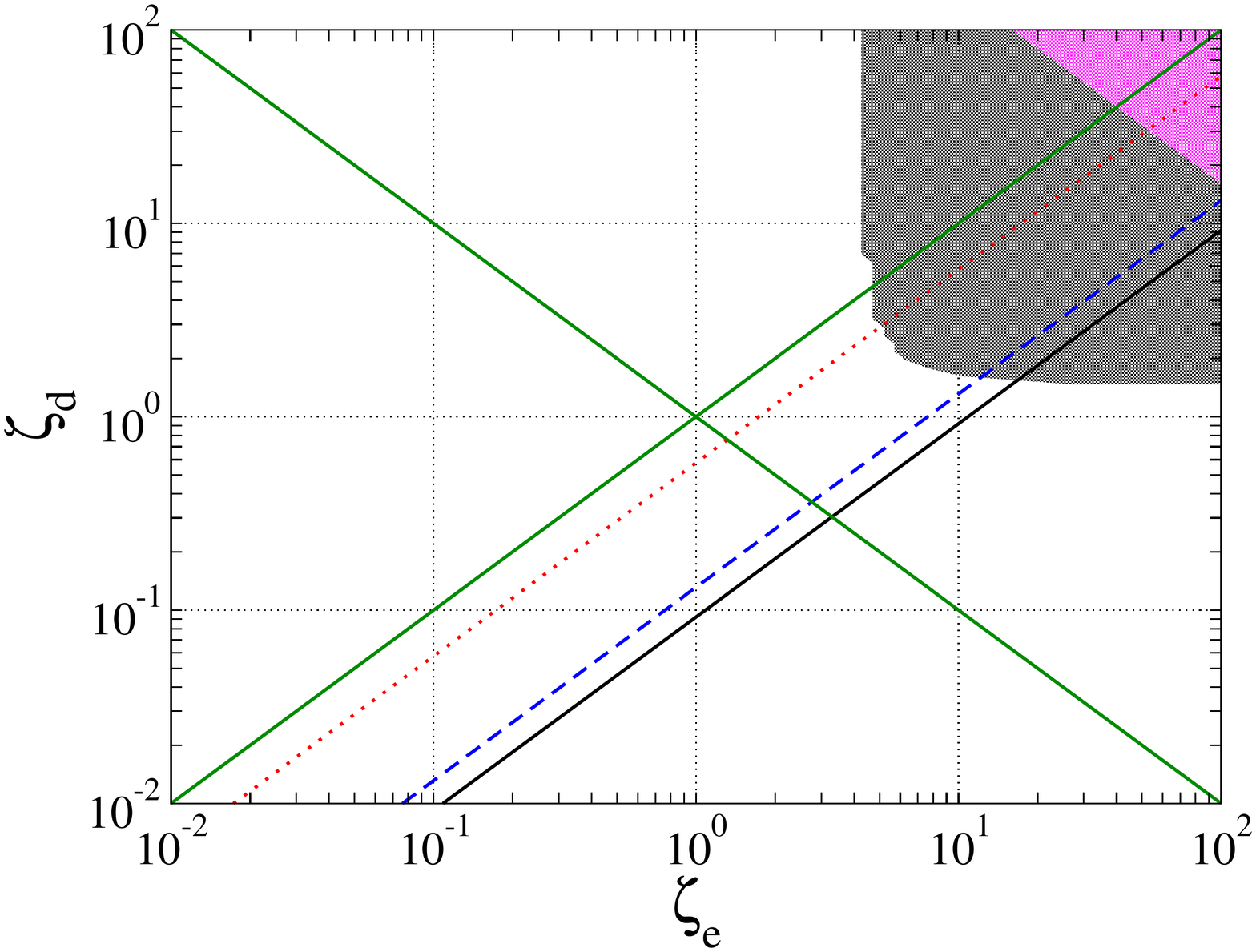}\hspace{-0.5cm}
\includegraphics[width=4.55cm]{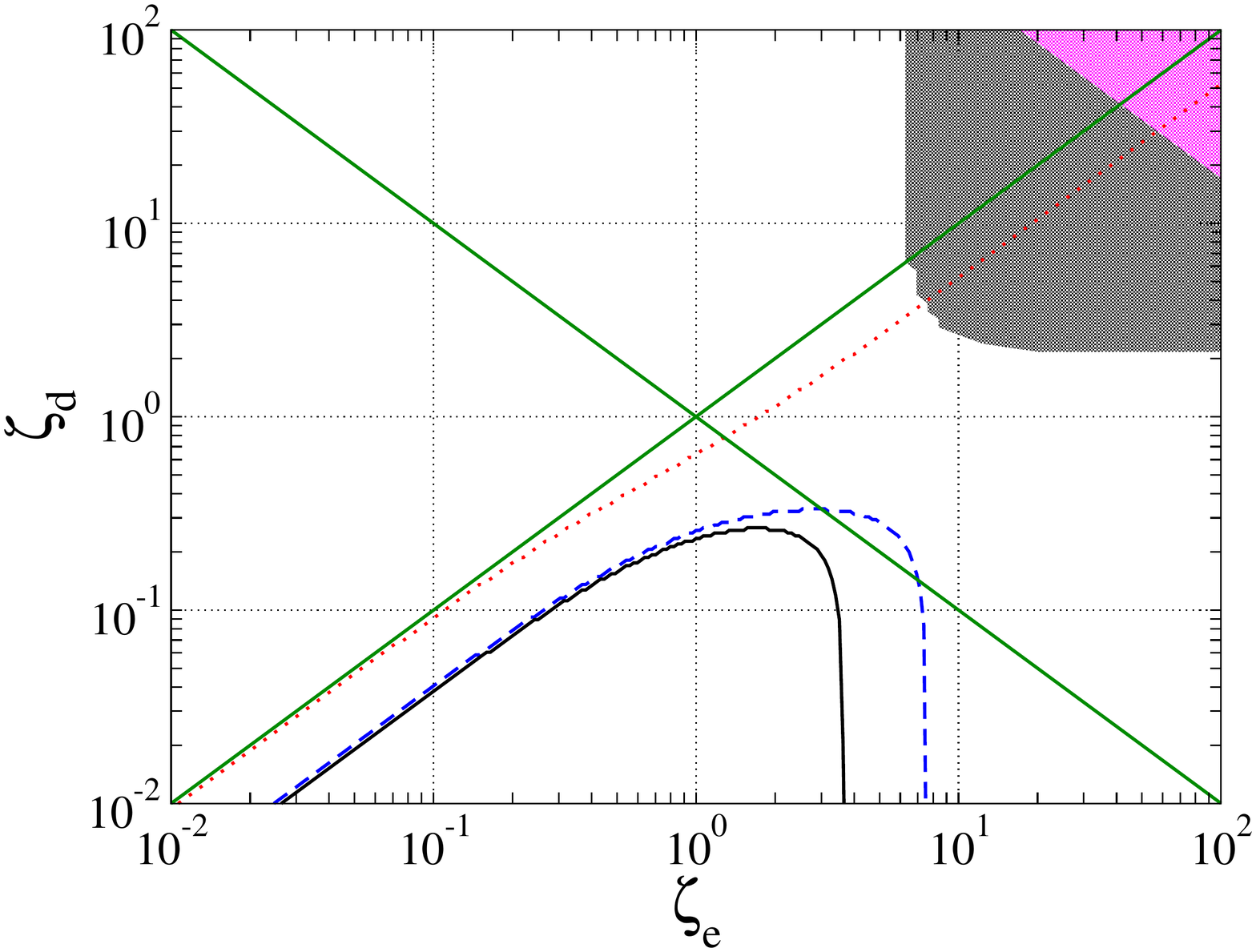}\hspace{-0.5cm}
}
\\
 \hspace{3.35cm}
{
\includegraphics[width=4.55cm]{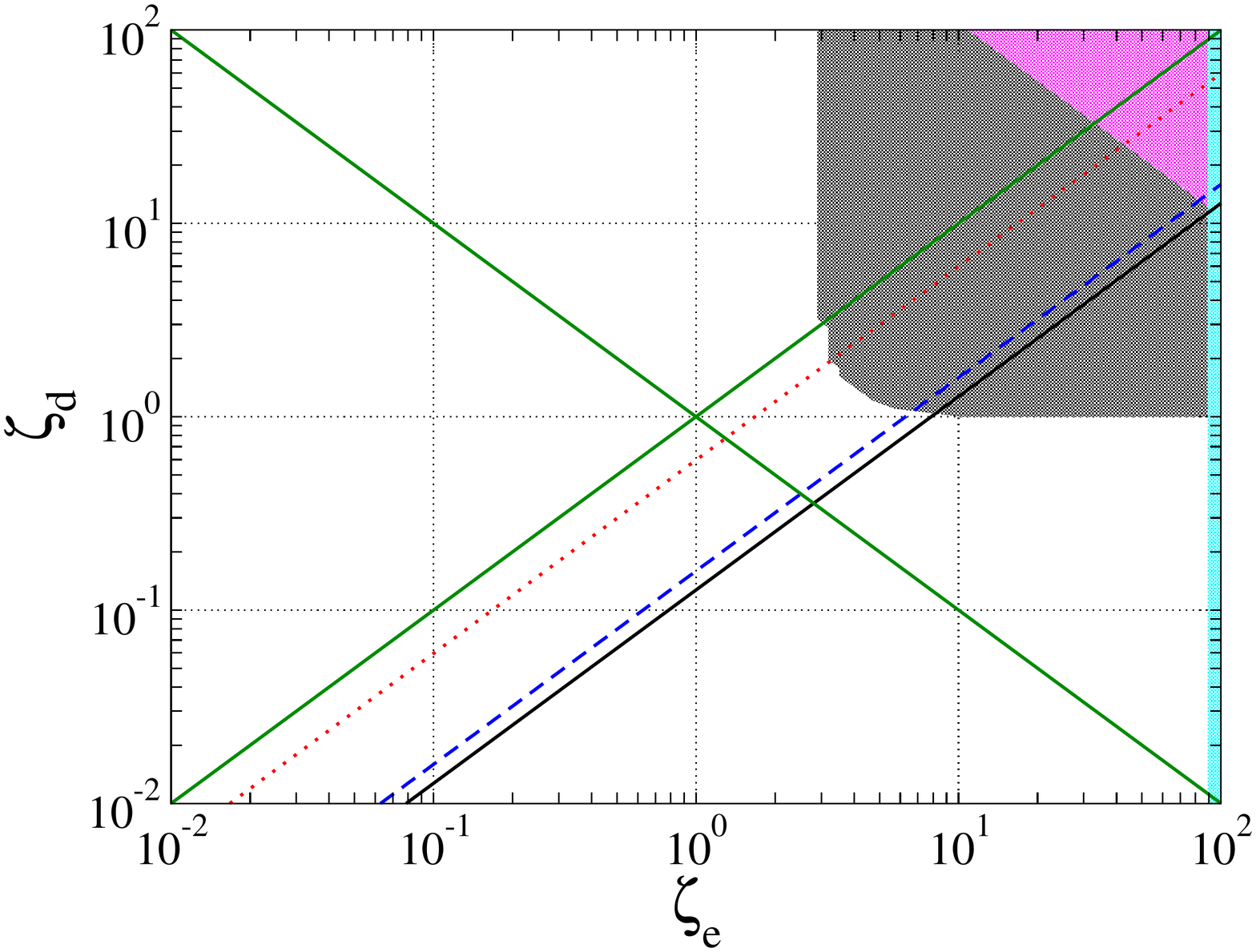}\hspace{-0.5cm}
\includegraphics[width=4.55cm]{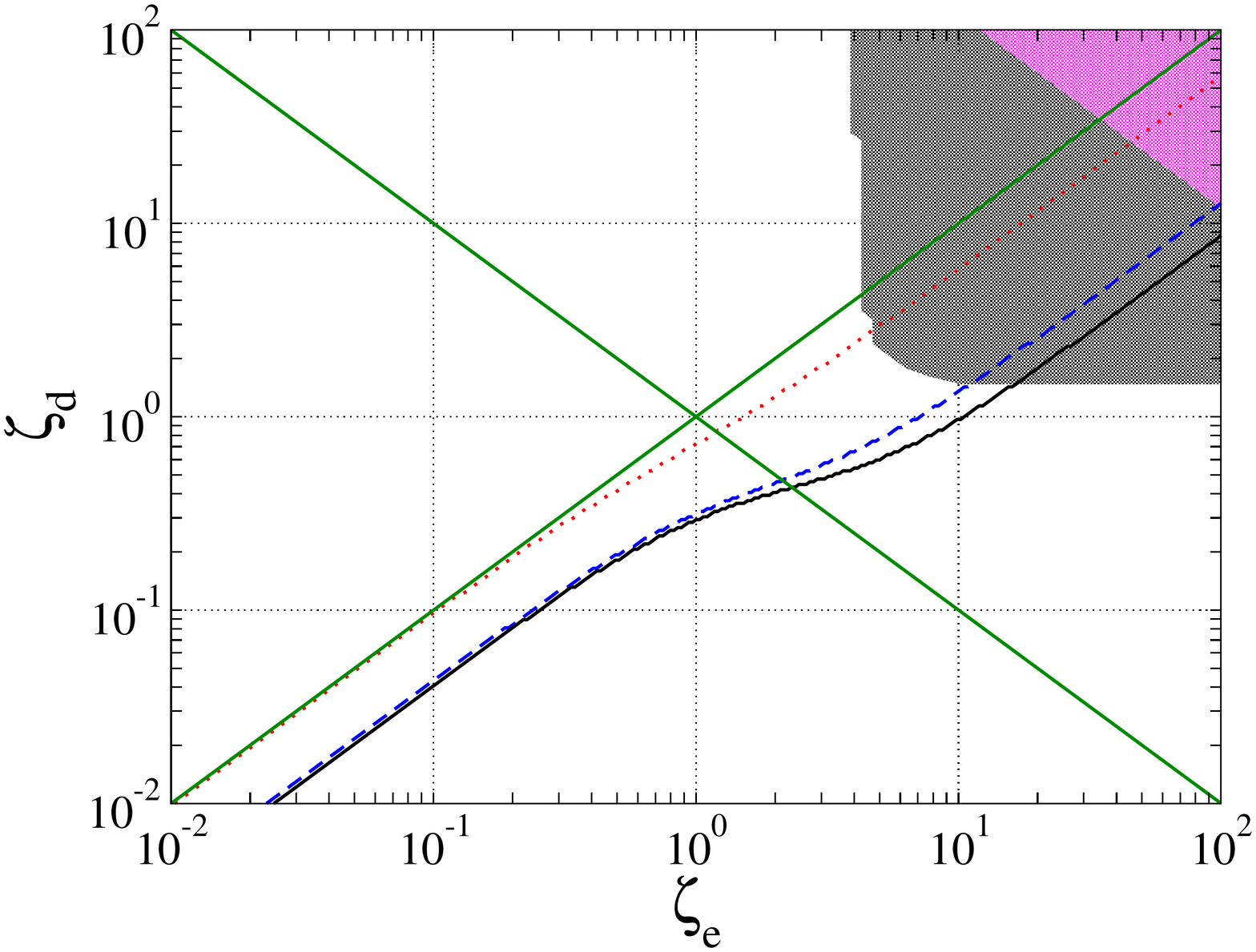}\hspace{-0.5cm}
\includegraphics[width=4.55cm]{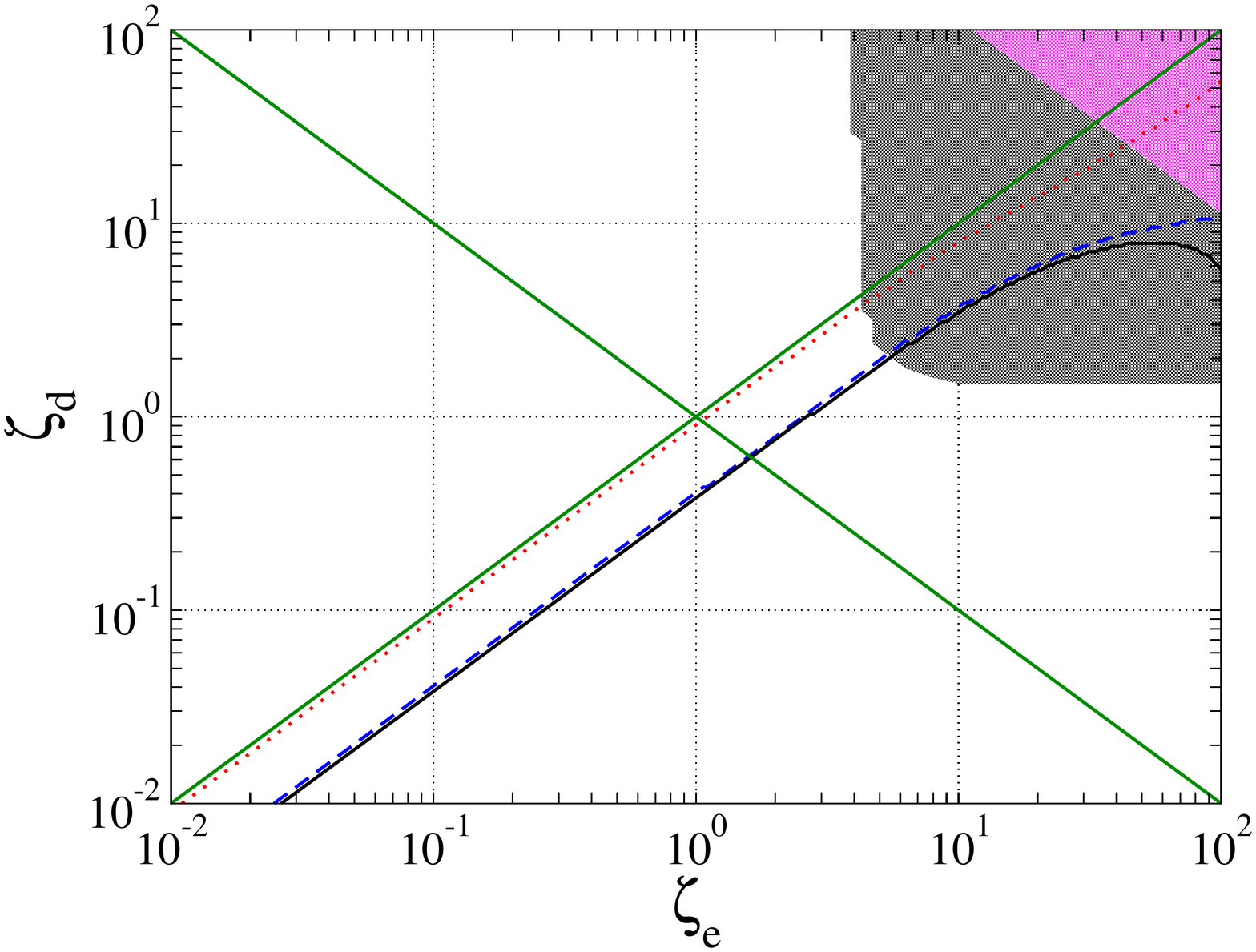}\hspace{-0.5cm}
}
\\
 \hspace{-0.7cm}
 {
 \includegraphics[width=4.55cm]{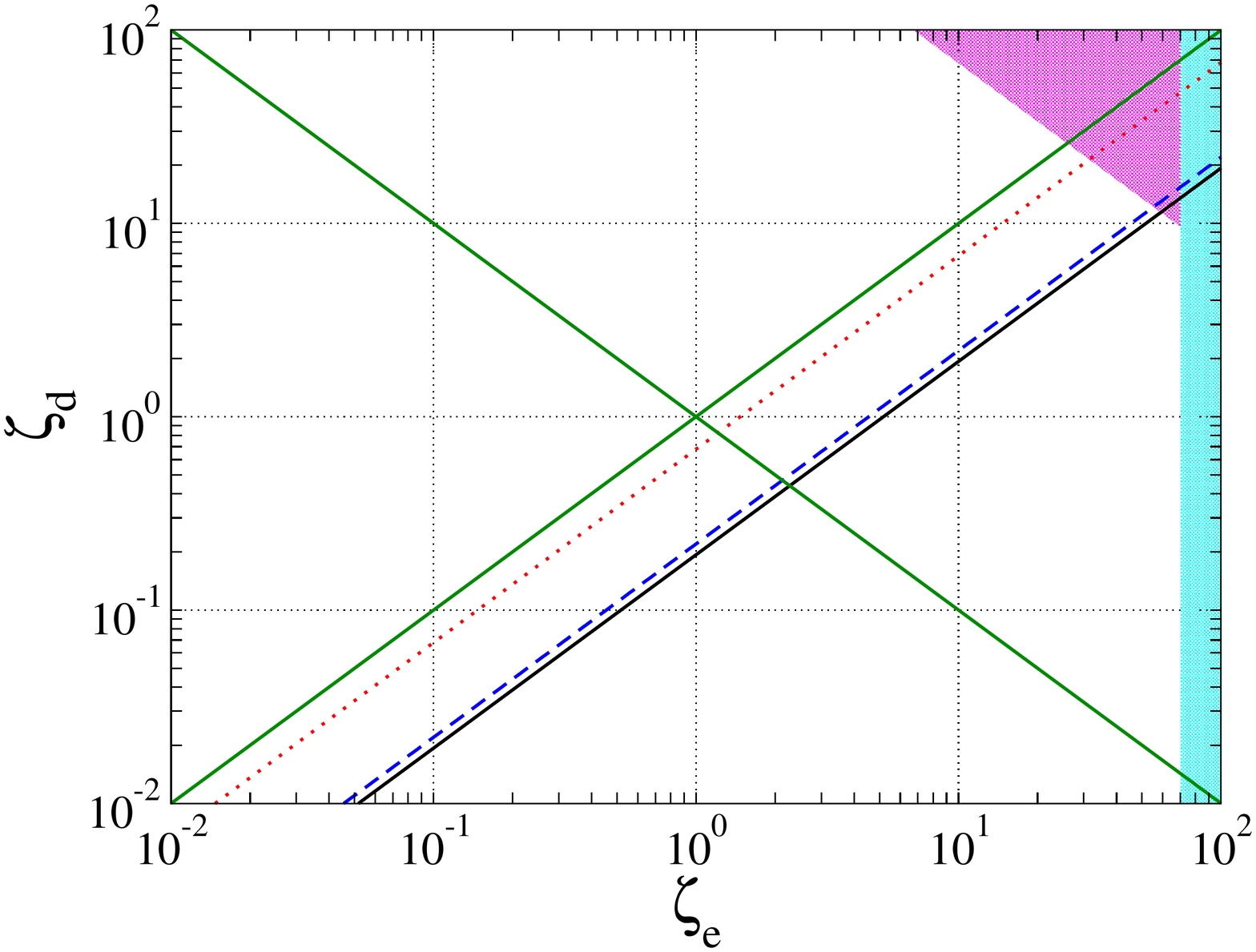}\hspace{-0.5cm}
 \includegraphics[width=4.55cm]{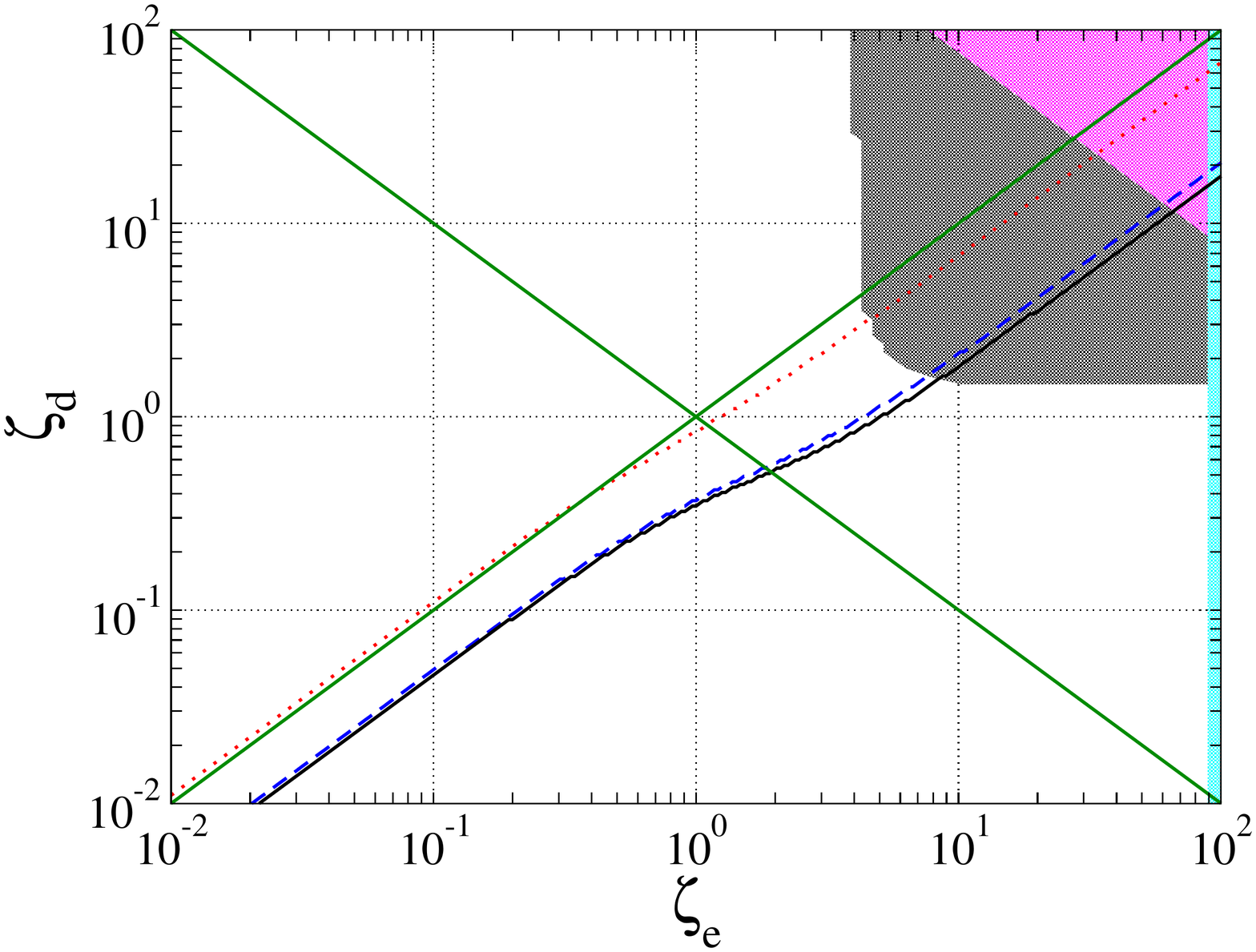}\hspace{-0.5cm}
 \includegraphics[width=4.55cm]{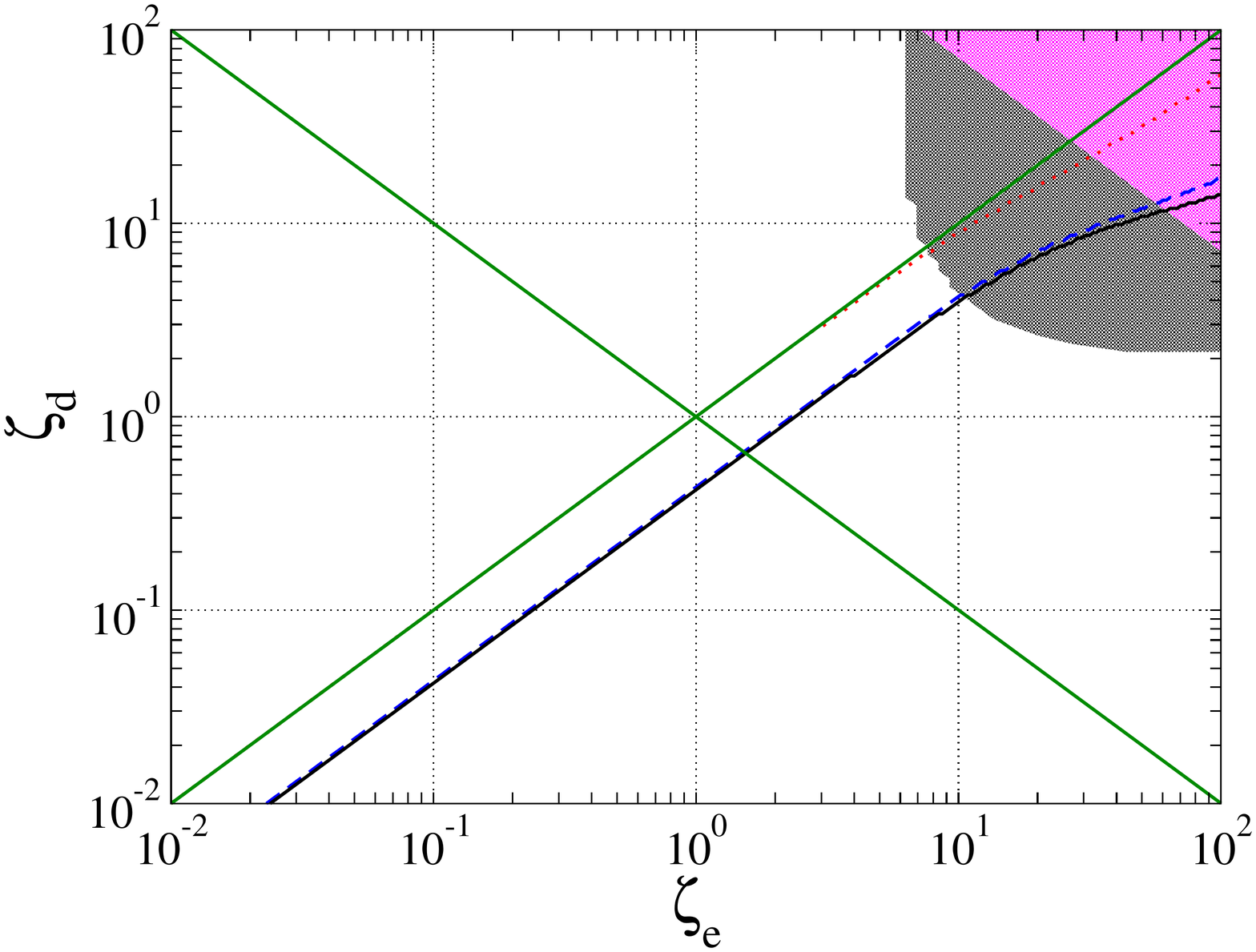}\hspace{-0.5cm}
 \includegraphics[width=4.55cm]{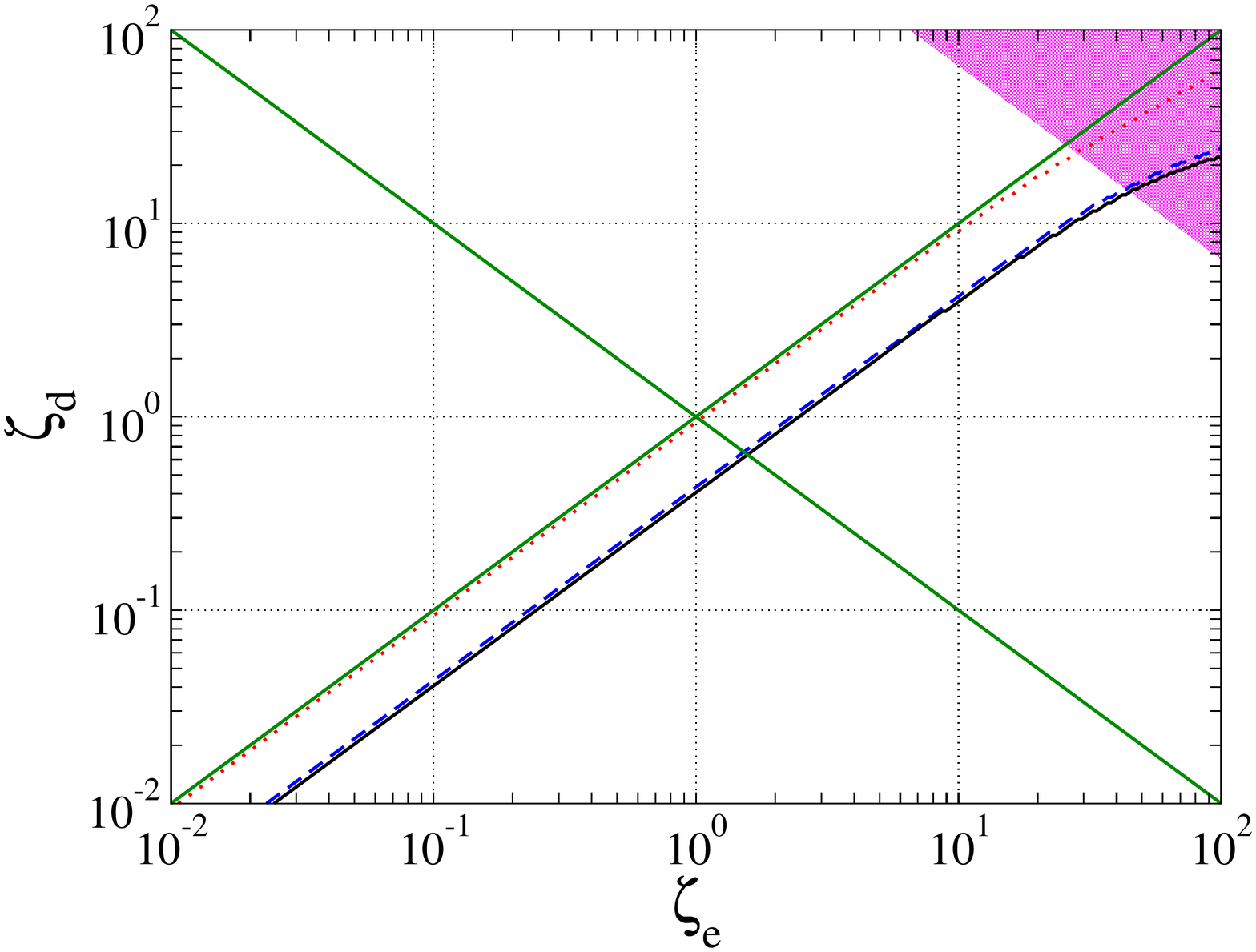}\hspace{-0.5cm}
}
\caption{The same plots as Fig.~\ref{fig:comb1} but for $\zeta_u = 0$ in the heavy $H^\pm$ scenario ($m_{H_2^0} \le m_{H_3^0}= m_{H^\pm}$). 
}
\label{fig:comb1_zetau0}
\end{figure}
\begin{figure}[h!]
\hspace{11.8cm} {\includegraphics[width=4.55cm]{./fig/330_330_330_combined_zetau0.pdf}~\hspace{-0.5cm}~\ }\\
\hspace{7.5cm} 
{
\includegraphics[width=4.55cm]{./fig/280_280_280_combined_zetau0.pdf}\hspace{-0.5cm}
\includegraphics[width=4.55cm]{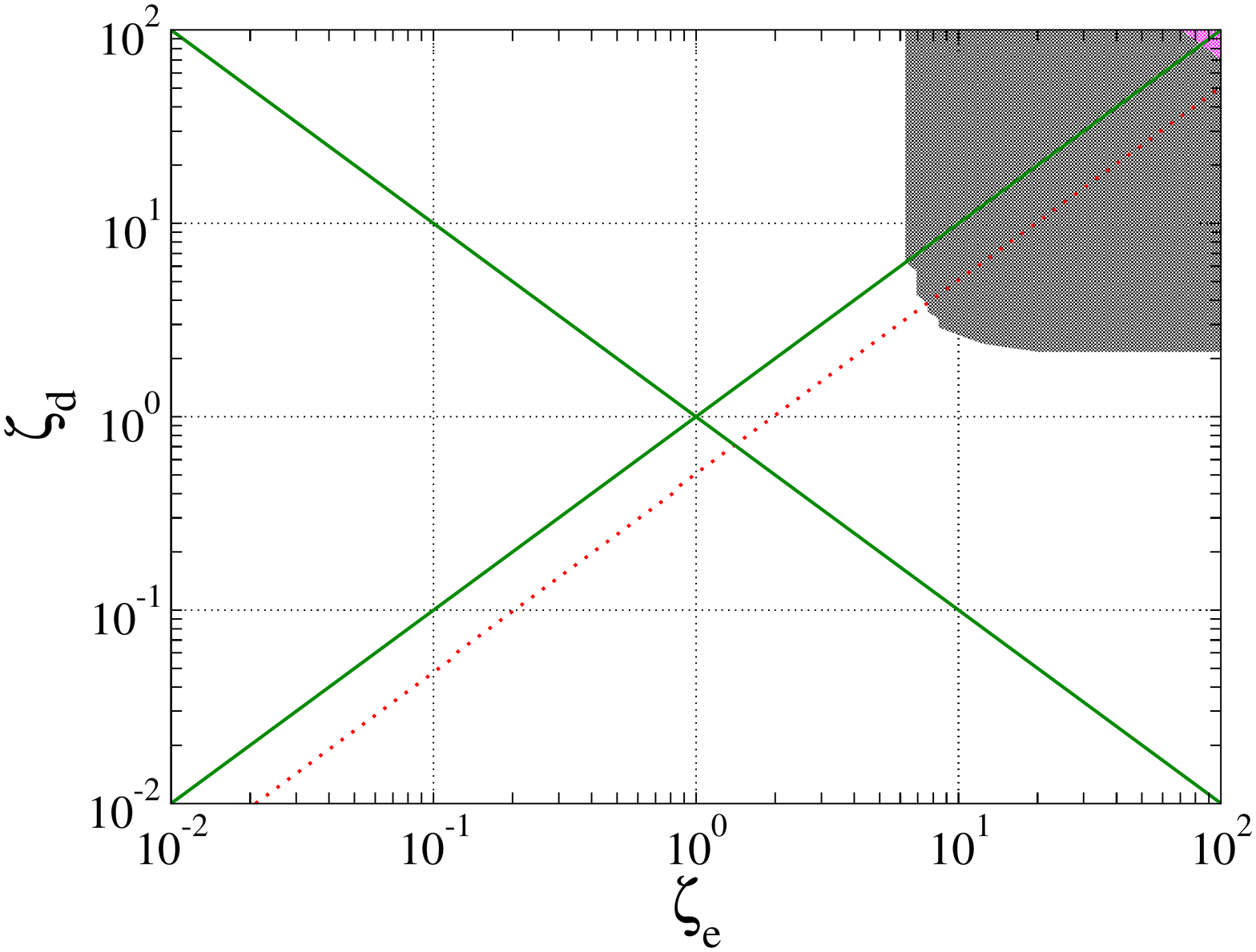}\hspace{-0.5cm}
}
\\
 \hspace{3.35cm}
{
\includegraphics[width=4.55cm]{./fig/230_230_230_combined_zetau0.pdf}\hspace{-0.5cm}
\includegraphics[width=4.55cm]{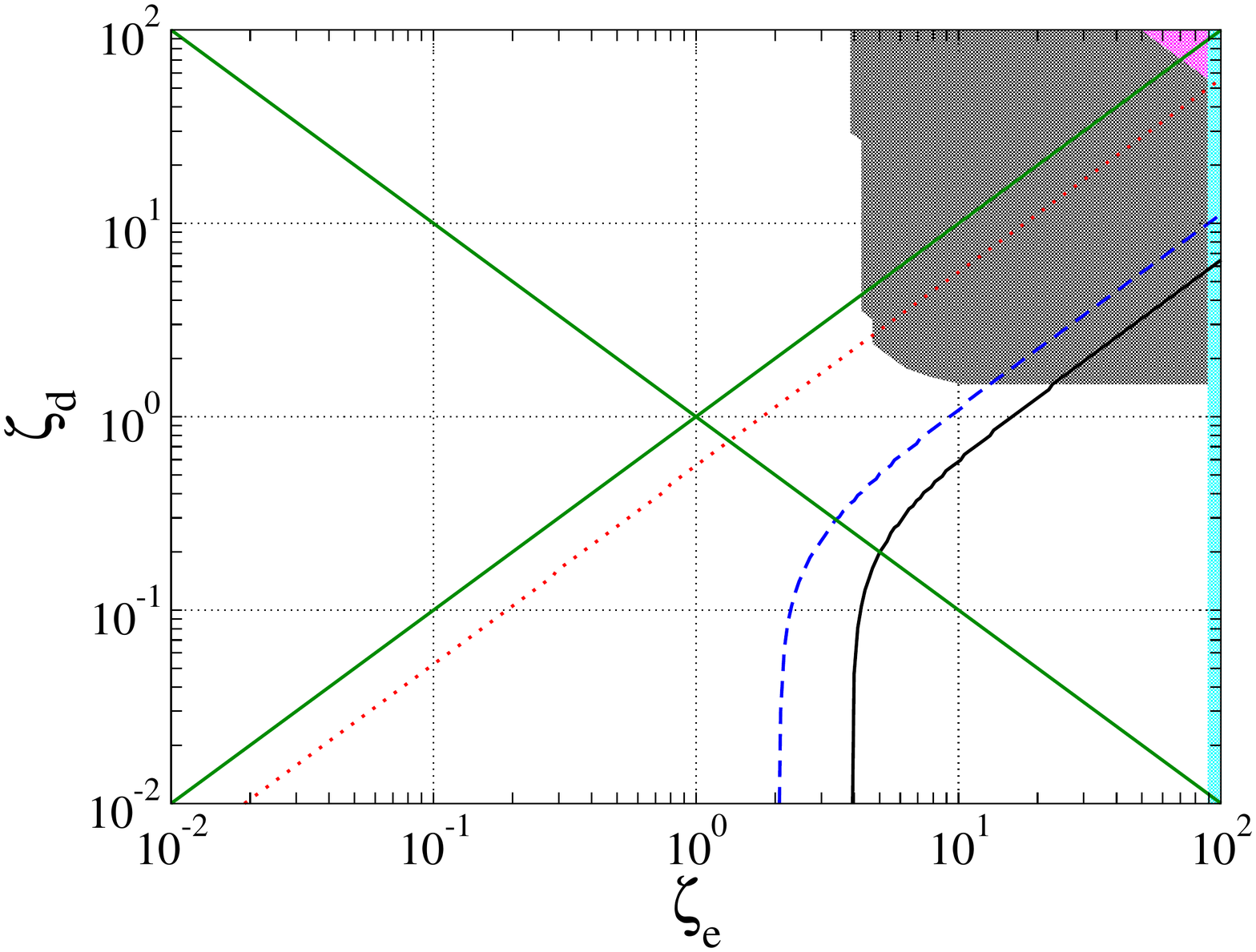}\hspace{-0.5cm}
\includegraphics[width=4.55cm]{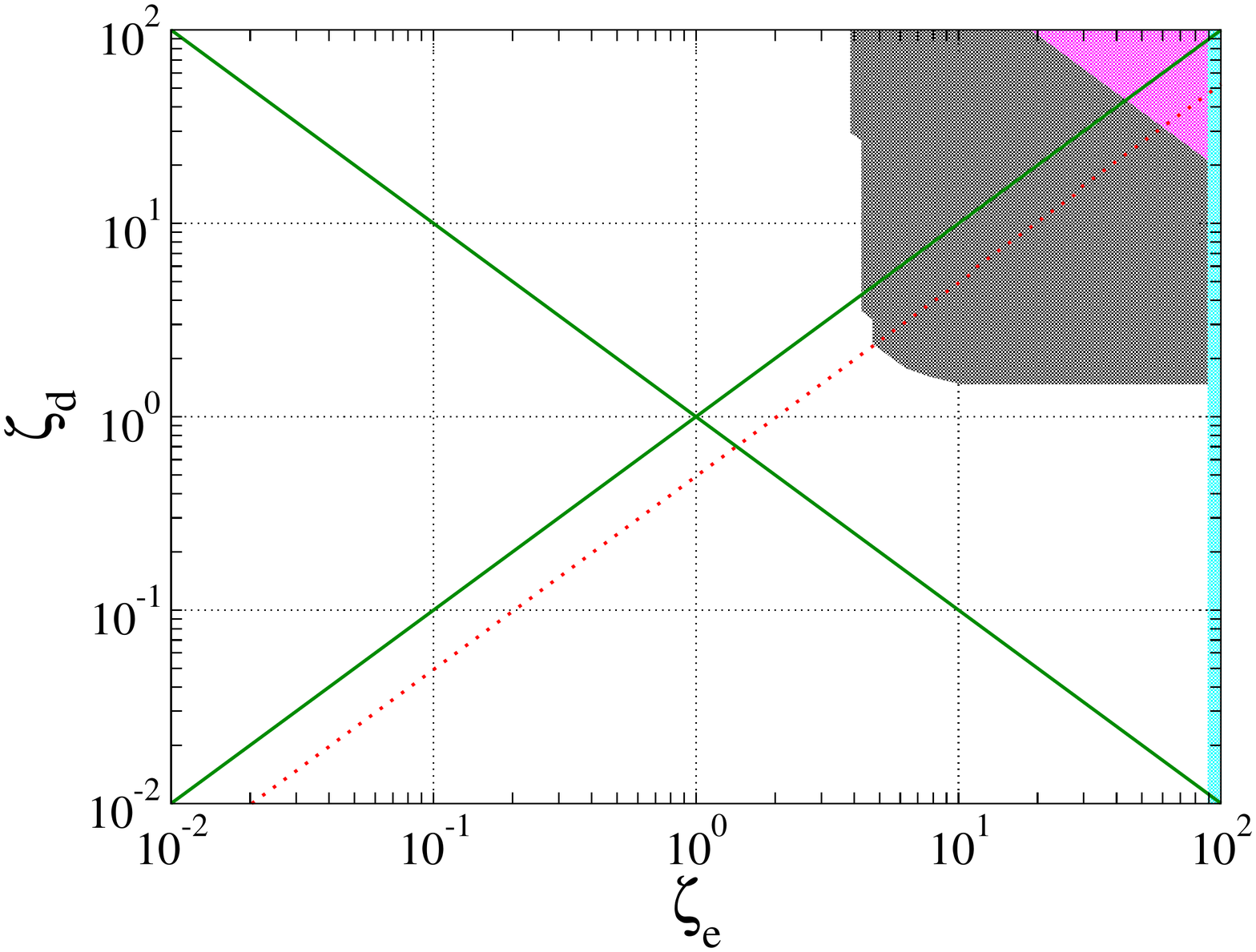}\hspace{-0.5cm}
}
\\
 \hspace{-0.7cm}
 {
 \includegraphics[width=4.55cm]{./fig/180_180_180_combined_zetau0.pdf}\hspace{-0.5cm}
 \includegraphics[width=4.55cm]{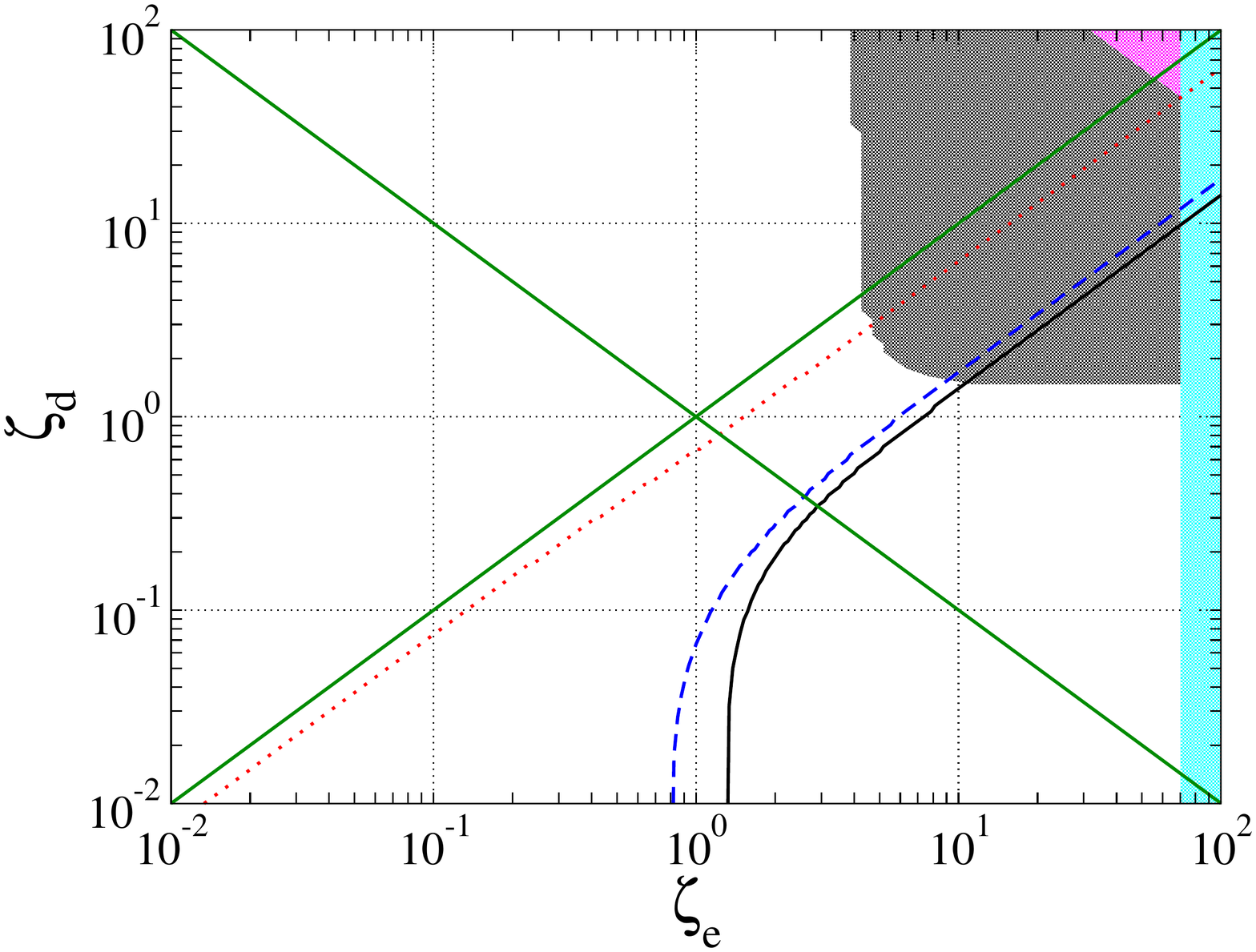}\hspace{-0.5cm}
 \includegraphics[width=4.55cm]{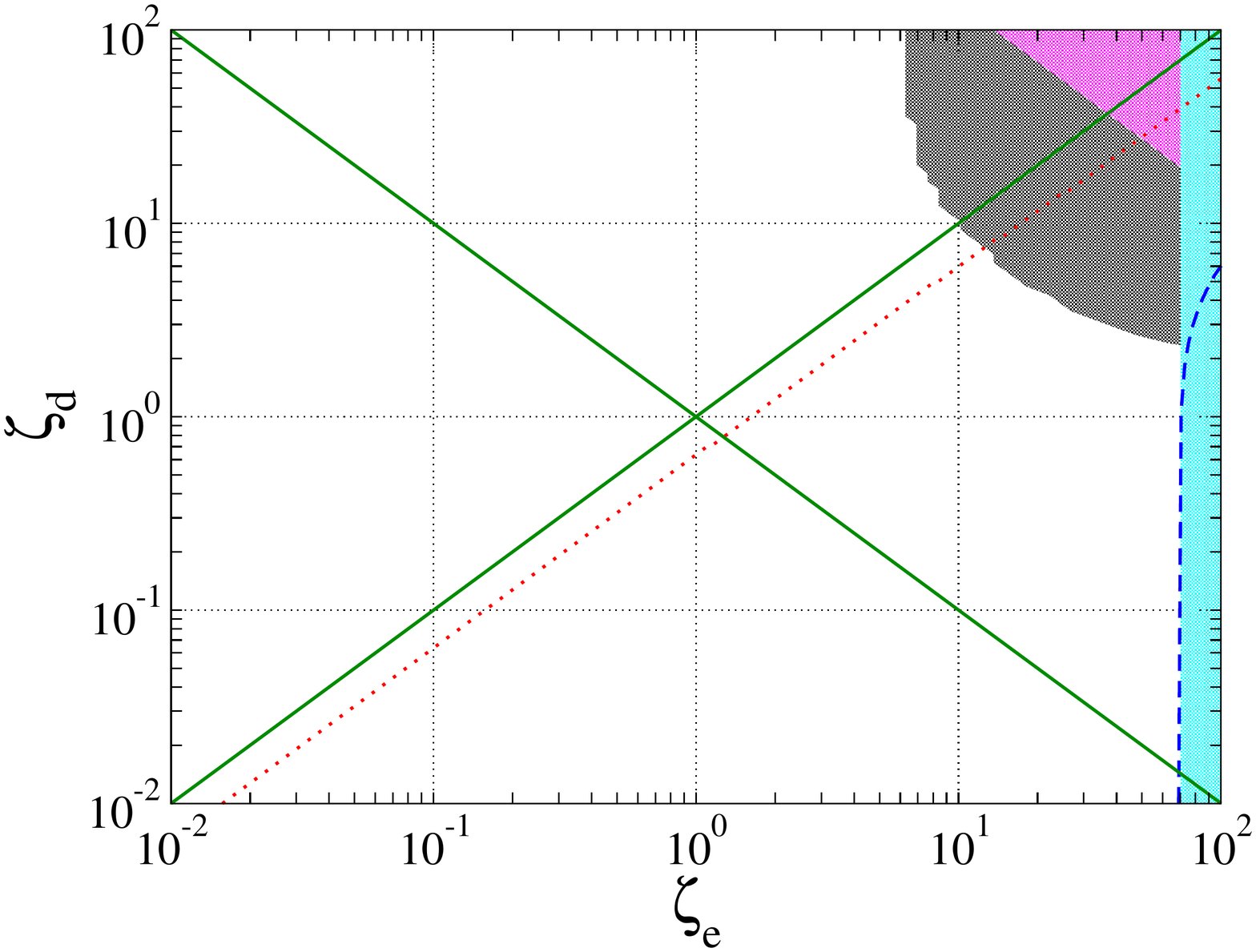}\hspace{-0.5cm}
 \includegraphics[width=4.55cm]{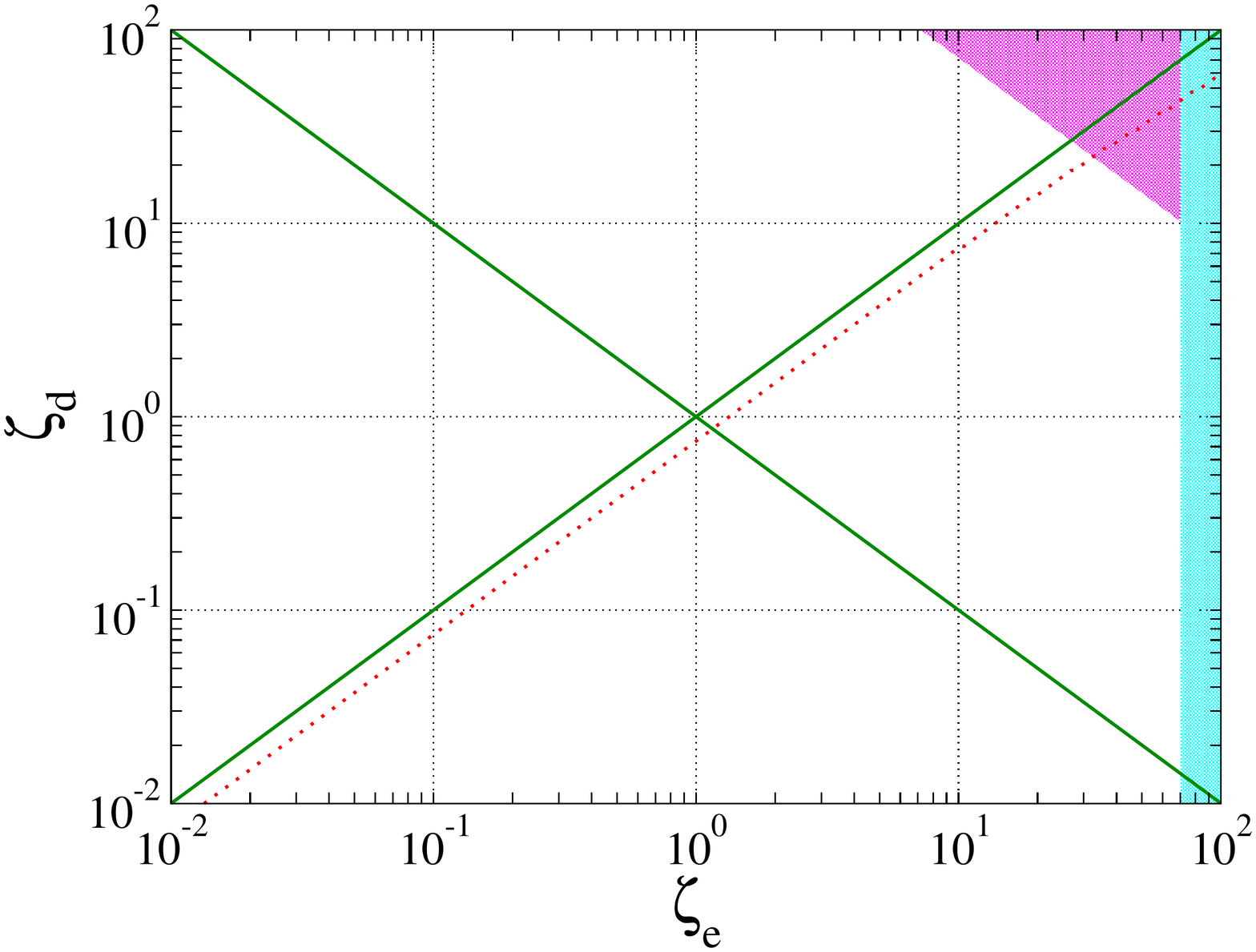}\hspace{-0.5cm}
}
\caption{The same plots as Fig.~\ref{fig:comb2} but for $\zeta_u = 0$ in the light $H^\pm$ scenario ($m_{H_2^0}= m_{H^\pm} \le m_{H_3^0}$). 
}
\label{fig:comb2_zetau0}
\end{figure}

\bibliography{references}

\end{document}